\documentclass[letterpaper,11pt]{article}
\pdfoutput=1

\usepackage{jheppub}
\usepackage{subfig}
\usepackage[dvipsnames]{xcolor}
\usepackage[normalem]{ulem}

\usepackage{mathtools}

\def\Fig#1{fig.~{\ref{#1}}}
\def\eq#1{eq.~(\ref{#1})}

\def\App#1{Appendix~\ref{#1}}
\def\ord#1{\mathcal{O}(#1)}
\DeclareRobustCommand{\Sec}[1]{sec.~\ref{#1}}
\DeclareRobustCommand{\Eq}[1]{eq.~(\ref{#1})}

\def\cN{\mathcal{N}}

\newcommand{\de}{\delta}

\newcommand{\eps}{\epsilon}

\def \as {\relax\ifmmode\alpha_s\else{$\alpha_s${ }}\fi}

\def\img{{\rm i}}
\newcommand{\df}{\mathrm{d}}
\newcommand{\rmd}{\mathrm{d}}
\newcommand{\nn}{\nonumber}

\definecolor{darkgreen}{rgb}{0.13,0.55,0.13}

\allowdisplaybreaks[2]

\begin{document}

\title{Multi-Collinear Splitting Kernels for Track Function Evolution}

\author[1]{Hao Chen,}
\author[2,3]{Max Jaarsma,}
\author[1]{Yibei Li,}
\author[4]{Ian Moult,}
\author[2,3]{Wouter Waalewijn,}
\author[1]{Hua Xing Zhu}
\affiliation[1]{Zhejiang Institute of Modern Physics, Department of Physics, Zhejiang University, Hangzhou, Zhejiang 310027, China}
\affiliation[2]{Nikhef, Theory Group,
	Science Park 105, 1098 XG, Amsterdam, The Netherlands}
\affiliation[3]{Institute for Theoretical Physics Amsterdam and Delta Institute for 
 Theoretical Physics, University of Amsterdam, Science Park 904, 1098 XH Amsterdam, The Netherlands}
\affiliation[4]{Department of Physics, Yale University, New Haven, CT 06511, USA\vspace{0.5ex}}

\abstract{Jets and their substructure play a central role in many analyses at the Large Hadron Collider (LHC). 
To improve the precision of measurements, as well as to enable measurement of jet substructure at increasingly small angular scales, tracking information is often used due to its superior angular resolution and robustness to pile-up. 
Calculations of track-based observables involve non-perturbative track functions, that absorb infrared divergences in perturbative calculations and describe the transition to charged hadrons. The infrared divergences are directly related to the renormalization group evolution (RGE), and can be systematically computed in perturbation theory.
Unlike the standard DGLAP evolution, the RGE of the track functions is non-linear, encoding correlations in the fragmentation process.
We compute the next-to-leading order (NLO) evolution of the track functions, which involves in its kernel the full $1\to3$ splitting function.
We discuss in detail how how we implement the evolution equation numerically, and illustrate the size of the NLO corrections. 
We also show that our equation can be viewed as a master equation for collinear evolution at NLO, by illustrating that by integrating out specific terms, one can derive the evolution for any $N$-hadron fragmentation function.
Our results provide a crucial ingredient for obtaining track-based predictions for generic measurements at the LHC, and for improving the description of the collinear dynamics of jets.
}

\maketitle

\section{Introduction}

Jets play a crucial role at hadron colliders, both for searches for new physics and measurements of Standard Model processes, as well as for studies of quantum chromodynamics (QCD) itself. This is increasingly true with the advent of jet substructure  (for reviews, see \cite{Larkoski:2017jix,Marzani:2019hun,Asquith:2018igt}), which allows detailed aspects of the structure of energy flow in jets to be measured. From the experimental perspective, jets are complicated final states, making it difficult to measure them, and in particular their internal structure, with high precision.

One approach to overcoming this is through the use of tracking information, which greatly improves the angular resolution of jet and jet substructure measurements, and also makes them more resilient to pile-up. For examples of measurements using tracking for jet substructure or fragmentation, see e.g.~\cite{ATLAS:2015ytt,CMS:2018ypj,ATLAS:2019mgf,ATLAS:2020bbn,ALICE:2021njq,ATLAS:2011myc,CMS:2014jjt,ALICE:2014dla,ATLAS:2017pgl,ALICE:2018ype,LHCb:2019qoc,ATLAS:2019dsv}. However, from the theoretical perspective, the use of tracking information significantly complicates first principles calculations, since track-based observables are no longer infrared and collinear safe, meaning that they cannot be computed purely in perturbation theory.

\begin{figure}
\begin{center}
\includegraphics[scale=0.35]{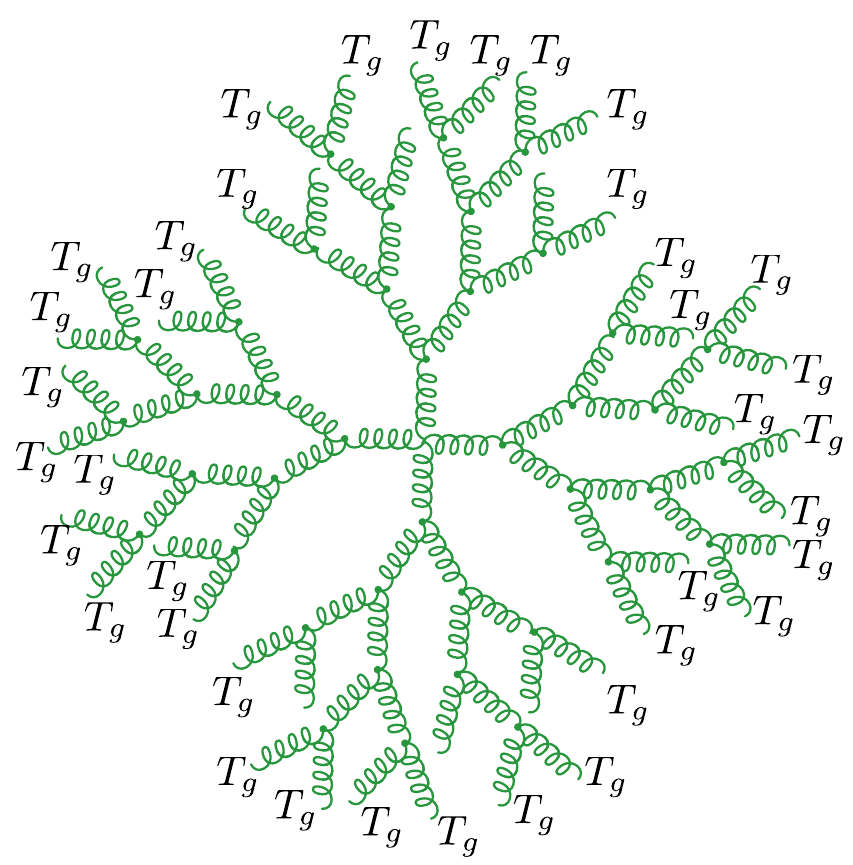}
\end{center}
\caption{Track functions, $T_i(x)$, allow an infrared and collinear safe observable to be computed on tracks, using the factorization  in \Eq{eq:intro_fact}. A track function is added to each particle in a perturbative final state, describing their transition to hadrons.}
\label{fig:intro_factorization}
\end{figure}

A systematic approach to the incorporation of tracks was derived in refs.~\cite{Chang:2013rca,Chang:2013iba}, where it was shown that the non-perturbative physics relevant for the description of measurements on tracks factorizes into a universal non-perturbative function called a track function, which we will denote $T_i(x)$. For a generic infrared and collinear safe (IRC safe) observable, $e$, we can write down a factorization theorem allowing for the calculation of the same observable measured on tracks, $\bar e$,
\begin{align}\label{eq:intro_fact}
\frac{\df \sigma}{\df e}&=\sum\limits_{N} \!\int\! \df \Pi_N \frac{\df \sigma_N}{\df \Pi_N} \,\delta(e-\hat e(\{p_i^\mu \}))\nn \\
\xrightarrow[]{\text{tracks}}
\frac{\df \sigma}{\df \bar e}&=\sum\limits_{N} \!\int\! \df \Pi_N\, \frac{\df \bar \sigma_N}{\df \Pi_N} \int  \prod\limits_{i=1}^N \df x_i T_i (x_i)\, \delta(\bar e-\hat e(\{ x_i p_i^\mu \}))  \,.
\end{align}
A track function for each parton in the perturbative state is included, that absorbs the IR divergences of the partonic cross section $\sigma_N$, resulting in the infrared-finite matching coefficient $\bar \sigma_N$. Here, $\df \Pi_N$ is the phase space associated with an $N$-particle final state. 
This factorization is illustrated schematically in \Fig{fig:intro_factorization}, where track functions are attached to each gluon in a complicated perturbative state, allowing for the measurement of a generic observable on a subset of hadrons in the final state.  As an explicit example sensitive to the full structure of the track function, one can consider the energy fraction in charged hadrons in $e^+e^-\to$ hadrons. While this cannot be computed in perturbation theory, it can be computed using the track function formalism, and its calculation at NNLO is presented in a companion paper. The calculation of generic observables on tracks at higher perturbative orders requires a calculation of the IR subtracted cross section $\bar \sigma$. This goes beyond what is currently possible with (N)NLO subtraction schemes, but there has been significant recent progress for identified hadrons or photons, whose cross sections are described by factorization theorems which share many similar features to \Eq{eq:intro_fact} \cite{Gehrmann:2022cih,Gehrmann:2022pzd}. While we focus on the renormalization group evolution (RGE) of track functions in this paper, we note that these are directly related to the infrared poles that will need to be subtracted.

Much like a fragmentation function, the RGE of track functions can be computed perturbatively, however, unlike the standard DGLAP equations governing the evolution of the fragmentation functions, the RG evolution for track functions is non-linear. While the LO evolution equation was computed in refs.~\cite{Chang:2013rca,Chang:2013iba}, a systematic understanding of the renormalization group properties of track functions beyond the leading order has been lacking.

There has recently been renewed interest in understanding the theoretical structure of track functions, due to advances in the measurement of jet substructure observables. In a series of papers, it was shown that for a particular class of observables, the energy correlators \cite{Basham:1978bw,Basham:1977iq,Basham:1979gh,Basham:1978zq}\footnote{For modern applications of energy correlators to jet substructure, see refs.~\cite{Dixon:2019uzg,Chen:2019bpb,Chen:2020vvp,Chen:2020adz,Holguin:2022epo,Komiske:2022enw,Lee:2022ige,Chen:2022swd,Ricci:2022htc,Liu:2022wop,Andres:2022ovj}.}, only moments of the track functions appear \cite{Chen:2020vvp}, and the evolution equations for the track functions moments were computed at next-to-leading order (NLO) \cite{Li:2021zcf,Jaarsma:2022kdd}. Furthermore, it was shown through explicit calculation that these equations reproduced the infrared poles in fixed order calculations at NLO, providing a highly non-trivial test of the track function formalism. However, for general observables, the complete evolution of the track function is required, not just its integer moments. A clean example for which the perturbative ingredients have been computed to high precision are jet-boson azimuthal decorrelations computed on tracks \cite{Chien:2020hzh,Chien:2022wiq}. Beyond the specific case of tracks, similar evolution equations appear for jet charge \cite{Waalewijn:2012sv,Krohn:2012fg}, or leading jets \cite{Scott:2019wlk,Neill:2021std}.

The derivation of the full evolution equation for the track functions is interesting more generally for understanding non-linear evolution equations for collinear dynamics beyond the leading order. Although we have phrased it for the particular application of track functions, it can be viewed as a master equation incorporating multi-hadron fragmentating functions. NLO evolution equations in the collinear sector have not yet appeared in the literature, as they combine the technical complication of combining the $1\to2$ and $1\to 3$ splitting functions into a single evolution equation. Thus, while the splitting function ingredients have long been available \cite{Catani:1999ss,Kosower:1999rx,Sborlini:2013jba}, they have not been combined into infrared finite evolution equations.\footnote{One interesting exception to this are the early extensions of the jet calculus \cite{Konishi:1978yx,Konishi:1979cb} beyond the leading order \cite{Kalinowski:1980ju,Kalinowski:1980wea,Gunion:1984xw,Gunion:1985pg}. It would be interesting to understand the connection of these evolution equations to our present work.} This is different from non-linear evolution equations for soft physics, which have received much attention \cite{Banfi:2002hw,Banfi:2021owj,Banfi:2021xzn}.

In this paper we give a detailed derivation and solution of the NLO RG equations for track functions. We derive the results both in the simplified case of $\cN=4$ SYM, as well as for both quark and gluon jets in QCD. We show how these equations can be reduced to the standard DGLAP \cite{Dokshitzer:1977sg,Gribov:1972ri,Altarelli:1977zs} equations, or multi-hadron fragmentation functions \cite{Konishi:1979cb,Sukhatme:1980vs,Sukhatme:1981ym,Majumder:2004wh} (which were previously unknown at NLO). We also show how they can be solved in an efficient and practical manner, and present results for the evolution of the track functions at NLO.

An outline of this paper is as follows: In \Sec{sec:track_review} we review the definition of track functions, and the general structure of their evolution. In \Sec{sec:strategy} we outline the strategy of our calculation of the track function kernels. In \Sec{sec:N4_kernel} we derive the structure of the NLO kernels in $\cN=4$ SYM as a warm-up to the full QCD calculation, which is presented in \Sec{sec:QCD_kernel} along with the explicit results for the quark and gluon track function evolution. In \Sec{sec:multihadron} we describe how the standard DGLAP equation, as well as the evolution equations for multi-hadron fragmentation functions can be derived from our master collinear equation. In \Sec{sec:num_imp} we present several approaches to solve the evolution equations for phenomenological applications, and present numerical results for the evolution of the track functions. We conclude in \Sec{sec:conclusions}.

\section{Track Functions and their Evolution}\label{sec:track_review}

Track functions were defined in refs.~\cite{Chang:2013rca,Chang:2013iba}, and describe the momentum fraction $x$ of an initial parton $i$ that is converted to a subset $R$ of the final-state hadrons. We assume that the subset $R\subset X$ of the final state $X$ is specified in terms of some quantum number, with electric charge being the canonical example. In light-cone gauge, the track functions are defined as~\cite{Chang:2013rca,Chang:2013iba}
\begin{align} \label{T_def}
T_q(x)&=\!\int\! \df y^+ \df ^{d-2} y_\perp e^{\img k^- y^+/2} \sum_X \delta \biggl( x\!-\!\frac{P_R^-}{k^-}\biggr)  \frac{1}{2N_c}
\text{tr} \biggl[  \frac{\gamma^-}{2} \langle 0| \psi(y^+,0, y_\perp)|X \rangle \langle X|\bar \psi(0) | 0 \rangle \biggr]\,,
 \\
T_g(x)&=\!\int\! \df y^+ \df^{d-2} y_\perp e^{\img k^- y^+/2} \sum_X \delta \biggl( x\!-\!\frac{P_R^-}{k^-}\biggr) \frac{\langle 0|G^a_{- \lambda}(y^+,0,y_\perp)|X\rangle \langle X|G^{\lambda,a}_- (0)|0\rangle}{(2\!-\!d)(N_c^2\!-\!1)k^-}\,. 
\nn 
\end{align} 
Here, $P_R^-$ is the large light-cone component of the total momentum of the particles in subset $R$ of $X$, $\psi$ is the quark field and $G$ the gluon field strength. 
The integral over $y^+$ fixes the large light-cone component of the momentum of initiating field to be $k^-$, and the $y_\perp$ integral sets its transverse momentum to zero.
Finally, the factor $2N_c$ and $(d-2)(N_c^2 - 1)$ in the denominator arises from averaging over spin and color.

These definitions imply that the track functions are normalized,
\begin{align}\label{T0}
\int_0^1 \df x \ T_i(x,\mu)=1\,.
\end{align}
The evolution equations are independent of the choice of $R$, and thus the evolution equations we derive in this paper are generic. When we numerically solve the RG equations in \Sec{sec:num_imp}, we will take the specific case where $R$ denotes the set of charged particles, but this choice only enters in the non-perturbative boundary conditions.

\begin{figure}
\begin{center}
\subfloat[]{
\includegraphics[scale=0.20]{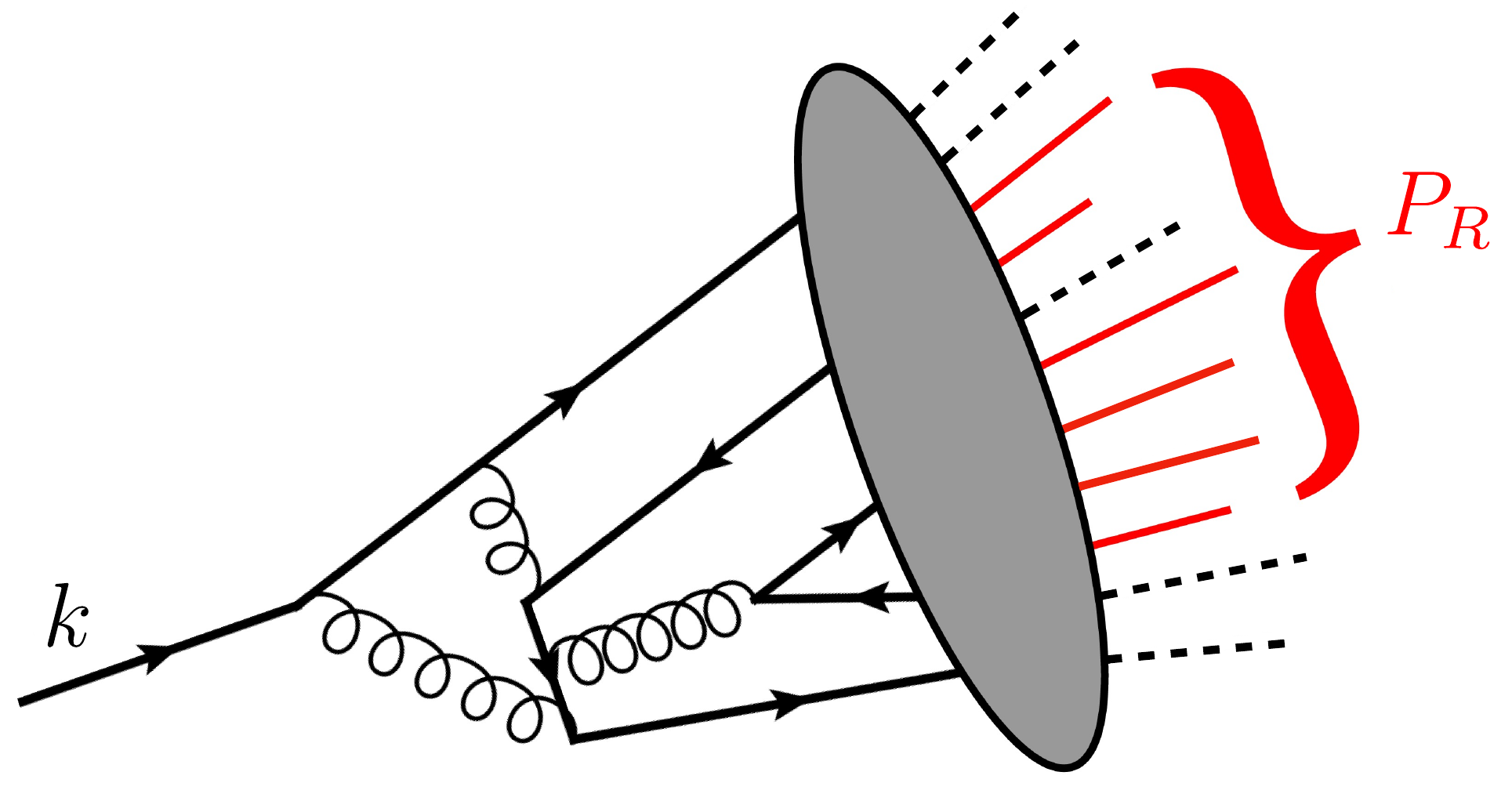}\label{fig:eflow_a}}\qquad
\subfloat[]{
\includegraphics[scale=0.10]{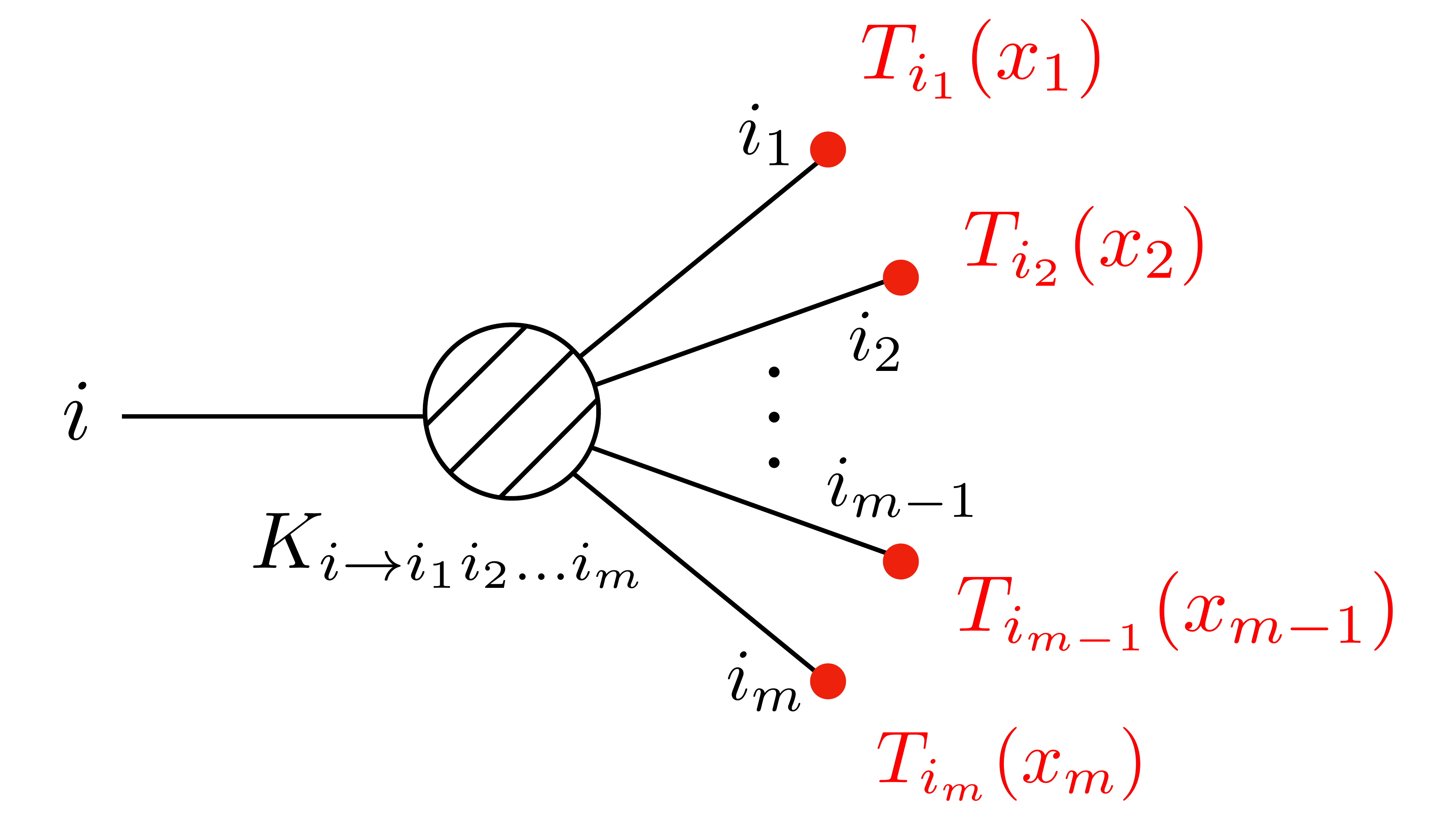}\label{fig:eflow_b}}\quad
\end{center}
\caption{(a) Fragmentation of a parton with momentum $k$ into a subset of hadrons with momentum $P_R$, is described by a universal non-perturbative track function, $T(x)$. (b) The scale evolution of $T(x)$ is determined by a non-linear evolution equation described by the kernels $K_{i\to i_1 i_2 \cdots i_m}$, which can be computed perturbatively from the multi-collinear splitting functions.}
\label{fig:nonlinear}
\end{figure}

Deriving the renormalization group equations for track functions serves two purposes. First, given a non-perturbative input function at some scale (for example as extracted from some experimental measurement), they allow this function to be evolved to other scales such that they can be used to make predictions about other measurements. This is of course familiar to the cases of parton distribution functions and fragmentation functions. The second application is that since they are specifically designed to absorb infrared divergences in perturbative calculations into non-perturbative functions, their renormalization group evolution also predicts the structure of infrared divergences in perturbative calculations of observables measured on tracks (or more general subsets of particles). Due to the non-trivial structure of the track function evolution equations, we review the necessary details here.

The evolution equations for track functions, in contrast to those for parton distributions functions (PDFs) and fragmentation functions (FFs), are non-linear. This is because track functions concern a subset $R$ of the final state hadrons, whereas PDFs and FFs only concern a single particle or final-state hadron. As a result, the evolution of track functions keeps track of all final state particles as is shown schematically in \Fig{fig:nonlinear}. The evolution of track functions can then schematically be written in the form
\begin{align}
    \frac{\df}{\df\ln\mu^2}T_i
    &=
    \sum_{\{i_f\}} \sum_{\ell=0}^\infty a_s^{\ell+1} 
    K^{(\ell)}_{i\rightarrow\{i_f\}} \otimes 
    \prod_{j\in\{i_f\}} T_j\,.
\end{align}
Here, $i$ denotes a parton species (e.g. $q$, $g$), $\{i_f\}$ denotes a set of parton species, $a_s=\alpha_s/(4\pi)$ and the arguments $x$ and $\mu$ have been suppressed for brevity. Additionally, the symbol $\otimes$ denotes a convolution in the momentum fractions, the specific form of which depends on the number of elements in $\{i_f\}$ or equivalently on the number of track functions that are involved in the convolution.

In this paper we will work up to NLO. At this order at most $1\to3$ splittings can occur, and so the evolution of track functions can be written more concretely as
\begin{align} \label{eq:T_RGE}
    \frac{\df}{\df\ln\mu^2}T_i
    &=
    a_s \Bigl[K^{(0)}_{i\to i}\, T_{i} 
        + K^{(0)}_{i\to i_1 i_2} \otimes T_{i_1} T_{i_2}\Bigr]
    \\
    &\qquad
    +a_s^2\Bigl[K^{(1)}_{i\to i}\, T_{i} 
        +K^{(1)}_{i\to i_1 i_2} \otimes T_{i_1} T_{i_2}
        +K^{(1)}_{i\to i_1 i_2 i_3} \otimes T_{i_1} T_{i_2} T_{i_3}\Bigr]\,,\nn
\end{align}
with a sum over all possible splittings implied. Here, the convolutions for the $1\to2$ and $1\to3$ terms explicitly read
\begin{align} \label{eq:T_conv}
    &K_{i\rightarrow i_1 i_2} \otimes T_{i_1} T_{i_2} (x)
    \\
    &=
    \int_0^1 \df x_1 \df x_2 \ T_{i_1}(x_1) T_{i_2}(x_2)
    \int_0^1 \df z_1 \df z_2 \ 
    \delta(1\!-\!z_1\!-\!z_2) \delta(x\!-\!z_1x_1\!-\!z_2x_2) 
    K_{i\rightarrow i_1 i_2}(z_1,z_2)\,,
    \nn
    \\[2ex]
    &K_{i\rightarrow i_1 i_2 i_3} \otimes T_{i_1} T_{i_2} T_{i_3} (x)
    \nn \\
    &=
    \int_0^1 \df x_1 \df x_2 \df x_3 \ 
    T_{i_1}(x_1) T_{i_2}(x_2) T_{i_3}(x_3)
    \nn
    \\
    &\qquad\times
    \int_0^1 \df z_1 \df z_2 \df z_3 \  
    \delta(1\!-\!z_1\!-\!z_2\!-\!z_3) 
    \delta(x\!-\!z_1x_1\!-\!z_2x_2\!-\!z_3x_3) 
    K_{i\rightarrow i_1 i_2 i_3}(z_1,z_2,z_3)\,.
    \nn
\end{align}
 
The LO kernels for the track function evolution were computed in refs.~\cite{Chang:2013rca,Chang:2013iba}, and a primary goal of this paper will be to extend this progress up to NLO.

The evolution kernels can be extracted from the radiative corrections to the (renormalized) track functions. Specifically, in dimensional regularization, the evolution kernels can be extracted from the infrared poles of the renormalized track function. To demonstrate this, we first use the form of the evolution to deduce that the bare and the renormalized track functions are related by
\begin{align}\label{eq:bareTUVpole}
    T^{\text{bare}}_i(x)
    &=
    T_i(x,\mu)+a_s\,\frac{1}{\epsilon}\biggl(K^{(0)}_{i\to i}\otimes T_i
    +K^{(0)}_{i\to i_1 i_2}\otimes T_{i_1}T_{i_2}\biggr)(x,\mu)\nn
    \\
    &\quad
    +\frac{a_s^2}{2}\Biggl\{
    \frac{1}{\epsilon}\biggl( K^{(1)}_{i\to i}\otimes T_i
    + K^{(1)}_{i\to i_1 i_2}\otimes T_{i_1}T_{i_2}
    + K^{(1)}_{i\to i_1 i_2 i_3}
        \otimes T_{i_1} T_{i_2} T_{i_3}\biggr) (x,\mu)\nn
    \\
    &\qquad\quad
    +\frac{1}{\epsilon^2}\biggl\{
    K^{(0)}_{i\to i}\Bigl(K^{(0)}_{i\to i}T_i+K^{(0)}_{i\to i_1 i_2}\otimes T_{i_1}T_{i_2}\Bigr)(x,\mu)\nn
    \\
    &\qquad\qquad\quad
    +K^{(0)}_{i\to i_1 i_2}\otimes
    \biggl[T_{i_1}\Bigl(K^{(0)}_{i_2\to i_2}\otimes T_{i_2}
        +K^{(0)}_{i_2\to j_1 j_2}\otimes T_{j_1}T_{j_2}\Bigr)\nn
    \\
    &\qquad\qquad\quad
    +\Bigl(K^{(0)}_{i_1\to i_1}\otimes T_{i_1}
        +K^{(0)}_{i_1\to j_1 j_2}\otimes T_{j_1}T_{j_2}\Bigr)T_{i_2}
    \biggr](x,\mu)\nn
    \\
    &\qquad\qquad\quad
    -\beta_0\Bigl(K^{(0)}_{i\to i}T_i+K^{(0)}_{i\to i_1 i_2}\otimes T_{i_1}T_{i_2}\Bigr)(x,\mu)
    \biggr\}
    \Biggr\}+\ord{a_s^3}\,.
\end{align}
Note that, since this is a relation between a bare and a renormalized quantity, the poles in $\epsilon$ are identified with UV divergences. Throughout this paper we carry out all the calculations in $d=4-2\epsilon$ dimensions and use the $\overline{\text{MS}}$ subtraction scheme. Next, we expand the renormalized track function in the strong coupling
\begin{align}\label{eq:renTeqbareT}
    T_i(x,\mu)=T_i^{(0)}(x)
    +a_s(\mu)\,T_i^{(1)}(x)
    +a_s^2(\mu)\,T_i^{(2)}(x)+\ord{a_s^3}\,. 
\end{align}
Lastly, we use the fact that track functions are scaleless, such that the radiative corrections to the bare track function formally vanish in dimensional regularization. Inserting the above expansion back into eq. \eqref{eq:bareTUVpole}, we then obtain
\begin{align}
    T^{(1)}_i(x)
    &=
    -\frac{1}{\epsilon} K^{(0)}_{i\to i}\otimes T_i^{(0)}(x)
    -\frac{1}{\epsilon} K^{(0)}_{i\to i_1 i_2}\otimes 
        T_{i_1}^{(0)}T_{i_2}^{(0)}(x)\,,
    \\
    T^{(2)}_i(x)
    &=
    -\frac{1}{2\epsilon}\biggl[K^{(1)}_{i\to i}\otimes T_i^{(0)}(x)
    +K^{(1)}_{i\to i_1 i_2}\otimes T_{i_1}^{(0)}T_{i_2}^{(0)}(x)
    +K^{(1)}_{i\to i_1 i_2 i_3}\otimes T_{i_1}^{(0)}T_{i_2}^{(0)}T_{i_3}^{(0)}(x) \biggr]\nn
    \\
    &\quad+\frac{1}{2\epsilon^2}\biggl\{
    K^{(0)}_{i\to i}\Bigl(K^{(0)}_{i\to i}T^{(0)}_i+K^{(0)}_{i\to i_1 i_2}\otimes T^{(0)}_{i_1}T^{(0)}_{i_2}\Bigr)(x)\nn
    \\
    &\qquad\quad\quad+
    K^{(0)}_{i\to i_1 i_2}\otimes\biggl[
    T_{i_1}^{(0)}\Bigl(K^{(0)}_{i_2\to i_2}\otimes T_{i_2}^{(0)}
    +K^{(0)}_{i_2\to j_1 j_2}\otimes T_{j_1}^{(0)}T_{j_2}^{(0)}\Bigr)\nn
    \\
    &\qquad\quad\quad
    +\Bigl(K^{(0)}_{i_1\to i_1}\otimes T_{i_1}^{(0)}
            +K^{(0)}_{i_1\to j_1 j_2}\otimes T_{j_1}^{(0)}T_{j_2}^{(0)}\Bigr)
        T_{i_2}^{(0)}
    \biggr](x)\nn
    \\
    &\qquad\quad\quad
    +\beta_0\Bigl(K^{(0)}_{i\to i}T^{(0)}_i+K^{(0)}_{i\to i_1 i_2}\otimes T^{(0)}_{i_1}T^{(0)}_{i_2}\Bigr)(x)
    \biggr\}\,.
\end{align}
where, in contrast to the UV poles of eq.~\eqref{eq:bareTUVpole}, the poles are now to be identified as IR divergences. The above result shows that the evolution kernels in eq.~\eqref{eq:T_RGE} can be extracted order by order in perturbation theory from the simple poles of the renormalized track functions.

We can't calculate the evolution kernels directly in this way because track functions are scaleless quantities, which renders the radiative corrections to the bare track function zero. More formally, in dimensional regularization, the UV and IR divergences cancel one another out. However, to renormalize the track functions, one must be able to distinguish between UV and IR divergences. In the next section, in order to circumvent this problem, we introduce a different object with an additional scale such that the radiative corrections do not vanish.

\section{Strategy for the Calculation of Track Function Kernels}\label{sec:strategy}

Since the calculation of the track functions is quite technical, in this section we provide a general overview of the strategy used for the calculation. In \Sec{sec:jet_func} we demonstrate how the track function evolution kernels can be obtained by calculating a jet function differential in the invariant mass $s$ of all particles and the momentum fraction $x$ of the (charged) subset $R$. We will use sector decomposition to avoid introducing multi-variable plus distributions,  as described in \Sec{sec:sectors}. The implementation of this approach for the specific cases of $\cN=4$ SYM and QCD are presented in \Sec{sec:N4_kernel} and \Sec{sec:QCD_kernel}, respectively.

\subsection{Track Jet Function Calculation}\label{sec:jet_func}

Track functions are scaleless in dimensional regularization, making it difficult to directly extract their evolution from the calculation of radiative corrections. The simplest way to extract their evolution then is to introduce an additional scale based on a physical measurement. We do this by considering a jet function on tracks, $J_{\text{tr},i}(s,x)$ differential in both the invariant mass $s$ of \emph{all} particles and the momentum fraction $x$ of the subset $R$ (e.g.~the charged hadrons), with $i$ denoting the flavor. The evolution of track functions is then inferred from the poles that remain after renormalization of the track jet function, which are infrared and must be absorbed by track functions. This approach is similar to that used for calculating the evolution of the moments of track functions~\cite{Li:2021zcf,Jaarsma:2022kdd}.

To extract the evolution equations for the track functions at $\mathcal{O}(\alpha_s^2)$, requires the calculation of the jet function at this order. The track jet function can be obtained using the same phase-space and squared collinear matrix elements as the standard invariant-mass jet function~\cite{Ritzmann:2014mka}, but where for each outgoing parton a track function of the corresponding flavor is attached,
\begin{align}\label{eq:barejetdef}
  J^{\text{bare}}_{\text{tr},i}(s,x)&=\sum_N\sum_{ \{i_f\} }\int \df\Phi_N^c\,\delta(s-s')\,\frac{1}{S_{\{i_f\}}}
  \sigma^c_{i\to \{i_f\} }\bigl(\{z_f\},\{s_{ff'}\},s'\bigr)\nn \\
  &\quad \times\int\biggl[\prod_{m=1}^N \df x_m T^{(0)}_{i_m}(x_m)\biggr]\delta\biggl(x-\sum_{m=1}^N x_mz_m\biggr)\,.
\end{align}
This involves a sum over $1\to N$ contributions, where total invariant mass $s'$ of the collinear phase-space $\df\Phi_N^c$ is fixed to the argument $s$ of the jet function by the delta function. Each squared collinear matrix element $\sigma^c$ is divided by an appropriate symmetry factor $S_{\{i_f\}}$ that depends on the final state $\{i_f\}$. The squared collinear matrix elements  depend on the momentum fraction $z_f$ and pair-wise invariant masses $s_{ff'}$ of the partons. They can be expressed in terms of splitting functions,
\begin{align}
  \sigma^c_{i\to i_1i_2}=&\left(\frac{\mu^2e^{\gamma_E}}{4\pi}\right)^{\epsilon}\frac{2g^2}{s^\prime}\,P_{i\to i_1i_2}\,,\\
  \sigma^c_{i\to i_1i_2i_3}=&\left(\frac{\mu^2e^{\gamma_E}}{4\pi}\right)^{2\epsilon}\frac{4g^4}{{s^\prime}^2}\,P_{i\to i_1i_2i_3}\,.
\end{align}
This requires both the tree-level $1\to 3$ splitting functions, as well as the one-loop $1\to 2$ splitting functions in dimensional regularization. These are well known, and can be found in ref.~\cite{Catani:1999ss} and ref.~\cite{Sborlini:2013jba}, respectively.

The collinear phase-space measure $\df\Phi_N^c$ is only needed for $N=2$ and $N=3$ in this work, which reads~\cite{Giele:1991vf,Gehrmann-DeRidder:1997fom} 
\begin{align}
  \df\Phi_2^c=& \ \df z\ \df s' \ \frac{[z(1-z)s']^{-\epsilon}}{(4\pi)^{2-\epsilon}\Gamma(1-\epsilon)}\,,\\
  \df\Phi_3^c=& \ \df z_1\df z_2\df z_3\delta(1-z_1-z_2-z_3)
  \ \df s' \df s_{12}\df s_{13}\df s_{23}\delta(s'-s_{12}-s_{13}-s_{23})\nn \\
  &\times\frac{4\Theta(-\Delta)(-\Delta)^{-\frac{1}{2}-\epsilon}}{(4\pi)^{5-2\epsilon}\Gamma(1-2\epsilon)}\,,
\end{align}
with the total invariant mass $s'\geq 0$ and where
\begin{align}
  \Delta=(z_3s_{12}-z_1s_{23}-z_2s_{13})^2-4z_1z_2s_{13}s_{23}\,.
\end{align}
For the $1\rightarrow3$ splittings, the phase-space integration over the invariant masses $s_{ff'}$ can be performed using the integrals presented in the appendix of ref.~\cite{Kosower:2003np}.

From the bare track jet function, we can obtain the renormalized track jet function using the renormalization kernel $Z_{J_i}(s)$ for the standard jet function. This was shown in ref.~\cite{Ritzmann:2014mka}, where a similar approach was used to study fragmentation in identified jets. The renormalization of the track jet function then reads
\begin{align}
  J_{\text{tr},i}(s,x,\mu)=\int_0^s \df s'\ Z_{J_i}(s',\mu) J^{\text{bare}}_{\text{tr},i}(s-s',x)\,,
\end{align}
upon renormalization of the coupling constant. The renormalization kernels for the quark and gluon jet functions can respectively be found in refs.~\cite{Becher:2006qw} and \cite{Becher:2010pd}. After renormalization, the track jet function can be written as a combination of $\delta(s)$ terms and terms involving the plus distributions of $s/\mu^2$ of the form $[\ln^k(s/\mu^2)/(s/\mu^2)]_+$ with $k \geq 0$. We will only need the $\delta(s)$ terms to extract the track function evolution kernels. The reason is that in the matching of the renormalized track jet function onto track functions (see \eq{eq:J_match}), the track function at order $a_s^2$ appears together with the tree-level matching coefficient which is proportional to $\de(s)$. For the remainder of this work we therefore solely focus on the $\de(s)$ terms needed for extracting the track function evolution, however, the plus distributions in $s/\mu^2$ can serve as a cross-check: the evolution up to order $a_s^\ell$ can be cross-checked against the plus distributions in the jet function at order $a_s^{\ell+1}$. 

After renormalization, any remaining poles in $\epsilon$ are associated with the infrared poles of the track functions. To extract the track function evolution kernels, we consider the matching of renormalized track jet function onto track functions, 
\begin{align} \label{eq:J_match}
    J_{\text{tr},i}(s,x,\mu)
    &=
    \delta(s)\sum_{\ell=0}^\infty \sum_{\{i_f\}} 
    \biggl[a_s^\ell \mathcal{J}^{(\ell)}_{i\rightarrow\{i_f\}}\otimes
        \prod_{j\in\{i_f\}} T_j\biggr](x,\mu)\,,
\end{align}
where $\mathcal{J}$ are the matching kernels. Inserting into the above expression eq.~\eqref{eq:renTeqbareT} for the renormalized track functions we obtain up to $\mathcal{O}(a_s^2)$, 
\begin{align}\label{eq:QCD_jet_renorm}
    &J_{\text{tr},i}(s,x,\mu)\nn\\
    &=
    \delta(s)\, T_{i}^{(0)}(x)
    +a_s(\mu)\,\delta(s)\,\biggl\{
    \Bigl(\mathcal{J}^{(1)}_{i\to i}
            -\frac{1}{\epsilon}K^{(0)}_{i\to i}\Bigr)
        \otimes T_{i}^{(0)}(x)
    +\Bigl(\mathcal{J}^{(1)}_{i\to i_1 i_2}
            -\frac{1}{\epsilon}K^{(0)}_{i\to i_1 i_2}\Bigr)
        \otimes T_{i_1}^{(0)} T_{i_2}^{(0)}(x)
    \biggr\} \nn
    \\
    &\quad
    +a_s^2(\mu)\,\delta(s)\,\biggl\{
    \Bigl(\mathcal{J}^{(2)}_{i\to i}
            -\frac{1}{2\epsilon}K^{(1)}_{i\to i}
            +\frac{\beta_0}{2\epsilon^2}K^{(0)}_{i\to i}\Bigr)
        \otimes T_{i}^{(0)}\nn
    \\
    &\qquad\qquad\quad
    +\Bigl(\mathcal{J}^{(2)}_{i\to i_1 i_2}
            -\frac{1}{2\epsilon}K^{(1)}_{i\to i_1 i_2}
            +\frac{\beta_0}{2\epsilon^2}K^{(0)}_{i\to i_1 i_2}\Bigr)
        \otimes T_{i_1}^{(0)} T_{i_2}^{(0)}\nn
    \\
    &\qquad\qquad\quad
    +\Bigl(\mathcal{J}^{(2)}_{i\to i_1 i_2 i_3}
            -\frac{1}{2\epsilon}K^{(1)}_{i\to i_1 i_2 i_3}\Bigr)
        \otimes T_{i_1}^{(0)} T_{i_2}^{(0)} T_{i_3}^{(0)}\nn
    \\
    &\qquad\qquad\quad
    +\Bigl(-\frac{1}{\epsilon} \mathcal{J}^{(1)}_{i\to i} 
            +\frac{1}{2\epsilon^2} K^{(0)}_{i\to i}\Bigr)
    \otimes\Bigl(K^{(0)}_{i\to i} \otimes T_i^{(0)}
            +K^{(0)}_{i\to i_1 i_2} \otimes T_{i_1}^{(0)} T_{i_2}^{(0)}\Bigr)\nn
    \\
    &\qquad\qquad\quad
    +\Bigl(-\frac{1}{\epsilon} \mathcal{J}^{(1)}_{i\to i_1 i_2} 
            +\frac{1}{2\epsilon^2} K^{(0)}_{i\to i_1 i_2}\Bigr)
    \otimes
    \biggl[
    T_{i_1}^{(0)}\Bigl(K^{(0)}_{i_2 \to i_2}\otimes T_{i_2}^{(0)}
            +K^{(0)}_{i_2 \to j_1 j_2}\otimes T_{j_1}^{(0)} T_{j_2}^{(0)}\Bigr)\nn
    \\
    &\qquad\qquad\qquad\quad
    +\Bigl(K^{(0)}_{i_1 \to i_1}\otimes T_{i_1}^{(0)}
            +K^{(0)}_{i_1 \to k_1 k_2}\otimes T_{k_1}^{(0)} T_{k_2}^{(0)}\Bigr)
        T_{i_2}^{(0)}
    \biggr]\biggr\}+\ord{a_s^3}\,,
\end{align}
where summation over all possible splittings is implied.  Writing the renormalized track jet function as an expansion in $a_s$ and $\epsilon$,
\begin{align}\label{eq:J_a&eps_expand}
    J_{\text{tr},i}(s,x,\mu)
    &=\delta(s)\sum_\ell a_s^\ell(\mu)J^{(\ell)}_{\text{tr},i}(x)
    =\delta(s)\sum_\ell \sum_m \frac{a_s^\ell(\mu)}{\epsilon^m} J^{(\ell,m)}_{\text{tr},i}(x) \ ,
\end{align}
the LO and NLO evolution kernels of the track function can be extracted from the simple pole terms as follows:
\begin{align} \label{eq:J_poles}
    J^{(1,1)}_{\text{tr},i}
    &=
    -K^{(0)}_{i\to i}\otimes T_i^{(0)}
    -K^{(0)}_{i\to i_1 i_2}\otimes T_{i_1}^{(0)} T_{i_2}^{(0)}\,,
    \\[2ex]
    J^{(2,1)}_{\text{tr},i}
    &=
    -\frac{1}{2}K^{(1)}_{i\to i}\otimes T_i^{(0)}
    -\frac{1}{2}K^{(1)}_{i\to i_1 i_2}\otimes T_{i_1}^{(0)} T_{i_2}^{(0)}
    -\frac{1}{2}K^{(1)}_{i\to i_1 i_2 i_3}\otimes 
        T_{i_1}^{(0)} T_{i_2}^{(0)} T_{i_3}^{(0)}
    \\
    &\quad
    -\mathcal{J}^{(1)}_{i\to i} \otimes
    \Bigl(K^{(0)}_{i\to i} \otimes T_{i}^{(0)}
        +K^{(0)}_{i\to i_1 i_2} \otimes T_{i_1}^{(0)} T_{i_2}^{(0)}\Bigr)
    \nn \\
    &\quad
    -\mathcal{J}^{(1)}_{i\to i_1 i_2}\otimes\biggl[
    T_{i_1}^{(0)}\Bigl(K^{(0)}_{i_2\to i_2} \otimes T_{i_2}^{(0)}
            +K^{(0)}_{i_2\to j_1 j_2} \otimes T_{j_1}^{(0)} T_{j_2}^{(0)}\Bigr)\nn
    \\
    &\qquad\qquad\quad
    +\Bigl(K^{(0)}_{i_1\to i_1} \otimes T_{i_1}^{(0)}
            +K^{(0)}_{i_1\to k_1 k_2} \otimes T_{k_1}^{(0)} T_{k_2}^{(0)}\Bigr)
        T_{i_2}^{(0)}\biggr]\,.\nn
\end{align}
This requires the known one-loop matching kernels $\mathcal{J}^{(1)}$~\cite{Jain:2011xz}\footnote{Strictly speaking these results were obtained for fragmentation in a jet whose invariant mass is measured. However, at this order there is at most a $1\!\to\!2$ splitting, so these results immediately carry over to our case.}, but the $\mathcal{J}^{(2)}$ only enter finite terms in \eq{eq:QCD_jet_renorm} and are not needed to determine the track function evolution at this order.
Full expressions for the sum over all splittings in QCD will be provided in section \Sec{sec:QCD_kernel}.

\subsection{Sector Decomposition}\label{sec:sectors}

The track function evolution kernels are infrared finite, but soft divergences, which occur as the momenta of one or more particles become soft, leave their imprints in the form of plus distributions. Explicitly, these are plus distribution of the momentum fractions of the splitting, which are the arguments of the evolution kernels $K$ in \eq{eq:T_RGE} (the arguments are written explicitly in \eq{eq:T_conv}). For $1\!\rightarrow\!2$ splittings, only one of the two momentum fractions can become soft, since $z_1 + z_2=1$. In this case soft divergences cancel out when the self-energy contribution is included, leaving behind a plus distribution. For $1\!\rightarrow\!3$ splittings, two of the three momentum fractions can become soft, complicating the structure of plus distributions, since they can become soft individually or simultaneously. When written in terms of the momentum fractions, this would necessitate a multi-variable plus distribution. The cancellation of soft divergences and introduction of plus distributions for the $1\!\rightarrow\!3$ kernel involves a non-trivial combination of the $1\!\rightarrow\!3$ splittings, the virtual corrections to $1\!\rightarrow\!2$ splittings, and the self-energy contribution. 

\begin{figure}
\begin{center}
\includegraphics[scale=0.27]{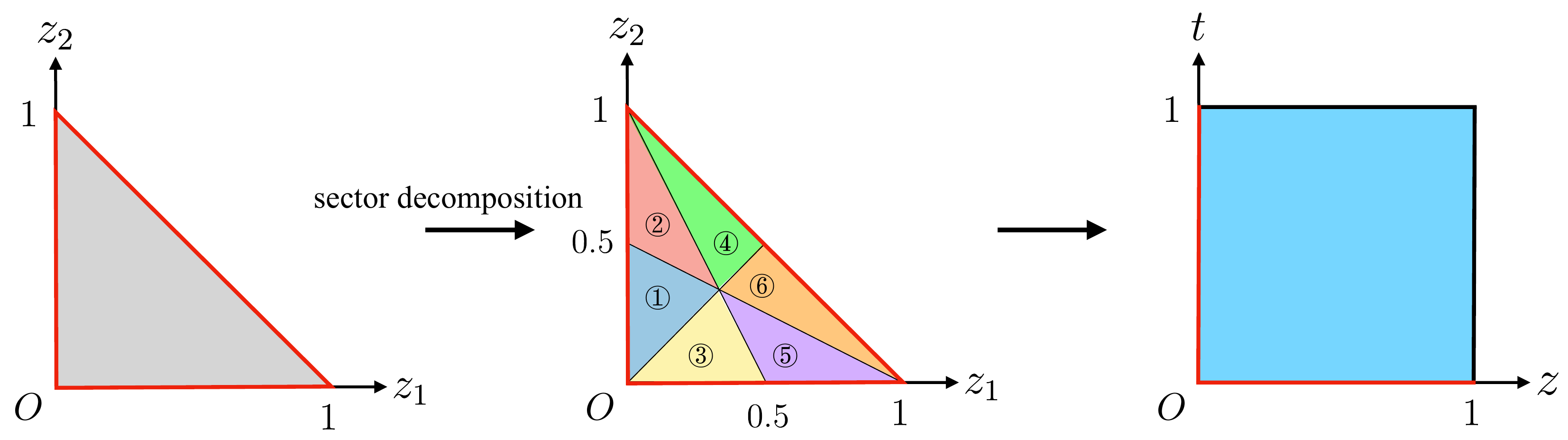}
\end{center}
\caption{The space of momentum fractions for a 3-particle final state is divided into six sectors. Each maps into a unit square after the coordinate transformations listed in Table~\ref{tab:sectors}. Red edges represent boundaries of phase space where soft divergences can be present.}
\label{fig:sector_decomposition}
\end{figure}

The standard approach to disentangling overlapping singularities is to use sector decomposition \cite{Binoth:2000ps,Binoth:2004jv,Anastasiou:2003gr}. Here, one divides the space of momentum fractions into different sectors, corresponding to different orderings in size. For the case of a 3-parton final state, the integration region can be divided into six sectors which we will label as follows:
\begin{align}\label{eq:sectors}
  &1:\quad z_1<z_2<z_3 \ ,\qquad 3:\quad z_2<z_1<z_3 \ ,\qquad 
  5:\quad z_2<z_3<z_1 \ ,\nn \\
  &2:\quad z_1<z_3<z_2 \ ,\qquad 4:\quad z_3<z_1<z_2 \ ,\qquad 
  6:\quad z_3<z_2<z_1\,,
\end{align}
as is illustrated in Fig.~\ref{fig:sector_decomposition}. 
For each sector, we denote the momentum fractions, ordered from smallest to largest, by $z_a$, $z_b$ and $z_c$.
We then apply a coordinate transformation to the momentum fractions based on this specific ordering, 
\begin{align}\label{eq:newcoord3}
  z_a=\frac{zt}{1+z+zt},\quad z_b=\frac{z}{1+z+zt},\quad 
  z_c=\frac{1}{1+z+zt} \ .
\end{align}
This explicitly written out for the sectors in \eq{eq:sectors} in table~\ref{tab:sectors}, and allows us to treat each sector in the same way.
As a result of this transformation, soft divergences only occur as $t\rightarrow0$ and $z\rightarrow0$ \emph{independently}, corresponding to $z_a$ becoming soft and $z_a$ and $z_b$ simultaneously becoming soft, respectively. Thus we will only need plus distributions in $z$ and $t$ and no multi-variable plus distributions. In principle, many other coordinate transformations could achieve the same disentangling of overlapping divergences. However, this choice of coordinates is convenient as the region of integration for each sector is simply the unit square bounded by $0 \leq z \leq 1$ and $0 \leq t \leq 1$. Moreover, this coordinate transformation can be straightforwardly extended to the general case of $1\!\rightarrow\!n$ splittings.

\renewcommand{\arraystretch}{1.5}
\begin{table}
\centering
\begin{tabular}{|c|c|c|c|c|}
\hline
\multicolumn{1}{|c|}{\textbf{sector}} & \multicolumn{1}{c|}{\textbf{ordering}} & \multicolumn{1}{c|}{${^n}\!z_1(z,t)$} & \multicolumn{1}{c|}{${^n}\!z_2(z,t)$} & \multicolumn{1}{c|}{${^n}\!z_3(z,t)$} \\ \hline\hline
sector 1 & $z_1<z_2<z_3$ & $\frac{z t}{1+z+z t}$ 
    & $\frac{z}{1+z+z t}$   & $\frac{1}{1+z+z t}$   \\ \hline
sector 2 & $z_1<z_3<z_2$ & $\frac{z t}{1+z+z t}$ 
    & $\frac{1}{1+z+z t}$   & $\frac{z}{1+z+z t}$   \\ \hline 
sector 3 & $z_2<z_1<z_3$ & $\frac{z}{1+z+z t}$   
    & $\frac{z t}{1+z+z t}$ & $\frac{1}{1+z+z t}$   \\ \hline
sector 4 & $z_3<z_1<z_2$ & $\frac{z}{1+z+z t}$   
    & $\frac{1}{1+z+z t}$   & $\frac{z t}{1+z+z t}$ \\ \hline
sector 5 & $z_2<z_3<z_1$ & $\frac{1}{1+z+z t}$   
    & $\frac{z t}{1+z+z t}$ & $\frac{z}{1+z+z t}$   \\ \hline
sector 6 & $z_3<z_2<z_1$ & $\frac{1}{1+z+z t}$   
    & $\frac{z}{1+z+z t}$   & $\frac{z t}{1+z+z t}$ \\ \hline
\end{tabular}
\caption{Our conventions for the ordering of the momentum fractions in different sectors. The index $n$ in ${^n}\!z_i$ denotes the sector.} \label{tab:sectors}
\end{table}
\renewcommand{\arraystretch}{1.0}

For the sake of clarity, we present the explicit expression for a $1\to3$ convolution in sector decomposed coordinates. In the standard momentum fractions, the convolution reads
\begin{align}
    &K_{i\rightarrow i_1 i_2 i_3} \otimes T_{i_1} T_{i_2} T_{i_3}(x)
    \\
    &=
    \int_0^1 \df x_1 \df x_2 \df x_3 \ 
    T_{i_1}(x_1) T_{i_2}(x_2) T_{i_3}(x_3)
    \nn
    \\
    &\qquad\times
    \int_0^1 \df z_1 \df z_2 \df z_3 \  
    \delta(1\!-\!z_1\!-\!z_2\!-\!z_3) 
    \delta(x\!-\!z_1x_1\!-\!z_2x_2\!-\!z_3x_3) 
    K_{i\rightarrow i_1 i_2 i_3}(z_1,z_2,z_3)\, .
\nn \end{align}
In these coordinates the kernel $K_{i\rightarrow i_1 i_2 i_3}(z_1,z_2,z_3)$ would contain multi-variable plus distributions, which are more cumbersome to work with. With sector decomposition applied, the convolution reads instead
\begin{align}
    &K_{i\rightarrow i_1 i_2 i_3} \otimes T_{i_1} T_{i_2} T_{i_3}(x)\\
    &=
    \sum_{n=1}^6
    \int_0^1 \df x_1 \df x_2 \df x_3 \ 
    T_{i_1}(x_1) T_{i_2}(x_2) T_{i_3}(x_3)
    \int_0^1 \df z \df t \ 
    \delta\bigl(x\!-\!{^n}\!z_1x_1\!-\!{^n}\!z_2x_2\!-\!{^n}\!z_3x_3\bigr)\
    {^n}\!K_{i\rightarrow i_1 i_2 i_3}(z,t) .\nn
\end{align}
Here, the sum on $n$ is over the sectors and ${^n}\!z_i$ can be read off from table~\ref{tab:sectors}. With sector decomposition applied, the kernels ${^n}\!K_{i\rightarrow i_1 i_2 i_3}(z,t)$ only involve plus distributions in $z$ and $t$ individually.

Note that for many configurations, due to symmetries of the particles in the final state, one can combine the kernels of different sectors. For example, for the $g\to ggg$ channel, all sectors give identical results. For all other channels present in QCD, only three sectors are independent. We will exploit these symmetries when presenting our results.


\section{Evolution Kernels for $\cN=4$ SYM}\label{sec:N4_kernel}

Before considering the more complicated case of real world QCD, we start by computing the NLO kernels in $\cN=4$ SYM. This is clearly not physical, as no particle states exist in this theory. However, we can let the track functions for all states in the theory be equal. This then provides a simplified setup where we can understand the structure of the calculation. In particular, we find that the $1\to3$ evolution kernels take a particularly simple form~\cite{Chen:2019bpb} after summing over all combinations of final state particles. Furthermore, in the process of performing sector decomposition, there is no need to consider sectors separately due to the symmetry between all the particles. This allows us to illustrate the calculation of the track function evolution kernels in the simplest possible setup.

\subsection{Setup and Calculation}\label{sec:N4_setup}

First, as all particles are identical in $\cN=4$ SYM, we introduce some simplified notation. We write the evolution of track function in \eq{eq:T_RGE} in the following form instead
\begin{align}\label{eq:N=4SYMevolution}
    \frac{\df}{\df\ln\mu^2}T(x,\mu)
    &=
    a \Bigl[K_{1\to 1}^{(0)}\otimes T(x,\mu)
        +K_{1\to2}^{(0)}\otimes TT(x,\mu)\Bigr]
    \\
    &\quad
    +a^2\Bigl[K_{1\to 1}^{(1)}\otimes T(x,\mu)
        +K_{1\to2}^{(1)}\otimes TT(x,\mu)
        +K_{1\to3}^{(1)}\otimes TTT(x,\mu)\Bigr]\,,\nn
\end{align}
and in analogy we now write the renormalized track jet function in \eq{eq:J_match} as
\begin{align}\label{eq:N4_jet_expand}
    J_{\text{tr}}(s,x,\mu)
    &=
    \delta(s)T(x,\mu)
    +a\,\delta(s)\,\Bigl[\mathcal{J}^{(1)}_{1\to 1}T(x,\mu)
        +\mathcal{J}^{(1)}_{1\to 2}\otimes TT(x,\mu)\Bigr]
    \\
    &\quad
    +a^2\,\delta(s)\,\Bigl[\mathcal{J}^{(2)}_{1\to 1}T(x,\mu)
        +\mathcal{J}^{(2)}_{1\to 2}\otimes TT(x,\mu)
        +\mathcal{J}^{(2)}_{1\to 3}\otimes TTT(x,\mu)\Bigr]\nn
    \\
    &\quad
    +\ord{a^3}\,,\nn
\end{align}
where $a=g^2N_c/(4\pi)^{2}$ with $N_c$ denoting the dimension of the fundamental representation of the $\text{SU}(N_c)$ gauge theory.

For the case of $\cN=4$ SYM, when applying sector decomposition it is not necessary to consider any sectors separately, as all track functions involved in the convolutions are identical. We therefore combine all sectors into one and drop the sector label on the kernels. To distinguish between kernels being represented in sector-decomposed coordinates and original momentum-fraction coordinates, we write the latter using all momentum fractions (e.g.~$z_1, z_2, z_3$) as arguments, while the former uses one less argument (e.g.~$z,t$).
\begin{align}\label{eq:secdecnotN=4SYM}
    &K_{1\to2} \otimes T T(x)
    \\
    &=
    \int_0^1 \df x_1 \df x_2\,T(x_1) T(x_2)
    \int_0^1 \df z_1 \df z_2\,\delta(1\!-\!z_1\!-\!z_2) 
    \delta(x\!-\!z_1x_1\!-\!z_2x_2)\,K_{1\to2}(z_1,z_2)
    \nn \\
    &=
    \int_0^1 \df x_1 \df x_2\,T(x_1) T(x_2)\int_0^1 \df z\,
    \delta\Bigl(x\!-\!\frac{z}{1+z}x_1\!-\!\frac{1}{1+z}x_2\Bigr)\,
    K_{1\to2}(z)\,,\nn
    \\[2ex]
    &K_{1\to3} \otimes TTT (x)
    \nn \\
    &=
    \int_0^1 \df x_1 \df x_2 \df x_3\,T(x_1) T(x_2) T(x_3)\nn
    \\
    &\qquad\times
    \int_0^1 \df z_1 \df z_2 \df z_3\,\delta(1\!-\!z_1\!-\!z_2\!-\!z_3) 
    \delta(x\!-\!z_1x_1\!-\!z_2x_2\!-\!z_3x_3)\,K_{1\to3}(z_1,z_2,z_3)
    \nn \\
    &=
    \int_0^1 \df x_1 \df x_2 \df x_3\,T(x_1) T(x_2) T(x_3)\nn
    \\
    &\qquad\times
    \int_0^1 \df z \df t\, 
    \delta\Bigl(x\!-\!\frac{z t}{1+z+z t}x_1\!-\!\frac{z}{1+z+z t}x_2
        \!-\!\frac{1}{1+z+z t}x_3\Bigr)\,
    K_{1\to3}(z,t)\,.\nn
\end{align}
Note that $K_{1 \to 2}(z)$ and $K_{1 \to 3}(z,t)$ includes a factor of 2 and 6, respectively, accounting for the number of sectors.

As explained in sec.~\ref{sec:jet_func}, the evolution kernels can be obtained from the simple pole terms of the renormalized track jet function. For the $\cN=4$ SYM track jet function, \eq{eq:QCD_jet_renorm} reduces to
\begin{align}\label{eq:RenormJform}
    &J_{\text{tr}}(s,x,\mu)\nn\\
    &=
    \delta(s)\, T_{i}^{(0)}(x)
    +a\,\delta(s)\,\biggl\{
    \Bigl(\mathcal{J}^{(1)}_{1\to1}
            -\frac{1}{\epsilon}K^{(0)}_{1\to1}\Bigr)
        \otimes T^{(0)}(x)
    +\Bigl(\mathcal{J}^{(1)}_{1\to2}
            -\frac{1}{\epsilon}K^{(0)}_{1\to2}\Bigr)
        \otimes T^{(0)} T^{(0)}(x)
    \biggr\} \nn
    \\
    &\quad
    +a^2\,\delta(s)\,\biggl\{
    \Bigl(\mathcal{J}^{(2)}_{1\to1}
            -\frac{1}{2\epsilon}K^{(1)}_{1\to1}
            -\frac{1}{\epsilon} \mathcal{J}^{(1)}_{1\to1} K^{(0)}_{1\to1}
            +\frac{1}{2\epsilon^2} K^{(0)}_{1\to1} K^{(0)}_{1\to1}\Bigr)
        \otimes T^{(0)}\nn
    \\
    &\qquad\qquad\quad
    +\Bigl(\mathcal{J}^{(2)}_{1\to2}
            -\frac{1}{2\epsilon}K^{(1)}_{1\to2}
            -\frac{1}{\epsilon} \mathcal{J}^{(1)}_{1\to1} K^{(0)}_{1\to2}
            +\frac{1}{2\epsilon^2} K^{(0)}_{1\to1} K^{(0)}_{1\to2}
    \\
    &\qquad\qquad\qquad\qquad\quad
            -\frac{2}{\epsilon} \mathcal{J}^{(1)}_{1\to2} K^{(0)}_{1\to1}
            +\frac{1}{\epsilon^2} K^{(0)}_{1\to2} K^{(0)}_{1\to1}\Bigr)
        \otimes T^{(0)} T^{(0)}\nn
    \\
    &\qquad\qquad\quad
    +\Bigl(\mathcal{J}^{(2)}_{1\to3}
            -\frac{1}{2\epsilon}K^{(1)}_{1\to3}\Bigr)
        \otimes T^{(0)} T^{(0)} T^{(0)}\nn
    \\
    &\qquad\qquad\quad
    +2\Bigl(-\frac{1}{\epsilon} \mathcal{J}^{(1)}_{1\to2} 
            +\frac{1}{2\epsilon^2} K^{(0)}_{1\to2}\Bigr)
    \otimes
    \Bigl[T^{(0)}\Bigl(K^{(0)}_{1\to2}\otimes T^{(0)} T^{(0)}\Bigr)\Bigr]
    \biggr\}
    +\ord{a^3}\,.\nn
\end{align}
The evolution kernels can then be extracted by focusing on the simple pole terms, for which the analogue of \eq{eq:QCD_jet_renorm} is
\begin{align}
    J^{(1,1)}_{\text{tr}}
    &=
    -K^{(0)}_{1\to1}\otimes T^{(0)}
    -K^{(0)}_{1\to2}\otimes T^{(0)} T^{(0)}\,,\label{eq:RenormJsimplepoles_1}
    \\[2ex]
    J^{(2,1)}_{\text{tr}}
    &=
    -\Bigl(\mathcal{J}^{(1)}_{1\to1}K^{(0)}_{1\to1}
        +\frac{1}{2}K^{(1)}_{1\to1}\Bigr)\otimes T^{(0)}
    -\Bigl(\mathcal{J}^{(1)}_{1\to1}K^{(0)}_{1\to2}
        +2\mathcal{J}^{(1)}_{1\to2}K^{(0)}_{1\to1}
        +\frac{1}{2}K^{(1)}_{1\to2}\Bigr)\otimes T^{(0)} T^{(0)}\nn
    \\
    &\quad
    -\frac{1}{2}K^{(1)}_{1\to3}\otimes T^{(0)} T^{(0)} T^{(0)}
    -2\mathcal{J}^{(1)}_{1\to2}\otimes
        \Bigl[T^{(0)}\Bigl(K^{(0)}_{1\to2} \otimes T^{(0)} T^{(0)}\Bigr)\Bigr]\,\label{eq:RenormJsimplepoles_2}
\end{align}
where $\mathcal{J}^{(1)}_{1\to 1}$ and $\mathcal{J}^{(1)}_{1\to 2}$ denote their $\mathcal{O}(\epsilon^0)$ terms in dimensional regularization. 

\subsection{Ingredients}

Here we collect the ingredients needed for calculating the track jet functions 
and the track function evolution in $\mathcal{N}=4$ SYM.

First of all, we need the $1\to 2$ splitting function at tree-level
\begin{align}
  P^{(0)}_{1\to2}(z)=2N_c\left(\frac{1}{z}+\frac{1}{1-z}\right)
\,,\end{align}
and one-loop order
\begin{align}
  P^{(1)}_{1\to2}(z)=2 \,\text{Re}\Bigl[r_S^{(1)}P^{(0)}_{1\to2}(z_1,z_2)\Bigr]
\,.\end{align}
The ratio  $r_S^{(1)}$ of the one-loop splitting amplitude to the tree-level one is
\begin{align}
  r^{(1)}_S(\epsilon; z,s)=& \,a\, \frac{e^{\epsilon\gamma_E }}{\epsilon^2}\biggl(\frac{\mu^2}{-s}\biggr)^\epsilon 
  \frac{\Gamma^2 (1-\epsilon ) \Gamma (1+\epsilon)}{\Gamma (1-2 \epsilon )}\\
    &\times \biggl[-\frac{\pi\epsilon}{\sin(\pi\epsilon)}\Bigl(\frac{1-z}{z}\Bigr)^\epsilon+2\sum_{k=0}^\infty\epsilon^{2k+1}\text{Li}_{2k+1}\Bigl(\frac{-z}{1-z}\Bigr)\biggr]
\,,\nn\end{align}
which can be obtained from refs.~\cite{Anastasiou:2003kj,Bern:2004cz}, with $\ln(-1)=-i\pi$ in the $\epsilon$-expansion of $[\mu^2/(-s)]^\epsilon$ for time-like kinematics. Finally, we also need the tree-level $1\to 3$ splitting function~\cite{Chen:2019bpb}
\begin{align}
  &P^{(0)}_{1\to 3}(z_1,z_2,z_3;s,s_{12},s_{13},s_{23})\\
  &\quad =N_c^2
  \biggl[\frac{s^2}{2 s_{13} s_{23}}\left(\frac{1}{z_1z_2}+\frac{1}{\left(1-z_1\right)
    \left(1-z_2\right)}\right)
    +\frac{s}{{s_{12} z_3}}\left(\frac{1}{z_1}+\frac{1}{1-z_1}\right)
+\text{(5 permutations)}\biggr]\,.\nn
\end{align}

To renormalize the bare track jet function requires 
the two-loop renormalization factor of the jet function in $\mathcal{N}=4$ theory. 
We derived the two-loop renormalization factor from the bare jet function\footnote{
We obtain the jet function $J(s)$ by 
integrating the track jet function $J(s,x)$ over the charged-hadron momentum fraction $x$,
 using the normalization condition for track functions $\int_0^1\df x\, T(x)=1$.}
 \begin{align}
  Z^{\mathcal{N}=4}_J(s,\mu)&=\delta(s)
  +a\biggl(-\frac{4}{\epsilon}\delta(s)+\frac{4}{\epsilon}\frac{1}{\mu^2}\biggl[\frac{1}{s/\mu^2}\biggr]_+ \biggr)
  +a^2\biggl[
    \delta(s)\biggl(\frac{8}{\epsilon^4}-\frac{\pi^2}{\epsilon^2}+\frac{8\zeta_3}{\epsilon}\biggr) + \dots \biggr] + \mathcal{O}(a_s^3)
\end{align}
The ``$\dots$" denote terms involving plus distributions in $s/\mu^2$, that are not needed to extract the track function evolution,
as discussed in sec.~\ref{sec:jet_func}.

\subsection{LO results} \label{sec:N_4_LO}

In $\cN=4$ SYM, the one-loop correction to the renormalized track jet function is given by
\begin{align}\label{eq:RenormJa1}
    J_{\text{tr}}^{(1)}(x,\mu)
    &= 
    \Bigl(-\frac{\pi^2}{3}\Bigr)T^{(0)}(x)
    +\biggl[-\frac{1}{\epsilon}\cdot 2\biggl(\biggl[\frac{1}{z_1}\biggr]_+ 
    +\biggl[\frac{1}{z_2}\biggr]_+\biggr)
    \\
    &+2\biggl(\biggl[\frac{\ln z_1}{z_1}\biggr]_+ +\biggl[\frac{\ln z_2}{z_2}\biggr]_+
    +\frac{\ln z_2}{z_1}+\frac{\ln z_1}{z_2}\biggr)\biggr]
    \otimes T^{(0)}T^{(0)}(x)
    +\ord{\epsilon}\,.\nn
\end{align}
Here, the convolution is written in traditional momentum fraction coordinates, with $z_1$ and $z_2$ acting as the momentum fractions of the final-state partons in $1\to 2$ splitting. Using \Eq{eq:RenormJform}, it is straightforward to extract from this
\begin{align}
    K^{(0)}_{1\to1}&=0\,,
    \label{eq:N4_K1to1_LO}\\
    K^{(0)}_{1\to2}(z_1,z_2)&= 
    2\biggl(\biggl[\frac{1}{z_1}\biggr]_+ +\biggl[\frac{1}{z_2}\biggr]_+\biggr)\,,\label{eq:N4_K1to2_LO}
\end{align}
and
\begin{align}
    \mathcal{J}^{(1)}_{1\to1}&=-\frac{\pi^2}{3}\,,
    \\
    \mathcal{J}^{(1)}_{1\to2}(z_1,z_2)&= 
    2\biggl(\biggl[\frac{\ln z_1}{z_1}\biggr]_+ +\frac{\ln z_2}{z_1}\biggr)
    +(z_1\leftrightarrow z_2)\,,
\end{align}
where the kernels are written in traditional momentum fraction coordinates. For consistency in notation we also provide $K^{(0)}_{1\to2}$ in sector decomposed coordinates, using the notation of \Eq{eq:secdecnotN=4SYM},
\begin{align}
    K^{(0)}_{1\to2}(z)&=4\biggl[\frac{1}{z}\biggr]_+\,.
\end{align}

\subsection{NLO results}

Next we present the results for the NLO evolution kernels. At this order, as $1\to3$ splittings are involved, we present all results in sector decomposed coordinates. In sector-decomposed coordinates (and summed over sectors), the simple pole term of the two-loop $\cN=4$ track jet function is given by
\begin{align}
    J_{\text{tr}}^{(2,1)}
    &=
    \frac{21\zeta_3}{2} T^{(0)}
    +\biggl\{-24\biggl[\frac{\ln ^2(z)}{z}\biggr]_+ 
        -\frac{24\ln^2(1+z)}{z}+\frac{56\ln(1+z)}{z}\ln z\biggr\}
    \otimes T^{(0)}T^{(0)}\nn
    \\
    &\quad
    +4\biggl\{
    -12\biggl[\frac{\ln z}{z}\biggr]_+ \biggl[\frac{1}{t}\biggr]_+
    \!+\!\biggl[\frac{1}{z}\biggr]_+ \biggl(-8\left[\frac{\ln t}{t}\right]_+ 
        \!+\frac{\ln t}{1\!+\!t}+\frac{3\ln(1\!+\!t)}{t}\biggr)
    +\frac{16\ln(1\!+\!z)}{z}\biggl[\frac{1}{t}\biggr]_+ \nn
    \\
    &\qquad\quad
    +\frac{\ln (t)}{(1+t) (1+tz)}
    -\frac{13 t^2 z^2+13 t^2 z+13 t z^2+28 t z+13 t+14 z+14}{t(1+t)(1+z)(1+tz)}
    \frac{\ln(1+z)}{z}
    \nn\\
    &\qquad\quad
    +\frac{\left(3 t^2 z^2+2 t^2 z+3 t z^2+4 t z+2 t+3 z+3\right) \ln (1+tz)}{t(1+t) z (1+z) (1+tz)}
    -\frac{(2+z+t z) \ln (1+t)}{(1+t) (1+z) (1+t z)}
    \nn\\
    &\qquad\quad
    +\frac{14 t^2 z^2+14 t^2 z+14 t z^2+30 t z+14 t+14 z+14}{t(1+t)(1+z)(1+tz)} \frac{\ln(1+z+zt)}{z}
    \biggr\}\nn
    \\
    &\qquad\quad
    \otimes T^{(0)} T^{(0)} T^{(0)}
    \,.
\end{align}
The precise form of the convolutions appearing in this expression are given in \eq{eq:secdecnotN=4SYM}.

To obtain the evolution kernels from the above expression, we first rewrite the double convolution term that appears in \Eq{eq:RenormJsimplepoles_2} in the same form as the convolution used  in sector-decomposed coordinates.
Thus 
\begin{align}
    &\mathcal{J}^{(1)}_{1\to2}\otimes
        T^{(0)}\Bigl(K^{(0)}_{1\to2} \otimes T^{(0)} T^{(0)}\Bigr)
    \\
    & \quad =
    \int_0^1 \df x_1 \df x_2 \df x_3\,T^{(0)}(x_1) T^{(0)}(x_2) T^{(0)}(x_3)
    \int_0^1 \df v \df w\, 
    \delta\bigl[x\!-\!v x_1\!-\!(1-v)w x_2\!-\!(1-v)(1-w)x_3\bigr]\nn
    \\
    &\qquad\times
    4\biggl(\biggl[\frac{\ln(v)}{v}\biggr]_+ +\biggl[\frac{\ln(1-v)}{1-v}\biggr]_+
    +\frac{\ln(1-v)}{v}+\frac{\ln(v)}{1-v}\biggr)
    \biggl(\biggl[\frac{1}{1-w}\biggr]_+ +\biggl[\frac{1}{w}\biggr]_+\biggr)
    \nn \\
    &\quad =
    E_{1\to3}\otimes T^{(0)} T^{(0)} T^{(0)} + E_{1\to2}\otimes T^{(0)} T^{(0)}
    + E_{1\to1}\otimes T^{(0)}\,.
\nn \end{align}
The formulae necessary to calculate the kernels on the right-hand side are provided in \App{sec:app} and the results are
\begin{align}
    E_{1\to1}
    &=
    \zeta_3\,,\nn
    \\
    E_{1\to2}(z)
    &=
    12\biggl[\frac{\ln^2 z}{z}\biggr]_+  
    +\frac{4 \ln^2(1+z)}{z}-\frac{24\ln(1+z)}{z}\ln z\,,\nn
    \\
    E_{1\to3}(z,t)
    &=
    8\biggl\{3\biggl[\frac{\ln z}{z}\biggr]_+ \biggl[\frac{1}{t}\biggr]_+
    +\biggl[\frac{1}{z}\biggr]_+\biggl(\left[\frac{\ln t}{t}\right]_+     
        +\frac{\ln(1+t)}{t}\biggr)
    -\frac{5\ln(1+z)}{z} \biggl[\frac{1}{t}\biggr]_+ \nn
    \\
    &\qquad\quad
    +\frac{6}{tz}\ln\Bigl(\frac{1+z}{1+z+z t}\Bigr)
    +\frac{\ln(1+tz)}{tz}
    \biggr\}\,.\nn
\end{align}

With the results in sec.~\ref{sec:N_4_LO} and the above in hand, we use \Eq{eq:RenormJsimplepoles_2} to obtain the next-to-leading order evolution kernels, 
\begin{align}
    K^{(1)}_{1\to1}
    &=
    -25\zeta_3\,,
    \label{eq:N4_K1to1}\\
    K^{(1)}_{1\to2}(z)
    &=
    \frac{8\pi^2}{3}\biggl[\frac{1}{z}\biggr]_+ 
    -\frac{16\ln(1+z)}{z}\ln z+\frac{32\ln^2(1+z)}{z}\,,
    \label{eq:N4_K1to2}\\
    K^{(1)}_{1\to3}(z,t)
    &=
    8\biggl\{
    \biggl[\frac{1}{z}\biggr]_+ \biggl(4\biggl[\frac{\ln t}{t}\biggr]_+ 
        -\frac{\ln t}{1+t}-\frac{7\ln(1+t)}{t}\biggr)
    +\frac{4\ln(1+z)}{z}\left[\frac{1}{t}\right]_+ 
    \label{eq:N4_K1to3}\\
    &\qquad\quad
    -\frac{7\ln(1+z t)}{z t}
    -\frac{10}{z t}\ln\Bigl(\frac{1+z}{1+z+z t}\Bigr)
    -\frac{1}{z}\frac{\ln(1+z)}{1+t}
    +\frac{1}{z}\frac{\ln(1+z t)}{(1+z)(1+t)}\nn
    \\
    &\qquad\quad 
    +\frac{\ln(1+z)}{(1+t)(1+z)}
    +\frac{\ln (1+t)}{(1+t)(1+z)}
    -\frac{z \ln (1+z)}{(1+z) (1+t z)}\nn
    \\
    &\qquad\quad 
    -\frac{1}{(1+t)(1+z t)}\ln\Bigl(\frac{t}{1+t}\Bigr)
    +\frac{2}{(1+z)(1+t)(1+z t)}\ln\Bigl(\frac{1+z t}{1+z+z t}\Bigr)
    \biggr\}\,.\nn
\end{align}
Note that the kernels are presented in sector decomposed coordinates, following the notation of \Eq{eq:secdecnotN=4SYM}. Interestingly, the above expressions exhibit some form of maximal transcendentality that we will discuss in more detail when we compare to our QCD results in \Sec{sec:compare}. In \Sec{sec:multihadron} we will show how the standard DGLAP evolution (which is known to be of uniform weight \cite{Kotikov:2004er}) can be obtained by integrating this equation.

\section{Evolution Kernels for QCD}\label{sec:QCD_kernel}

We now present the calculation for the more complicated case of QCD. The ingredients necessary for this calculation are provided in \Sec{sec:ingredients}. The LO results for the evolution kernels are presented in \Sec{sec:LOresults} and the NLO results in \Sec{sec:NLOresults}.

In QCD the track function evolution equations for quarks and gluons read
\begin{align}
    \frac{\df}{\df\ln\mu^2}T_q
    &=
    a_s \Bigl[K^{(0)}_{q\to q} \otimes T_q 
        + K^{(0)}_{q\to q g} \otimes T_q T_g\Bigr]
    +a_s^2\biggl[K^{(1)}_{q\to q} \otimes T_q
        +K^{(1)}_{q\to q g} \otimes T_q T_g
    \\
    &\qquad\qquad\quad
        +K^{(1)}_{q\to q g g} \otimes T_q T_g T_g
        +K^{(1)}_{q\to q q \bar{q}} \otimes T_q T_q T_{\bar{q}}
        +\sum_Q K^{(1)}_{q\to q Q \bar{Q}} \otimes T_q T_Q T_{\bar{Q}}\biggr]\,,\nn
    \\[2ex]
    \frac{\df}{\df\ln\mu^2}T_g
    &=
    a_s \biggl[K^{(0)}_{g\to g} \otimes T_g 
        +K^{(0)}_{g\to g g} \otimes T_g T_g
        +\sum_q K^{(0)}_{g\to q \bar{q}} \otimes T_q T_{\bar{q}}\biggr]
    \\
    &\quad
    +a_s^2\biggl[K^{(1)}_{g\to g} \otimes T_g 
        +K^{(1)}_{g\to g g} \otimes T_g T_g
        +\sum_q K^{(1)}_{g\to q \bar{q}} \otimes T_q T_{\bar{q}}\nn
    \\
    &\qquad\qquad\quad
        +K^{(1)}_{g\to g g g} \otimes T_g T_g T_g
        +\sum_q K^{(1)}_{g\to g q \bar{q}} \otimes T_g T_q T_{\bar{q}}\biggr]\,,\nn
\end{align}
where we have dropped the arguments $x$ and $\mu$ for brevity. As mentioned before, the specific form of the convolution depends on the number of track functions involved. For the $1\!\rightarrow\!1$ terms the convolution is simply multiplication by a constant while for the $1\!\rightarrow\!2$ and $1\!\rightarrow\!3$ terms the convolutions are respectively given by
\begin{align}
    &K_{i\rightarrow i_1 i_2} \otimes T_{i_1} T_{i_2} (x)
    \\
    &=
    \int_0^1 \df x_1 \df x_2 \ T_{i_1}(x_1) T_{i_2}(x_2)
    \int_0^1 \df z_1 \df z_2 \ 
    \delta(1\!-\!z_1\!-\!z_2) \delta(x\!-\!z_1x_1\!-\!z_2x_2) 
    K_{i\rightarrow i_1 i_2}(z_1,z_2)
    \nn
    \\
    &=
    \sum_{n=1}^2
    \int_0^1 \df x_1 \df x_2 \ T_{i_1}(x_1) T_{i_2}(x_2)
    \int_0^1 \df z \ 
    \delta\bigl(x\!-\!{^n}\!z_1x_1\!-\!{^n}\!z_2x_2\bigr) \ 
    {^n}\!K_{i\rightarrow i_1 i_2}(z) \ ,
    \nn
    \\[2ex]
    &K_{i\rightarrow i_1 i_2 i_3} \otimes T_{i_1} T_{i_2} T_{i_3} (x)
    \\
    &=
    \int_0^1 \df x_1 \df x_2 \df x_3 \ 
    T_{i_1}(x_1) T_{i_2}(x_2) T_{i_3}(x_3)
    \nn
    \\
    &\qquad\times
    \int_0^1 \df z_1 \df z_2 \df z_3 \  
    \delta(1\!-\!z_1\!-\!z_2\!-\!z_3) 
    \delta(x\!-\!z_1x_1\!-\!z_2x_2\!-\!z_3x_3) 
    K_{i\rightarrow i_1 i_2 i_3}(z_1,z_2,z_3)
    \nn
    \\
    &=
    \sum_{n=1}^6
    \int_0^1 \df x_1 \df x_2 \df x_3 \ 
    T_{i_1}(x_1) T_{i_2}(x_2) T_{i_3}(x_3)
    \int_0^1 \df z \df t \ 
    \delta\bigl(x\!-\!{^n}\!z_1x_1\!-\!{^n}\!z_2x_2\!-\!{^n}\!z_3x_3\bigr)\
    {^n}\!K_{i\rightarrow i_1 i_2 i_3}(z,t) ,
    \nn
\end{align}
with the argument $\mu$ suppressed for brevity. For both expressions, the top line expresses the convolution in standard momentum fraction coordinates and the bottom line in sector-decomposed coordinates which are specified in Table~\ref{tab:sectors} for ${^n}\!K_{i\to i_1i_2i_3}$. In order to make the expressions for the evolution equations in uniform coordinates, we order the two momentum fractions, $z_1$ and $z_2$, in $1\to 2$ channels as well, as is shown in Table~\ref{tab:sectors_1to2}. 
For the sector-decomposed kernels, symmetry reduces the number of independent sectors. Specifically, for the $1\to2$ kernels we have
\begin{align}\label{eq:kernel_sym_1}
    {^1}\!K_{g\to g g} = {^2}\!K_{g\to g g}\,,
    \qquad\text{and}\qquad
    {^1}\!K_{g\to q\bar{q}} = {^2}\!K_{g\to q\bar{q}}\,,
\end{align}
while for the $1\to3$ kernels we have
\begin{align}\label{eq:kernel_sym_2}
    {^1}\!K_{q\to q Q\bar{Q}} &= {^2}\!K_{q\to q Q\bar{Q}}\,, &  
    {^3}\!K_{q\to q Q\bar{Q}} &= {^4}\!K_{q\to q Q\bar{Q}}\,, &    
    {^5}\!K_{q\to q Q\bar{Q}} &= {^6}\!K_{q\to q Q\bar{Q}}\,, &
    \nn\\
    {^1}\!K_{q\to q q\bar{q}} &= {^3}\!K_{q\to q q\bar{q}}\,, &    
    {^2}\!K_{q\to q q\bar{q}} &= {^5}\!K_{q\to q q\bar{q}}\,, &    
    {^4}\!K_{q\to q q\bar{q}} &= {^6}\!K_{q\to q q\bar{q}}\,, &
    \nn \\
    {^1}\!K_{q\to q g g} &= {^2}\!K_{q\to q g g}\,, &    
    {^3}\!K_{q\to q g g} &= {^4}\!K_{q\to q g g}\,, &    
    {^5}\!K_{q\to q g g} &= {^6}\!K_{q\to q g g}\,, &
    \nn\\
    {^1}\!K_{g\to g q\bar{q}} &= {^2}\!K_{g\to g q\bar{q}}\,, &    
    {^3}\!K_{g\to g q\bar{q}} &= {^4}\!K_{g\to g q\bar{q}}\,,&    
    {^5}\!K_{g\to g q\bar{q}} &= {^6}\!K_{g\to g q\bar{q}}\,, &
\end{align}
and
\begin{align}\label{eq:kernel_sym_3}
    {^1}\!K_{g\to g g g}={^2}\!K_{g\to g g g}={^3}\!K_{g\to g g g}={^4}\!K_{g\to g g g}={^5}\!K_{g\to g g g}={^6}\!K_{g\to g g g}\,.
\end{align}

\renewcommand{\arraystretch}{2.0}
\begin{table}
  \centering
  \begin{tabular}{|c|c|c|c|}
  \hline
  \multicolumn{1}{|c|}{\textbf{sector}} & \multicolumn{1}{c|}{\textbf{ordering}} & \multicolumn{1}{c|}{${^n}\!z_1(z)$} & \multicolumn{1}{c|}{${^n}\!z_2(z)$} 
  \\ \hline\hline
  sector 1 & $z_1<z_2$ & $\frac{z}{1+z}$ 
      & $\frac{1}{1+z}$      \\ \hline
  sector 2 & $z_2<z_1$ & $\frac{1}{1+z}$ 
      & $\frac{z}{1+z}$      \\ \hline 
  \end{tabular}
  \caption{The conventions for the ordering of the momentum fractions in different sectors, for $1\to 2$ channels.}
  \label{tab:sectors_1to2}
\end{table}
\renewcommand{\arraystretch}{1.0}

\subsection{Ingredients}\label{sec:ingredients}

The method of calculation was outlined in \Sec{sec:strategy}. Here we list the ingredients necessary for calculating the track jet functions and the track function evolution kernels in QCD. 

For the calculation of the NLO evolution kernels, we need the tree-level $1\to 3$ splitting functions and the one-loop $1\to 2$ splitting functions. The tree-level $1\to 2$ splitting functions for QCD are given by
\begin{align}
    P^{(0)}_{q\to qg}(z)&=C_F\biggl(\frac{1+z^2}{1-z}-\epsilon(1-z)\biggr)\ ,
    \\
    P^{(0)}_{g\to q\bar{q}}(z)&=T_F\biggl(1-\frac{2z(1-z)}{1-\epsilon}\biggr)\ ,
    \\
    P^{(0)}_{g\to gg}(z)&=2C_A\biggl(\frac{z}{1-z}+\frac{1-z}{z}+z(1-z)\biggr)\ ,
\end{align}
while the one-loop corrections to the splitting functions, in the conventional dimensional regularization scheme, are \cite{Kosower:1999rx,Sborlini:2013jba}\footnote{For the $n_f T_F$ term in the one-loop $g\to q\bar q$ splitting, we find a typo in \cite{Sborlini:2013jba}. The denominator should be $4(\epsilon-2)\epsilon+3$, as compared to $4(\epsilon-2)\epsilon-3$.}
\begin{align}
    P^{(1)}_{q\to qg}(z;s)
    &=
    2a_s\Bigl(\frac{\mu^2e^{\gamma_E}}{s}\Bigr)^\epsilon
    \frac{\pi\ \Gamma(1-\epsilon)}{\epsilon\tan{(\pi\epsilon)}\Gamma(1-2\epsilon)}\
    \biggl\{P^{(0)}_{q\to qg}(z)\biggl[C_F
    +(C_F-C_A)\Bigl(1-\frac{\epsilon^2}{1-2\epsilon}\Bigr)\nn
    \\
    &\qquad\qquad
    +(C_A-2C_F)\,{_2}F_1\Bigl(1,-\epsilon;1-\epsilon;\frac{z-1}{z}\Bigr)
    -C_A\,{_2}F_1\Bigl(1,-\epsilon;1-\epsilon;\frac{z}{z-1}\Bigr)\biggr]\nn
    \\
    &\qquad+C_F(C_F-C_A)\frac{z(1+z)}{1-z}\frac{\epsilon^2}{1-2\epsilon}
    \biggr\}
    \\[2ex]
    P^{(1)}_{g\to q \bar{q}}(z;s)
    &=
    2a_s\Bigl(\frac{\mu^2e^{\gamma_E}}{s}\Bigr)^\epsilon
    \frac{\pi\ \Gamma(1-\epsilon)}{\epsilon\tan{(\pi\epsilon)}\Gamma(1-2\epsilon)}\
    P^{(0)}_{g\to q\bar{q}}(z)
    \biggl\{
    C_A \biggl[\frac{[2 (\epsilon -2) \epsilon +2 \epsilon -3]
    \epsilon ^2+3}{(3-2 \epsilon )(\epsilon -1)(2\epsilon-1)}\nn
    \\
    &\qquad\qquad+2
    -\,{_2}F_1\Bigl(1,-\epsilon;1-\epsilon;\frac{z-1}{z}\Bigr)
    -\,{_2}F_1\Bigl(1,-\epsilon;1-\epsilon;\frac{z}{z-1}\Bigr)\biggr]
    \\
    &\qquad+C_F\,\frac{\epsilon  \left(2 \epsilon ^2-3 \epsilon +3\right)-2}{(\epsilon -1) (2 \epsilon -1)}
    +n_f T_F\,\frac{4 (\epsilon -1) \epsilon}{4
    (\epsilon -2) \epsilon +3}\biggr\}\ ,\nn
    \\[2ex]
    P^{(1)}_{g\to gg}(z;s)
    &=
    2a_s\Bigl(\frac{\mu^2e^{\gamma_E}}{s}\Bigr)^\epsilon
    \frac{\pi\ \Gamma(1-\epsilon)}{\epsilon\tan{(\pi\epsilon)}\Gamma(1-2\epsilon)}\
    C_A \biggl\{\frac{\epsilon^2 [1-2z(1-z)\epsilon] [C_A(\epsilon-1)+2n_fT_F]}
        {(1-\epsilon)(\epsilon-1)(2\epsilon-3) (2\epsilon-1)}\nn
    \\
    &\qquad+P^{(0)}_{g\to gg}(z)\biggl[1
    -\,{_2}F_1\Bigl(1,-\epsilon;1-\epsilon;\frac{z-1}{z}\Bigr)
    -\,{_2}F_1\Bigl(1,-\epsilon;1-\epsilon;\frac{z}{z-1}\Bigr)\biggr]\biggr\}\ .
\end{align}
The tree-level $1\to3$ splitting functions are given by \cite{Campbell:1997hg,Catani:1998nv}
\begin{align}
    &P^{(0)}_{q\to qQ\bar{Q}}(z_1,z_2,z_3;s,s_{12},s_{13},s_{23})
    \\
    &=C_F T_F \frac{s}{2s_{23}}\biggl\{-\frac{1}{s\ s_{23}}
    \biggl[\frac{s_{23}(z_2-z_3)}{z_2+z_3}
    +\frac{2(s_{13}z_2-s_{12}z_3)}{z_2+z_3}\biggr]^2\nn
    +(1-2\epsilon)\Bigl(-\frac{s_{23}}{s}+z_2+z_3\Bigr)\nn
    \\
    &\qquad\quad
    +\frac{(z_2-z_3)^2+4z_1}{z_2+z_3}\biggr\}\ ,\nn
    \\[2ex]
    &P^{(0)}_{q\to qq\bar{q}}(z_1,z_2,z_3;s,s_{12},s_{13},s_{23})
    \\
    &=C_F\bigl(C_F-\tfrac{1}{2}C_A\bigr)\biggl\{
    -\frac{s^2}{2s_{23}s_{13}} z_3\biggl[\frac{1+z_3^2}{(1-z_2)(1-z_1)}
    -\epsilon\biggl(\frac{2(1-z_2)}{1-z_1}+1\biggr)-\epsilon^2\biggr]\nn
    \\
    &\qquad\quad
    +\frac{s}{s_{23}}\biggl[\frac{1+z_3^2}{1-z_2}
    -\epsilon\biggl(1+z_3+\frac{(1-z_1)^2}{1-z_2}-\frac{2z_2}{1-z_1}\biggr)
    -\frac{2z_2}{1-z_1}-\epsilon^2(1-z_1)\biggr]\nn
    \\
    &\qquad\quad
    +(1-\epsilon)\Bigl(\frac{2s_{12}}{s_{23}}-\epsilon\Bigr)\biggr\}
    +P^{(0)}_{q\to qQ\bar{Q}}(z_1,z_2,z_3;s,s_{12},s_{13},s_{23})
    +(1\leftrightarrow2)\ ,\nn
    \\[2ex]
    &P^{(0)}_{q\to qgg}(z_1,z_2,z_3;s,s_{12},s_{13},s_{23})
    \\
    &=C_F^2\biggl\{
    \frac{s^2}{2s_{13}s_{12}} z_1\biggl[\frac{1+z_1^2}{z_2 z_3}
    -\epsilon\frac{z_2^2+z_3^2}{z_2 z_3}-\epsilon(1+\epsilon)\biggr]
    +(1-\epsilon)\Bigl(\epsilon-(1-\epsilon)\frac{s_{12}}{s_{13}}\Bigr)\nn
    \\
    &\qquad\quad
    +\frac{s}{s_{13}}\biggl[\frac{(1-z_3)z_1+(1-z_2)^3}{z_2 z_3}
    -\epsilon\frac{(1-z_2)(z_2^2+z_2 z_3+z_3^2)}{z_2 z_3}
    +(1+z_1)\epsilon^2\biggr]
    \biggr\}\nn
    \\
    &\quad
    +C_A C_F\biggl\{\frac{s^2}{2s_{23} s_{13}}
    \biggl[\frac{z_2^2 (1-\epsilon)+2(1-z_2)}{1-z_1}
    +\frac{2z_1+(1-\epsilon)(1-z_1)^2}{z_2}\biggr]\nn
    \\
    &\qquad\quad
    -\frac{s^2}{4s_{12} s_{13}} z_1 
    \biggl[\frac{(1-z_1)^2(1-\epsilon)+2z_1}{z_2 z_3}
    +\epsilon(1-\epsilon)\biggr]\nn
    \\
    &\qquad\quad
    +\frac{s}{2s_{23}}\biggl[(1-\epsilon)
    \frac{z_3(z_3^2-2 z_3+2)-z_2(z_2^2-6 z_2+6)}{z_2(1-z_1)}
    +2\epsilon\frac{z_1(z_3-2 z_2) - z_2}{z_2 (1-z_1)}\biggr]\nn
    \\
    &\qquad\quad
    +\frac{s}{2s_{13}}\biggl[\epsilon(1-z_2)
    \Bigl(\frac{z_2^2+z_3^2}{z_2 z_3}-\epsilon\Bigr)
    -\epsilon\Bigl(z_2-z_3+\frac{2(1-z_2)(z_2-z_1)}{z_2 (1-z_1)}\Bigr)\nn
    \\
    &\qquad\qquad\quad
    -\frac{(1-z_3)z_1+(1-z_2)^3}{z_2 z_3}
    +(1-\epsilon)\frac{(1-z_2)^3-z_2+z_1^2}{z_2(1-z_1)}\biggr]\nn
    \\
    &\qquad\quad
    +(1-\epsilon)\biggl[\frac{1}{4s_{23}^2}\biggl(\frac{s_{23}(z_3-z_2)}{z_2+z_3}
    +\frac{2(s_{12}z_3-s_{13} z_2)}{z_2+z_3}\biggr)^2
    -\frac{\epsilon}{2}+\frac{1}{4}\biggr]\biggr\}+(2\leftrightarrow3)\nn
    \\[2ex]
    &P^{(0)}_{g\to gq\bar{q}}(z_1,z_2,z_3;s,s_{12},s_{13},s_{23})
    \\
    &=C_F T_F\biggl\{
    \frac{s^2}{s_{12} s_{13}}\Bigl(z_1^2-\frac{z_1+2z_2 z_3}{1-\epsilon}+1\Bigr)
    -\frac{s}{s_{12}}\Bigl(2z_1-\frac{2(z_1+z_2)}{1-\epsilon}+1+\epsilon\Bigr)
    -(1-\epsilon)\frac{s_{23}}{s_{12}}-1\biggr\}\nn
    \\
    &\quad+
    C_A T_F\biggl\{-\frac{s^2}{2 s_{12} s_{13}}
    \biggl[1+z_1^2-\frac{z_1+2 z_2 z_3}{1-\epsilon}\biggr]
    +\frac{s^2}{2s_{13} s_{23}}z_3\biggl[\frac{(1-z_1)^3-z_1^3}{(1-z_1) z_1}
    -\frac{2z_3 (1-2z_1 z_2-z_3)}{(1-z_1) z_1 (1-\epsilon )}\biggr]\nn
    \\
    &\qquad\quad
    +\frac{s}{2s_{13}}(1\!-\!z_2)\biggl[\frac{1+z_1-z_1^2}{(1\!-\!z_1) z_1}
    -\frac{2 (1\!-\!z_2)z_2}{(1\!-\!z_1) z_1 (1\!-\!\epsilon)}\biggr]
    -\frac{1}{4s_{23}^2}\biggl[\frac{2 (s_{13}z_2-s_{12} z_3)}{z_2+z_3}
    +\frac{s_{23}(z_2-z_3)}{z_2+z_3}\biggr]^2\nn
    \\
    &\qquad\quad
    +\frac{s}{2s_{23}}\biggl[\frac{z_1^3+1}{(1-z_1) z_1}
    +\frac{z_1 (z_3-z_2)^2-2 (z_1+1)z_2 z_3}{(1-z_1) z_1 (1-\epsilon)}\biggr]
    +\frac{\epsilon }{2}-\frac{1}{4}\biggr\}+(2\leftrightarrow3)\ ,\nn
    \\[2ex]
    &P^{(0)}_{g\to ggg}(z_1,z_2,z_3;s,s_{12},s_{13},s_{23})
    \\
    &=
    C_A^2\biggl\{
    \frac{s^2}{s_{12} s_{13}}
    \biggl[\frac{2z_1(1\!+\!z_1)+1}{2(1\!-\!z_2)(1\!-\!z_3)}
    +\frac{1\!-\!2(1\!-\!z_1)z_1}{2z_2 z_3}
    +\frac{z_1 z_2(1\!-\!z_2)(1\!-\!2z_3)}{(1\!-\!z_3)z_3}
    +\frac{z_1}{2}(1\!+\!2z_1)\!+\!z_2 z_3\!-\!2\biggr]\nn
    \\
    &\qquad\quad
    +\frac{s}{s_{12}}\biggl[\frac{4 (z_1 z_2-1)}{1-z_3}+\frac{z_1 z_2-2}{z_3}
    +\frac{[1-z_3(1-z_3)]^2}{(1-z_1) z_1 z_3}
    +\frac{5z_3}{2}+\frac{3}{2}\biggr]\nn
    \\
    &\qquad\quad
    +\frac{1}{4s_{12}^2}(1-\epsilon) 
    \biggl[\frac{s_{12} (z_1-z_2)}{z_1+z_2}
    +\frac{2(s_{23} z_1-s_{13} z_2)}{z_1+z_2}\biggr]^2
    +\frac{3 (1-\epsilon)}{4}\biggr\}+(\text{all permutations})\ .\nn
\end{align}
With these, and with the integrals listed in the appendix of ref.~\cite{Kosower:2003np}, we calculate the bare track jet function to order $a_s^2$.

To renormalize the track jet function at this order, the renormalization kernels for the jet function are needed to two-loop order. The renormalization kernel for the quark jet function was calculated in ref.~\cite{Becher:2006qw} and is given by. 
\begin{align}
    Z_{J_q}(s)\nn
    &=
    \delta(s)
    +a_s C_F\biggl\{-\Bigl(\frac{4}{\epsilon^2}
    \!+\!\frac{3}{\epsilon}\Bigr)\delta(s)
    +\frac{4}{\epsilon} \frac{1}{\mu^2}\mathcal{L}_0\Bigl(\frac{s}{\mu^2}\Bigr)
    \biggr\}\nn
    \\
    &\quad+a_s^2 C_F\biggl\{\biggl[
    C_F\Bigl(\frac{8}{\epsilon^4}\!+\!\frac{12}{\epsilon^3}
    \!+\!\frac{9}{2\epsilon^2}\!-\!\frac{4\pi^2}{3\epsilon^2}
    \!-\!\frac{12\zeta_3}{\epsilon}\!-\!\frac{3}{4\epsilon}
    \!+\!\frac{\pi^2}{\epsilon}\Bigr)
    +n_f T_F\Bigl(-\frac{4}{\epsilon^3}\!+\!\frac{2}{9\epsilon^2}
    \!+\!\frac{121}{27\epsilon}\!+\!\frac{2\pi^2}{9\epsilon}\Bigr)\nn
    \\
    &\qquad\qquad\quad
    +C_A\Bigl(\frac{11}{\epsilon^3}
    \!+\!\frac{\pi^2}{3\epsilon^2}\!-\!\frac{35}{18\epsilon^2}
    \!+\!\frac{20\zeta_3}{\epsilon}\!-\!\frac{1769}{108\epsilon}
    \!-\!\frac{11\pi^2}{18\epsilon}\Bigr)\biggr]\delta(s) 
    \\
    &\qquad\quad
    +\biggl[-C_A\Bigl(\frac{22}{3\epsilon^2}
        \!-\!\frac{134}{9\epsilon}\!+\!\frac{2\pi^2}{3\epsilon}\Bigr)
    -C_F\Bigl(\frac{16}{\epsilon^3}\!+\!\frac{12}{\epsilon^2}\Bigr)
    +n_f T_F\Bigl(\frac{8}{3\epsilon^2}\!-\!\frac{40}{9\epsilon}\Bigr)\biggr]
    \frac{1}{\mu^2}\mathcal{L}_0\Bigl(\frac{s}{\mu^2}\Bigr)\nn
    \\
    &\qquad\quad
    +C_F\frac{16}{\epsilon^2}
    \frac{1}{\mu^2}\mathcal{L}_1\Bigl(\frac{s}{\mu^2}\Bigr)
    \biggr\}\ .\nn
\end{align}
The renormalization factor for the gluon was calculated in Laplace space in ref.~\cite{Becher:2010pd}. Laplace transforming this result, which converts logarithms of the Laplace variable into plus distributions $\mathcal{L}_n(s/\mu^2)\equiv[\ln^n(s/\mu^2)/(s/\mu^2)]_+$,  the renormalization kernel for the gluon is 
\begin{align}
    Z_{J_g}(s)
    &=
    \delta(s)+a_s\biggl\{
    \biggl[C_A\Bigl(-\frac{4}{\epsilon^2}\!-\!\frac{11}{3\epsilon}\Bigr)
    +n_f T_F \frac{4}{3\epsilon}\biggr]\delta(s)
    +C_A\frac{4}{\epsilon}\frac{1}{\mu^2}\mathcal{L}_0\Bigl(\frac{s}{\mu^2}\Bigr)
    \biggr\}
    \\
    &\quad
    +a_s^2\biggl\{\biggl[C_A^2\Bigl(\frac{8\zeta_3}{\epsilon}
    \!+\!\frac{8}{\epsilon^4}\!+\!\frac{77}{3\epsilon^3}
    \!-\!\frac{\pi^2}{\epsilon^2}\!+\!\frac{6}{\epsilon^2}
    \!+\!\frac{11\pi^2}{18\epsilon}\!-\!\frac{548}{27\epsilon}\Bigr)
    \nn \\
    &\qquad\qquad\quad
    +C_A n_f T_F\Bigl(-\frac{28}{3\epsilon^3}\!-\!\frac{68}{9\epsilon^2}
    \!+\!\frac{184}{27\epsilon}\!-\!\frac{2\pi^2}{9\epsilon}\Bigr)
    +C_F n_f T_F\frac{2}{\epsilon}
    +n_f^2 T_F^2\frac{16}{9\epsilon^2}\biggr]\delta(s)\nn
    \\
    &\qquad\quad
    +\biggl[C_A^2\Bigl(-\frac{16}{\epsilon^3}\!-\!\frac{22}{\epsilon^2}
    \!-\!\frac{2\pi^2}{3\epsilon}\!+\!\frac{134}{9\epsilon}\Bigr)
    +C_A n_f T_F\Bigl(\frac{8}{\epsilon^2}\!-\!\frac{40}{9\epsilon}\Bigr)\biggr]
    \frac{1}{\mu^2}\mathcal{L}_0\Bigl(\frac{s}{\mu^2}\Bigr)\nn
    \\
    &\qquad\quad
    +C_A^2\frac{16}{\epsilon^2}
    \frac{1}{\mu^2}\mathcal{L}_1\Bigl(\frac{s}{\mu^2}\Bigr)\biggr\}\ .\nn
\end{align}

\subsection{QCD Evolution Kernels at LO}\label{sec:LOresults}

The $\delta(s)$ term of the one-loop renormalized quark track jet function, up to order-$\epsilon^0$, is given by
\begin{align}
    J_{\text{tr},q}^{(1)}(s,x)
    &=
    \Bigl(\mathcal{J}^{(1)}_{q\to q}-\frac{1}{\epsilon}K^{(0)}_{q\to q}\Bigr)
        \otimes T_q^{(0)}(x)
    +\Bigl(\mathcal{J}^{(1)}_{q\to qg}-\frac{1}{\epsilon}K^{(0)}_{q\to qg}\Bigr)
        \otimes T_q^{(0)}T_g^{(0)}(x)\,,
    \\
    J_{\text{tr},g}^{(1)}(s,x)
    &=
    \Bigl(\mathcal{J}^{(1)}_{g\to g}-\frac{1}{\epsilon}K^{(0)}_{g\to g}\Bigr)
        \otimes T_g^{(0)}(x)
    +\Bigl(\mathcal{J}^{(1)}_{g\to gg}-\frac{1}{\epsilon}K^{(0)}_{g\to gg}\Bigr)
        \otimes T_g^{(0)}T_g^{(0)}(x)\nn
    \\
    &\qquad\qquad
    +\sum_q\Bigl(\mathcal{J}^{(1)}_{g\to q\bar{q}}
            -\frac{1}{\epsilon}K^{(0)}_{g\to q\bar{q}}\Bigr)
        \otimes T_q^{(0)}T_{\bar{q}}^{(0)}(x)\,.
\end{align}
The track function evolution kernels can be extracted from the simple pole terms,
\begin{align}
    J^{(1,1)}_{\text{tr},q}
    &=
    -K^{(0)}_{q\to q}\otimes T_q^{(0)}
    -K^{(0)}_{q\to qg}\otimes T_q^{(0)} T_g^{(0)}\,,
    \\[2ex]
    J^{(1,1)}_{\text{tr},g}
    &=
    -K^{(0)}_{g\to g}\otimes T_g^{(0)}
    -K^{(0)}_{g\to gg}\otimes T_g^{(0)} T_g^{(0)} 
    -\sum_q K^{(0)}_{g\to q\bar{q}}\otimes T_q^{(0)} T_{\bar{q}}^{(0)}\,.
\end{align}

Since the LO evolution kernels only involve $1\to 2$ splittings, it is not strictly necessary to apply sector decomposition. However, for consistency in notation we provide the LO results both in standard momentum fraction coordinates and in sector-decomposed coordinates. In the traditional coordinates the LO evolution kernels read
\begin{align}
    K^{(0)}_{q\to q}&=3C_F\,,
    \\
    K^{(0)}_{g\to g}&=\frac{11 C_A}{3}-\frac{4 n_f T_F}{3}\,,
\end{align}
and
\begin{align}
    K^{(0)}_{q\to qg}(z_1,z_2)
    &=
    C_F\biggl(4\biggl[\frac{1}{1-z_1}\biggr]_+ - 2(1+z_1)\biggr)\,,
    \\
    K^{(0)}_{g\to gg}(z_1,z_2)
    &=
    C_A\biggl(2\left[\frac{1}{z_1}\right]_+ +z_1 z_2 -2\biggr) 
    +(z_1 \leftrightarrow z_2)\,,
    \\
    K^{(0)}_{g\to q\bar{q}}(z_1,z_2)
    &=
    2T_F(1-2z_1 z_2)\,.
\end{align}
In sector decomposed coordinates, for which the conventions can be found in \App{sec:app}, the results read
\begin{align}
    {^1}\!K^{(0)}_{q\to qg}(z)
    &=
    2C_F \frac{1+2z+2z^2}{(1+z)^3}\,,
    \\
    {^2}\!K^{(0)}_{q\to qg}(z)
    &=
    C_F\biggl(4\biggl[\frac{1}{z}\biggr]_+ -\frac{8+10z+4z^2}{(1+z)^3}\biggr)\,,
    \\&\nn\\
    {^1}\!K^{(0)}_{g\to gg}(z)
    &=
    2C_A \biggl(\biggl[\frac{1}{z}\biggr]_+ - \frac{2+3z+2z^2}{(1+z)^4}\biggr)\,,
    \\
    {^1}\!K^{(0)}_{g\to q\bar{q}}(z)
    &=
    2T_F\frac{1+z^2}{(1+z)^4}\,.
\end{align}
The LO $1\to1$ kernels are the same in both coordinate systems (this is no longer the case at NLO).

For later reference, we also present the other ingredients that comprise the one-loop track jet functions. The $1\to 1$ terms read 
\begin{align}
    \mathcal{J}^{(1)}_{q\to q}
    &=-\frac{\pi^2}{3}C_F\ ,
    \\
    \mathcal{J}^{(1)}_{g\to g}
    &=-\frac{\pi^2}{3}C_A\ .
\end{align}
The $1\to 2$ terms can again be represented either in standard momentum fractions or in sector-decomposed coordinates. In the remainder of this work, we only need these one-loop matching kernels in the former, in which they read
\begin{align}
    \mathcal{J}^{(1)}_{q\to qg}(z_1,z_2)
    &=2C_F\biggl(2\biggl[\frac{\ln z_2}{z_2}\biggr]_+ 
        +\frac{1+z_1^2}{1-z_1}\ln z_1 - (1+z_1)\ln z_2 + z_2\biggr)\ ,
    \\
    \mathcal{J}^{(1)}_{g\to gg}(z_1,z_2)
    &=2C_A\biggl(\biggl[\frac{\ln z_1}{z_1}\biggr]_+ +\frac{\ln z_1}{z_2}
        +(z_1 z_2 - 2)\ln z_1\biggr)+(z_1\leftrightarrow z_2)\ ,
    \\
    \mathcal{J}^{(1)}_{g\to q\bar{q}}(z_1,z_2)
    &=2T_F\biggl(2z_1 z_2 + (1-2z_1 z_2)\ln(z_1 z_2)\biggr)\ .
\end{align}

\subsection{QCD Evolution Kernels at NLO}\label{sec:NLOresults}

Let us now proceed to extract the evolution kernels at NLO. The results for the two-loop renormalized track jet functions can be written in the form
\begin{align}
    J_{\text{tr},q}^{(2)}
    &=
    \Bigl(\mathcal{J}^{(2)}_{q\to q}
            -\frac{1}{2\epsilon}K^{(1)}_{q\to q}
            +\frac{\beta_0}{2\epsilon^2}K^{(0)}_{q\to q}\Bigr)
        \otimes T_q^{(0)}\nn
    \\
    &\quad
    +\Bigl(\mathcal{J}^{(2)}_{q\to q g}
            -\frac{1}{2\epsilon}K^{(1)}_{q\to q g}
            +\frac{\beta_0}{2\epsilon^2}K^{(0)}_{q\to q g}\Bigr)
        \otimes T_q^{(0)} T_g^{(0)}\nn
    \\
    &\quad
    +\Bigl(\mathcal{J}^{(2)}_{q\to q g g}
            -\frac{1}{2\epsilon}K^{(1)}_{q\to q g g}\Bigr)
        \otimes T_q^{(0)} T_g^{(0)} T_g^{(0)}
    +\Bigl(\mathcal{J}^{(2)}_{q\to q q \bar{q}}
            -\frac{1}{2\epsilon}K^{(1)}_{q\to q q \bar{q}}\Bigr)
        \otimes T_q^{(0)} T_q^{(0)} T_{\bar{q}}^{(0)}\nn
    \\
    &\quad
    +\sum\!\mathop{}_Q\Bigl(\mathcal{J}^{(2)}_{q\to q Q \bar{Q}}
            -\frac{1}{2\epsilon}K^{(1)}_{q\to q Q \bar{Q}}\Bigr)
        \otimes T_q^{(0)} T_Q^{(0)} T_{\bar{Q}}^{(0)}
    \\
    &\quad
    +\Bigl(-\frac{1}{\epsilon} \mathcal{J}^{(1)}_{q\to q} 
            +\frac{1}{2\epsilon^2} K^{(0)}_{q\to q}\Bigr)
    \otimes\Bigl(K^{(0)}_{q\to q} \otimes T_q^{(0)}
            +K^{(0)}_{q\to q g} \otimes T_q^{(0)} T_g^{(0)}\Bigr)\nn
    \\
    &\quad
    +\Bigl(-\frac{1}{\epsilon} \mathcal{J}^{(1)}_{q\to q g} 
            +\frac{1}{2\epsilon^2} K^{(0)}_{q\to q g}\Bigr)
    \otimes
    \biggl[
    \Bigl(K^{(0)}_{q\to q}\otimes T_q^{(0)}
            +K^{(0)}_{q\to q g}\otimes T_q^{(0)} T_g^{(0)}\Bigr)T_g^{(0)}\nn
    \\
    &\qquad\quad
    +T_q^{(0)}\Bigl(K^{(0)}_{g\to g}\otimes T_g^{(0)}
        +K^{(0)}_{g\to g g}\otimes T_g^{(0)} T_g^{(0)}
        +K^{(0)}_{g\to q\bar{q}}\otimes T_q^{(0)} T_{\bar{q}}^{(0)}
        +\sum\!\mathop{}_Q K^{(0)}_{g\to q\bar{q}}
            \otimes T_Q^{(0)} T_{\bar{Q}}^{(0)}\Bigr)\nn
    \biggr]\,,\nn
\end{align}
for the case of the quark track jet function, and
\begin{align}
    J_{\text{tr},g}^{(2)}
    &=
    \Bigl(\mathcal{J}^{(2)}_{g\to g}
            -\frac{1}{2\epsilon}K^{(1)}_{g\to g}
            +\frac{\beta_0}{2\epsilon^2}K^{(0)}_{g\to g}\Bigr)
        \otimes T_g^{(0)}\nn
    \\
    &\quad
    +\Bigl(\mathcal{J}^{(2)}_{g\to g g}
            -\frac{1}{2\epsilon}K^{(1)}_{g\to g g}
            +\frac{\beta_0}{2\epsilon^2}K^{(0)}_{g\to g g}\Bigr)
        \otimes T_g^{(0)} T_g^{(0)}\nn
    \\
    &\quad
    +\sum\!\mathop{}_q\Bigl(\mathcal{J}^{(2)}_{g\to q\bar{q}}
            -\frac{1}{2\epsilon}K^{(1)}_{g\to q\bar{q}}
            +\frac{\beta_0}{2\epsilon^2}K^{(0)}_{g\to q\bar{q}}\Bigr)
        \otimes T_q^{(0)} T_{\bar{q}}^{(0)}\nn
    \\
    &\quad
    +\Bigl(\mathcal{J}^{(2)}_{g\to g g g}
            -\frac{1}{2\epsilon}K^{(1)}_{g\to g g g}\Bigr)
        \otimes T_g^{(0)} T_g^{(0)} T_g^{(0)}\nn
    +\sum\!\mathop{}_q\Bigl(\mathcal{J}^{(2)}_{g\to g q \bar{q}}
            -\frac{1}{2\epsilon}K^{(1)}_{g\to g q \bar{q}}\Bigr)
        \otimes T_g^{(0)} T_q^{(0)} T_{\bar{q}}^{(0)}\nn
    \\
    &\quad
    +\Bigl(-\frac{1}{\epsilon} \mathcal{J}^{(1)}_{g\to g} 
            +\frac{1}{2\epsilon^2} K^{(0)}_{g\to g}\Bigr)
    \otimes\Bigl(K^{(0)}_{g\to g} \otimes T_g^{(0)}
            +K^{(0)}_{g\to g g} \otimes T_g^{(0)} T_g^{(0)}
            +\sum\!\mathop{}_q K^{(0)}_{g\to q\bar{q}} 
                \otimes T_q^{(0)} T_{\bar{q}}^{(0)}\Bigr)\nn
    \\
    &\quad
    +2\Bigl(-\frac{1}{\epsilon} \mathcal{J}^{(1)}_{g\to g g} 
            +\frac{1}{2\epsilon^2} K^{(0)}_{g\to g g}\Bigr)
    \otimes\biggl[T_g^{(0)}\Bigl(K^{(0)}_{g \to g}\otimes T_g^{(0)}
    \\
    &\qquad\quad
            +K^{(0)}_{g\to g g}\otimes T_g^{(0)} T_g^{(0)}
            +\sum\!\mathop{}_q K^{(0)}_{g\to q\bar{q}}
                \otimes T_q^{(0)} T_{\bar{q}}^{(0)}\Bigr)\nn
    \biggr]
    \\
    &\quad
    +\sum\!\mathop{}_q\Bigl(-\frac{1}{\epsilon} \mathcal{J}^{(1)}_{g\to q\bar{q}} 
            +\frac{1}{2\epsilon^2} K^{(0)}_{g\to q\bar{q}}\Bigr)
    \otimes
    \biggl[
    T_q^{(0)}\Bigl(K^{(0)}_{\bar{q}\to\bar{q}}\otimes T_{\bar{q}}^{(0)}
            +K^{(0)}_{\bar{q}\to\bar{q}g}
                \otimes T_{\bar{q}}^{(0)}T_g^{(0)}\Bigr)\nn
    \\
    &\qquad\quad
    +\Bigl(K^{(0)}_{q\to q}\otimes T_q^{(0)}
            +K^{(0)}_{q\to q g}\otimes T_q^{(0)} T_g^{(0)}\Bigr)T_{\bar{q}}^{(0)}
    \biggr]\,,\nn
\end{align}
for the case of the gluon track jet function. To extract the evolution kernels we specifically focus on the simple poles in $\epsilon$, which can be rearranged as
\begin{align}
    J^{(2,1)}_{\text{tr},q}
    &=
    -\Bigl(\frac{1}{2}K^{(1)}_{q\to q}
            +\mathcal{J}^{(1)}_{q\to q}K^{(0)}_{q\to q}\Bigr)
        \otimes T_q^{(0)}\nn
    \\
    &\quad
    -\Bigl(\frac{1}{2}K^{(1)}_{q\to qg}
            +\mathcal{J}^{(1)}_{q\to q} K^{(0)}_{q\to qg}
            +\mathcal{J}^{(1)}_{q\to qg} K^{(0)}_{q\to q}
            +\mathcal{J}^{(1)}_{q\to qg} K^{(0)}_{g\to g}\Bigr)
        \otimes T_q^{(0)}T_g^{(0)}
    \\
    &\quad
    -\frac{1}{2}K^{(1)}_{q\to qgg}\otimes T_q^{(0)}T_g^{(0)}T_g^{(0)}
    -\frac{1}{2}K^{(1)}_{q\to qq\bar{q}}
        \otimes T_q^{(0)}T_q^{(0)}T_{\bar{q}}^{(0)}
    -\frac{1}{2}\sum\!\mathop{}_Q K^{(1)}_{q\to qQ\bar{Q}}
        \otimes T_q^{(0)}T_Q^{(0)}T_{\bar{Q}}^{(0)}\nn
    \\
    &\quad
    -\mathcal{J}^{(1)}_{q\to q g}\otimes
    \biggl[\Bigl(K^{(0)}_{q\to q g}\otimes T_q^{(0)} T_g^{(0)}\Bigr)T_g^{(0)}
    +T_q^{(0)}\Bigl(K^{(0)}_{g\to g g}\otimes T_g^{(0)} T_g^{(0)}\Bigr)\nn
    \\
    &\qquad\qquad\quad
    +T_q^{(0)}\Bigl(K^{(0)}_{g\to q\bar{q}}
        \otimes T_q^{(0)} T_{\bar{q}}^{(0)}\Bigr)
    +\sum\!\mathop{}_Q T_q^{(0)}\Bigl(K^{(0)}_{g\to q\bar{q}}
            \otimes T_Q^{(0)} T_{\bar{Q}}^{(0)}\Bigr)\nn
    \biggr]\,,\nn
\end{align}
for the quark case, and
\begin{align}
    J^{(2,1)}_{\text{tr},g}
    &=
    -\Bigl(\frac{1}{2}K^{(1)}_{g\to g}
            +\mathcal{J}^{(1)}_{g\to g}K^{(0)}_{g\to g}\Bigr)
        \otimes T_g^{(0)}\nn
    \\
    &\quad
    -\Bigl(\frac{1}{2}K^{(1)}_{g\to gg}
            +\mathcal{J}^{(1)}_{g\to g}K^{(0)}_{g\to gg}
            +2\mathcal{J}^{(1)}_{g\to gg}K^{(0)}_{g\to g}\Bigr)
        \otimes T_g^{(0)}T_g^{(0)}\nn
    \\
    &\quad
    -\sum\!\mathop{}_q\Bigl(\frac{1}{2}K^{(1)}_{g\to q\bar{q}}
            +\mathcal{J}^{(1)}_{g\to g}K^{(0)}_{g\to q\bar{q}}
            +\mathcal{J}^{(1)}_{g\to q\bar{q}}K^{(0)}_{q\to q}
            +\mathcal{J}^{(1)}_{g\to q\bar{q}}K^{(0)}_{\bar{q}\to\bar{q}}\Bigr)
        \otimes T_q^{(0)}T_{\bar{q}}^{(0)}\nn
    \\
    &\quad
    -\frac{1}{2}K^{(1)}_{g\to g g g}\otimes T_g^{(0)}T_g^{(0)}T_g^{(0)}\nn
    -\frac{1}{2}\sum\!\mathop{}_q K^{(1)}_{g\to gq\bar{q}}
        \otimes T_g^{(0)}T_q^{(0)}T_{\bar{q}}^{(0)}
    \\
    &\quad
    -2\mathcal{J}^{(1)}_{g\to gg}\otimes
    \biggl[T_g^{(0)}\Bigl(K^{(0)}_{g\to gg}\otimes T_g^{(0)}T_g^{(0)}\Bigr)
    +\sum\!\mathop{}_q T_g^{(0)}\Bigl(K^{(0)}_{g\to q\bar{q}}
            \otimes T_q^{(0)}T_{\bar{q}}^{(0)}\Bigr)\biggr]\nn
    \\
    &\quad
    -\sum\!\mathop{}_q\mathcal{J}^{(1)}_{g\to q\bar{q}}\otimes
    \biggl[T_q^{(0)}\Bigl(K^{(0)}_{\bar{q}\to\bar{q}g}
        \otimes T_g^{(0)} T_{\bar{q}}^{(0)}\Bigr)
    +\Bigl(K^{(0)}_{q\to qg}\otimes T_q^{(0)}T_g^{(0)}\Bigr)T_{\bar{q}}^{(0)}
    \biggr]\,,\nn
\end{align}
for the gluon case.

To extract the evolution kernels from the simple pole terms of the track jet function, one first needs to subtract the double convolutions. To subtract these terms it is necessary to first rewrite these double convolutions as a single convolution as follows,
\begin{align}
    \mathcal{J}_{i\to i_1 i_2'} \otimes T_{i_1} 
        \Bigl(K_{i_2'\to i_2 i_3} \otimes T_{i_2}T_{i_3}\Bigr)
    &=
    E_{i\to i_1 i_2 i_3}\bigl[\mathcal{J},K\bigr]\otimes T_{i_1}T_{i_2}T_{i_3}.
\end{align}
Note that $1\to2$ and $1\to1$ terms are also allowed, but are for convenience absorbed into a single $1\to3$ kernel. Finding the kernel $E_{i\to i_1 i_2 i_3}$, however, is not trivial due to the fact that the kernels on both sides of the equation contain plus distributions, so that one cannot simply apply a coordinate transformation to get from one side of the equation to the other without introducing multi-variable plus distributions. In \App{sec:app} we provide the necessary machinery to obtain the effective $1\to 3$ kernel $E[f,g]$ given two kernels $f$ and $g$.

After having subtracted the double convolution terms from the simple pole of the track jet function, only the NLO evolution kernels remain. As a way of organizing the kernels, all terms involving delta functions of momentum fractions are absorbed into evolution kernels of a lower splitting. We present the evolution kernels below, which are organized by flavor content and correspond to independent sectors according to the symmetry relations between kernels, eqs.~\eqref{eq:kernel_sym_1} - \eqref{eq:kernel_sym_3}.

\subsection*{$\boldsymbol{1\to1}$ kernels}

\begin{align}
    K^{(1)}_{q\to q}
    &=
    C_F^2\Bigl(\frac{3}{2}-12\zeta_2+24\zeta_3\Bigr)
    +C_A C_F \Bigl(\frac{3337}{54}-\frac{274}{9}\ln(2)-\frac{44}{3}\ln^2(2)
        -49\zeta_3\Bigr)
    \\
    &\quad
    +C_F n_f T_F\Bigl(-\frac{532}{27}+\frac{92}{9}\ln(2)+\frac{16}{3}\ln^2(2)\Bigr)
    \nn\,,
    \\&\nn\\
    K^{(1)}_{g\to g}
    &=
    C_A^2\Bigl(\frac{1880}{27}-\frac{44}{3}\zeta_2-\frac{274}{9}\ln(2)
        -\frac{44}{3}\ln^2(2)-25\zeta_3\Bigr)
    \\
    &\quad
    +C_A n_f T_F\Bigl(-\frac{658}{27}+\frac{16}{3}\zeta_2+\frac{92}{9}\ln(2)
        +\frac{16}{3}\ln^2(2)\Bigr)
    -4C_F n_f T_F
    \,.\nn
\end{align}

\subsection*{$\boldsymbol{1\to2}$ kernels}

\subsubsection*{$\boldsymbol{q\to qg}$}

\begin{align}
    &{^1}\!K^{(1)}_{q\to qg}(z)
    \\
    &=
    4C_F^2\biggl\{\nn
    \frac{1+2z+2z^2}{(z+1)^3}
    \biggl[-2\text{Li}_2\Bigl(\frac{1}{1+z}\Bigr)
    +\ln^2\Bigl(\frac{z}{(1+z)^2}\Bigr)+\ln^2(1+z)\biggr]\nn
    \\
    &\qquad\quad
    +\frac{2}{(z+1)^3}\ln\Bigl(\frac{z}{1+z}\Bigr)
    -\frac{1}{(z+1)^2}
    \biggr\}\nn
    \\
    &\quad
    +4C_A C_F\biggl\{
    \frac{1+2z+2z^2}{(z+1)^3}
    \biggl[
    2\biggl(\text{Li}_2(-z)+\frac{\pi^2}{3}\biggr)
    -\ln^2(z)+4\ln(z+1)\ln(z)
    \biggr]\nn
    \\
    &\qquad\quad
    -\frac{2}{(z+1)^3}\ln(z)
    +\frac{1}{(z+1)^2}
    \biggr\}\nn
    \\&\nn\\
    &{^2}\!K^{(1)}_{q\to qg}(z)
    \\
    &=
    4C_F^2\biggl\{
    \frac{4+4z+2z^2}{z (z+1)^3}
    \biggl[\text{Li}_2(-z)+2\ln ^2(z+1)-\ln(z)\ln(1+z)\biggr]\nn
    \\
    &\qquad\quad
    -\frac{2z}{(z+1)^3}\ln (z+1)
    -\frac{1}{(z+1)^2}
    \biggr\}\nn
    \\
    &\quad
    +4C_A C_F\biggl\{
    \frac{2\pi^2}{3} \biggl[\frac{1}{z}\biggr]_+
    -\frac{4+4z+2z^2}{z (z+1)^3}\text{Li}_2(-z)
    -\frac{\pi^2}{3}\,\frac{2z^2+5z+4}{(z+1)^3}
    +\frac{1}{(z+1)^2}
    \biggr\}\nn
\end{align}

\subsubsection*{$\boldsymbol{g\to q\bar{q}}$}

\begin{align}
    &{^1}\!K^{(1)}_{g\to q\bar{q}}(z)
    \\
    &=
    4C_F T_F\biggl\{
    \frac{1+z^2}{(1+z)^4}\biggl[\pi^2-\ln^2(z)+4\ln(z)\ln(1+z)+6\ln(1+z)\biggr]\nn
    \\
    &\qquad\quad
    -\frac{3z^2+4z+3}{(z+1)^4}\ln(z)
    -\frac{8+6z+8z^2}{(z+1)^4}
    \biggr\}\nn
    \\
    &\quad
    +4C_A T_F\biggl\{
    \frac{1+z^2}{(z+1)^4}
    \biggl[-\frac{2\pi^2}{3}+\ln^2(z)-6\ln(z)\ln(1+z)+4\ln^2(1+z)\biggr]\nn
    \\
    &\qquad\quad
    +\frac{11 z^2+12 z+11}{3(z+1)^4}\ln(z)
    -\frac{22+12z+22z^2}{3(z+1)^4}\ln(1+z)
    +\frac{76+66z+76z^2}{9(z+1)^4}
    \biggr\}\nn
    \\
    &\quad
    -16n_f T_F^2\biggl\{
    \frac{1+z^2}{3 (z+1)^4}\ln\biggl(\frac{z}{(1+z)^2}\biggr)
    +\frac{5+6z+5z^2}{9(z+1)^4}
    \biggr\}\nn
\end{align}

\subsubsection*{$\boldsymbol{g\to gg}$}

\begin{align}
    &{^1}\!K^{(1)}_{g\to gg}(z)
    \\
    &=
    2C_A^2\biggl\{
    \frac{2\pi^2}{3}\biggl[\frac{1}{z}\biggr]_+
    -\frac{\pi^2}{3}\,\frac{4+6z+4z^2}{(1+z)^4}
    +\frac{1}{3(1+z)^2}
    -\frac{4(1+z+z^2)^2}{(1+z)^4}\,
    \frac{\ln(1+z)}{z}\ln\biggl(\frac{z}{(1+z)^2}\biggr)
    \biggr\}\nn
   \\
   &\quad
   -C_A n_f T_F\frac{4}{3(1+z)^2}\nn
\end{align}

\subsection*{$\boldsymbol{1\to3}$ kernels}

\subsubsection*{$\boldsymbol{q\to qQ\bar{Q}}$}

\begin{align}
    &{^1}\!K^{(1)}_{q\to qQ\bar{Q}}(z,t)
    \\
    &=
    C_F T_F \biggl\{
    \biggl[\frac{8z(1+z^2)}{(1\!+\!z\!+\!z t)^2 (1\!+\!z)^3}
    \!-\!\frac{4z(1+z^2)}{(1\!+\!z\!+\!z t)^3 (1\!+\!z)^2}
    \!-\!\frac{8z(1+z^2)}{(1\!+\!z)^4 (1\!+\!z\!+\!z t)}
    \biggr]\log\biggl(\frac{t (1\!+\!z)^2}{1\!+\!z\!+\!z t}\biggr)\nn
    \\
    &\qquad\quad
    -\frac{8z-8z^2+8z^3}{(1+z)^2 (1\!+\!z\!+\!z t)^3}
    +\frac{8z-48z^2+8z^3}{(1+z)^3 (1\!+\!z\!+\!z t)^2}
    -\frac{8z-48z^2+8z^3}{(1+z)^4 (1\!+\!z\!+\!z t)}
    \biggr\}\nn
    \\&\nn\\
    &{^3}\!K^{(1)}_{q\to qQ\bar{Q}}(z,t)
    \\
    &=C_F T_F\biggl\{
    \biggl[\frac{4 z-8 t}{(1\!+\!z\!+\!z t)^3}
    +\frac{32 t^2+32 t+8}{(1\!+\!z\!+\!z t)^2}
    -\frac{88 t^3+144 t^2+72 t+16}{1\!+\!z\!+\!z t}\nn
    \\
    &\qquad\qquad\quad
    +\frac{16}{(1+z t)^4}+\frac{32 t-16}{(1+z t)^3}+\frac{56 t^2+8}{(1+z t)^2}
    +\frac{88 t^3+56 t^2+16 t}{1+z t}\biggr]
    \log\biggl(\frac{t(1\!+\!z\!+\!z t)}{(1+z t)^2}\biggr)\nn
    \\
    &\qquad\quad
    +\frac{24 t-8 z}{(1\!+\!z\!+\!z t)^3}
    -\frac{112 t^2 + 112 t+8}{(1\!+\!z\!+\!z t)^2}
    +\frac{328 t^3+544 t^2+232 t+16}{1\!+\!z\!+\!z t}\nn
    \\
    &\qquad\quad
    -\frac{64}{(1+z t)^4}
    -\frac{128 t-64}{(1+z t)^3}
    -\frac{216 t^2+8}{(1+z t)^2}
    -\frac{328 t^3+216 t^2+16}{1+z t}
    \biggr\}\nn
    \\&\nn\\
    &{^5}\!K^{(1)}_{q\to qQ\bar{Q}}(z,t)
    \\
    &=C_F T_F\biggl\{
    8\biggl[\frac{1}{z}\biggr]_+ 
    \biggl[\frac{1+t^2}{(1+t)^4}\log\biggl(\frac{t}{(1+t)^2}\biggr)
    -\frac{1-6t+t^2}{(1+t)^4}\biggr]
    +8\frac{1+t^2}{(1+t)^4}\frac{\log(1\!+\!z\!+\!z t)}{z}\nn\nn
    \\
    &\qquad\quad
    +\biggl[\frac{8z^3+8z^2+4z}{(1\!+\!z\!+\!z t)^3}
    +\frac{32z^3+24z^2+8z}{(1\!+\!z\!+\!z t)^2}
    +\frac{88z^3+56z^2+16z}{1\!+\!z\!+\!z t}\nn
    \\
    &\qquad\qquad\quad
    +\frac{56 z}{(1+t)^2}
    -\frac{88z^2+56z}{1+t}
    -16\frac{1+t^2}{(1+t)^3}\biggr]
    \log\biggl(\frac{t(1\!+\!z\!+\!z t)}{(1+t)^2}\biggr)\nn
    \\
    &\qquad\quad
    -\frac{24z^3+24z^2+8z}{(1\!+\!z\!+\!z t)^3}
    -\frac{112z^3+88z^2+8z}{(1\!+\!z\!+\!z t)^2}
    -\frac{328z^3+216z^2+16z}{1\!+\!z\!+\!z t}\nn
    \\
    &\qquad\quad
    +\frac{128}{(t+1)^3}
    -\frac{216z+128}{(t+1)^2}
    +\frac{328z^2+216z+16}{1+t}
    \biggr\}\nn
\end{align}

\subsubsection*{$\boldsymbol{q\to qq\bar{q}}$}

\begin{align}
    &{^1}\!K^{(1)}_{q\to qq\bar{q}}(z,t)
    \\
    &=
    4C_F\bigl(C_F-\tfrac{1}{2}C_A\bigr)\biggl\{
    \biggl[-\frac{1}{(z+1) (1\!+\!z\!+\!z t)^3}
    +\frac{2+z}{(z+1) (1\!+\!z\!+\!z t)^2}
    -\frac{3+z}{(z+1) (1\!+\!z\!+\!z t)}\nn
    \\
    &\qquad\qquad\quad
    +\frac{t}{(1\!+\!z\!+\!z t)^2}
    +\frac{-t^2-2 t}{1\!+\!z\!+\!z t}
    +\frac{2}{(z+1) (1+z t)}
    +\frac{t^2+t}{1+z t}\biggr]
    \ln\biggl(\frac{(1+z) (1+z t)}{1\!+\!z\!+\!z t}\biggr)\nn
    \\
    &\qquad\quad
    +\frac{1}{(z+1) (1\!+\!z\!+\!z t)^3}
    -\frac{2+3z+3z^2}{(z+1)^2 (1\!+\!z\!+\!z t)^2}
    -\frac{3t}{(1\!+\!z\!+\!z t)^2}
    +\frac{1+6t+5t^2}{1\!+\!z\!+\!z t}\nn
    \\
    &\qquad\quad
    -\frac{2 t}{(1+z t)^2}
    -\frac{5 t^2+t}{1+z t}
    \biggr\}\nn
    \\
    &\quad
    +4C_F T_F\biggl\{\nn
    \biggl[\frac{z^2}{(z+1)^2 (1\!+\!z\!+\!z t)^3}
    +\frac{1+4z+3z^2+2z^3}{(z+1)^3 (1\!+\!z\!+\!z t)^2}
    -\frac{2+9z+12z^2+9z^3+2z^4}{(z+1)^4 (1\!+\!z\!+\!z t)}\nn
    \\
    &\qquad\qquad\quad
    -\frac{t}{(1\!+\!z\!+\!z t)^3}
    +\frac{4t^2+4t}{(1\!+\!z\!+\!z t)^2}
    -\frac{11 t^3+18 t^2+9 t}{1\!+\!z\!+\!z t}\nn
    \\
    &\qquad\qquad\quad
    +\frac{2}{(1+z t)^4}
    -\frac{2-4 t}{(1+z t)^3}
    +\frac{7t^2+1}{(1+z t)^2}
    +\frac{11t^3+7t^2+2t}{1+z t}\biggr] 
    \ln\biggl(\frac{t(1\!+\!z\!+\!z t)}{(1+z t)^2}\biggr)\nn
    \\
    &\qquad\quad
    +\biggl[\frac{z+z^3}{(z+1)^2 (1\!+\!z\!+\!z t)^3}
    -\frac{2z+2z^3}{(z+1)^3 (1\!+\!z\!+\!z t)^2}
    +\frac{2z+2z^3}{(z+1)^4 (1\!+\!z\!+\!z t)}
    \biggr] 
    \ln\biggl(\frac{1\!+\!z\!+\!z t}{(1+z)(1+z t)}\biggr)\nn
    \\
    &\qquad\quad
    -\frac{2z+z^2+2z^3}{(z+1)^2 (1\!+\!z\!+\!z t)^3}
    -\frac{1+2z+9z^2}{(z+1)^3 (1\!+\!z\!+\!z t)^2}
    +\frac{1+3z+12z^2+3z^3+z^4}{(z+1)^4 (1\!+\!z\!+\!z t)}\nn
    \\
    &\qquad\quad
    +\frac{3t}{(1\!+\!z\!+\!z t)^3}
    +\frac{-14 t^2-14 t}{(1\!+\!z\!+\!z t)^2}
    +\frac{41 t^3+68 t^2+29 t}{1\!+\!z\!+\!z t}\nn
    \\
    &\qquad\quad
    -\frac{8}{(1+z t)^4}
    +\frac{8-16 t}{(1+z t)^3}
    +\frac{-27 t^2-1}{(1+z t)^2}
    +\frac{-41 t^3-27 t^2-2 t}{1+z t}
    \biggr\}\nn
    \\&\nn\\
    &{^2}\!K^{(1)}_{q\to qq\bar{q}}(z,t)
    \\
    &=
    4C_F\bigl(C_F-\tfrac{1}{2}C_A\bigr)\biggl\{\nn
    \biggl[\frac{z^3}{(1\!+\!z\!+\!z t)^3}
    +\frac{z^3}{(1\!+\!z\!+\!z t)^2}
    +\frac{z+z^3}{1\!+\!z\!+\!z t}
    -\frac{1+z^2}{t+1}
    \biggr]
    \frac{1}{1+z}\ln\biggl(\frac{1\!+\!z\!+\!z t}{(1\!+\!z)(1\!+\!t)}\biggr)\nn
    \\
    &\qquad\quad
    +\frac{z^3}{(z+1) (1\!+\!z\!+\!z t)^3}
    +\frac{6z^2+7z^3+3z^4}{(z+1)^2 (1\!+\!z\!+\!z t)^2}
    +\frac{5 z^2+z}{1\!+\!z\!+\!z t}
    +\frac{2}{(t+1)^2}
    +\frac{-5 z-1}{t+1}
    \biggr\}\nn
    \\
    &\quad
    +4C_F T_F\biggl\{\nn
    \biggl[\frac{1}{z}\biggr]_+
    \biggl[\frac{1+t^2}{(t+1)^4}\ln\biggl(\frac{t}{(1+t)^2}\biggr)
    -\frac{1-6t+t^2}{(t+1)^4}\biggr]
    +\frac{1+t^2}{(t+1)^4}\frac{\ln(1\!+\!z\!+\!z t)}{z}\nn
    \\
    &\qquad\quad
    +\biggl[\frac{z/2+z^2+z^3}{(1\!+\!z\!+\!z t)^3}
    +\frac{z+3z^2+4z^3}{(1\!+\!z\!+\!z t)^2}
    +\frac{2z+7z^2+11z^3}{1\!+\!z\!+\!z t}\nn
    \\
    &\qquad\qquad\quad
    -\frac{4}{(t+1)^3}
    +\frac{4+7z}{(t+1)^2}
    -\frac{11z^2+7z+2}{t+1}
    \biggr]
    \ln\biggl(\frac{t(1\!+\!z\!+\!z t)}{(1+t)^2}\biggr)\nn
    \\
    &\qquad\quad
    -\biggl[\frac{z+z^3}{2(z+1)^2 (1\!+\!z\!+\!z t)^3}
    -\frac{z+z^3}{(z+1)^3 (1\!+\!z\!+\!z t)^2}
    +\frac{z+z^3}{(z+1)^4 (1\!+\!z\!+\!z t)}\nn
    \biggr]
    \ln\biggl(\frac{t(1+z)^2}{1\!+\!z\!+\!z t}\biggr)\nn
    \\
    &\qquad\quad
    -\frac{8z+5z^3+3z^4+3z^5}{(z+1)^2 (1\!+\!z\!+\!z t)^3}
    -\frac{20z^2+49z^3+76z^4+53z^5+14z^6}{(z+1)^3 (1\!+\!z\!+\!z t)^2}\nn
    \\
    &\qquad\quad
    -\frac{3z+29z^2+162z^3+334z^4+356z^5+191z^6+41z^7}{(z+1)^4 (1\!+\!z\!+\!z t)}\nn
    \\
    &\qquad\quad
    +\frac{16}{(t+1)^3}
    -\frac{27 z+16}{(t+1)^2}
    -\frac{-41 z^2-27 z-2}{t+1}\biggr\}\nn
    \\&\nn\\
    &{^4}\!K^{(1)}_{q\to qq\bar{q}}(z,t)
    \\
    &=
    4C_F\bigl(C_F-\tfrac{1}{2}C_A\bigr)\biggl\{\nn
    \biggl[-\frac{z^2+2 z+1}{(1\!+\!z\!+\!z t)^3}
    -\frac{z+z^2-t}{(1\!+\!z\!+\!z t)^2}
    -\frac{1+z+z^2+t^2}{1\!+\!z\!+\!z t}\nn
    \\
    &\qquad\qquad\quad
    +\frac{t+t^3}{(t+1) (1+z t)}
    +\frac{1+z}{t+1}
    \biggr] 
    \ln\biggl(\frac{t(1\!+\!z\!+\!z t)}{(1+t)(1+z t)}\biggr)\nn
    \\
    &\qquad\quad
    -\frac{1+2z+z^2}{(1\!+\!z\!+\!z t)^3}
    +\frac{2-z-3z^2+3t}{(1\!+\!z\!+\!z t)^2}
    -\frac{1+z+5z^2+6t+5t^2}{1\!+\!z\!+\!z t}\nn
    \\
    &\qquad\quad
    +\frac{2 t}{(1+z t)^2}
    +\frac{5 t^2+t}{1+z t}
    -\frac{2}{(t+1)^2}
    +\frac{5 z+1}{t+1}
    \biggr\}\nn
    \\
    &\quad
    +4C_F T_F\biggl\{
    \biggl[\frac{1}{z}\biggr]_+
        \biggl[\frac{1+t^2}{(1+t)^4}\ln\biggl(\frac{t}{(1+t)^2}\biggr)
            -\frac{1-6t+t^2}{(t+1)^4}\biggr]
    +\frac{1+t^2}{(t+1)^4}\frac{\ln(1\!+\!z\!+\!z t)}{z}\nn
    \\
    &\qquad\quad
    +\biggl[
    +\frac{z^3+z^2+z/2}{(1\!+\!z\!+\!z t)^3}
    +\frac{4z^3+3z^2+z}{(1\!+\!z\!+\!z t)^2}
    +\frac{11z^3+7z^2+2z}{1\!+\!z\!+\!z t}\nn
    \\
    &\qquad\qquad\quad
    -\frac{4}{(t+1)^3}
    +\frac{7z+4}{(t+1)^2}
    -\frac{11z^2+7z+2}{t+1}
    \biggr]
    \ln\biggl(\frac{t(1\!+\!z\!+\!z t)}{(1+t)^2}\biggr)\nn
    \\
    &\qquad\quad
    +\biggl[
    +\frac{z/2-t}{(1\!+\!z\!+\!z t)^3}
    +\frac{1+4t+4t^2}{(1\!+\!z\!+\!z t)^2}
    -\frac{2+9t+18t^2+11t^3}{1\!+\!z\!+\!z t}\nn
    \\
    &\qquad\qquad\quad
    +\frac{2}{(1+z t)^4}
    -\frac{2-4t}{(1+z t)^3}
    +\frac{1+7t^2}{(1+z t)^2}
    +\frac{2t+7t^2+11t^3}{1+z t}
    \biggr]
    \ln\biggl(\frac{t(1\!+\!z\!+\!z t)}{(1+z t)^2}\biggr)\nn
    \\
    &\qquad\quad
    -\frac{2z+3z^2+3z^3-3t}{(1\!+\!z\!+\!z t)^3}
    -\frac{1+z+11z^2+14z^3+14t+14t^2}{(1\!+\!z\!+\!z t)^2}\nn
    \\
    &\qquad\quad
    +\frac{2-2z-27z^2-41z^3+29t+68t^2+41t^3}{1\!+\!z\!+\!z t}\nn
    \\
    &\qquad\quad
    -\frac{8}{(1+z t)^4}
    +\frac{8-16 t}{(1+z t)^3}
    +\frac{-27 t^2-1}{(1+z t)^2}
    +\frac{-41 t^3-27 t^2-2 t}{1+z t}\nn
    \\
    &\qquad\quad
    +\frac{16}{(t+1)^3}
    +\frac{-27 z-16}{(t+1)^2}
    +\frac{41 z^2+27 z+2}{t+1}
    \biggr\}\nn
\end{align}

\subsubsection*{$\boldsymbol{q\to qgg}$}

\begin{align}
    &{^1}\!K^{(1)}_{q\to qgg}(z,t)
    \\
    &=
    2C_F^2\biggl\{\biggl[\frac{2+4z+3z^2}{(1\!+\!z\!+\!z t)^3}
    -\frac{4+6z}{(1\!+\!z\!+\!z t)^2}
    +\frac{5+t^2}{1\!+\!z\!+\!z t}
    -\frac{t}{(1+z t)^2}
    +\frac{t-t^2}{1+z t}\biggr] 
    \ln\biggl(\frac{t(1\!+\!z\!+\!z t)}{(1+z t)^2}\biggr)\nn
    \\
    &\qquad\quad
    +\biggl[\frac{3+4z+2z^2}{(1\!+\!z\!+\!z t)^3}
    -\frac{6+4z}{(1\!+\!z\!+\!z t)^2}
    +\frac{6+2z+z^2}{1\!+\!z\!+\!z t}
    -\frac{2+z}{t+1}
    +\frac{1}{(t+1)^2}
    \biggr]
    \ln\biggl(\frac{t(1\!+\!z\!+\!z t)}{(1+t)^2}\biggr)\nn
    \\
    &\qquad\quad
    -\frac{3+4z+3z^2}{(1\!+\!z\!+\!z t)^3}
    +\frac{2-3z-5z^2+5t}{(1\!+\!z\!+\!z t)^2}
    +\frac{1-3z-7z^2-6t-7t^2}{1\!+\!z\!+\!z t}\nn
    \\
    &\qquad\quad
    +\frac{2t}{(1+z t)^2}
    -\frac{t-7t^2}{1+z t}
    -\frac{2}{(t+1)^2}
    +\frac{3+7z}{t+1}
    \biggr\}\nn
    \\
    &\quad
    +2C_A C_F\biggl\{
    -\biggl[\frac{3+8z+12z^2+8z^3+3z^4}{(z+1)^2 (1\!+\!z\!+\!z t)^3}
    -\frac{6+16z+24z^2+16z^3+6z^4}{(z+1)^3 (1\!+\!z\!+\!z t)^2}\nn
    \\
    &\qquad\qquad\quad
    +\frac{6+16z+24z^2+16z^3+6z^4}{(z+1)^4 (1\!+\!z\!+\!z t)}
    \biggr]
    \ln\biggl(\frac{t(1+z)^2}{1\!+\!z\!+\!z t}\biggr)\nn
    \\
    &\qquad\quad
    +\biggl[
    \frac{1}{(1\!+\!z\!+\!z t)^3}
    -\frac{2}{(1\!+\!z\!+\!z t)^2}
    +\frac{2}{1\!+\!z\!+\!z t}
    \biggr]
    \frac{1}{1+z}
    \ln\biggl(\frac{1\!+\!z\!+\!z t}{(1+t)(1+z)}\biggr)\nn
    \\
    &\qquad\quad
    +\biggl[
    \frac{z^3}{(1\!+\!z\!+\!z t)^3}
    -\frac{2z^2}{(1\!+\!z\!+\!z t)^2}
    +\frac{2z}{1\!+\!z\!+\!z t}
    \biggr]
    \frac{1}{1+z}
    \ln\biggl(\frac{1\!+\!z\!+\!z t}{(1+z)(1+z t)}\biggr)\nn
    \\
    &\qquad\quad
    -\frac{2+3z+6z^2+3z^3+2z^4}{(z+1)^2 (1\!+\!z\!+\!z t)^3}
    -\frac{2z-8z^2+2z^3}{(z+1)^3 (1\!+\!z\!+\!z t)^2}
    +\frac{2z-8z^2+2z^3}{(z+1)^4 (1\!+\!z\!+\!z t)}
    \biggr\}\nn
    \\&\nn\\
    &{^3}\!K^{(1)}_{q\to qgg}(z,t)
    \\
    &=
    2C_F^2\biggl\{
    \frac{6+12z+12z^2}{(z+1)^3}\,
    \frac{1}{t}\ln\biggl(\frac{1\!+\!z\!+\!z t}{(1+z)(1+t)}\biggr)\nn
    \\
    &\qquad\quad
    +\biggl[\frac{2z^2+z^3}{(z+1) (1\!+\!z\!+\!z t)^3}
    +\frac{2z+4z^2+z^3}{(z+1)^2 (1\!+\!z\!+\!z t)^2}
    +\frac{z-z^2}{1\!+\!z\!+\!z t}\nn
    \\
    &\qquad\qquad\quad
    -\frac{1}{(t+1)^2}
    +\frac{z-1}{t+1}\biggr]
    \ln\biggl(\frac{t(1\!+\!z\!+\!z t)}{(1+t)^2}\biggr)\nn
    \\
    &\qquad\quad
    +\biggl[\frac{2z+4z^2+6z^3}{(z+1) (1\!+\!z\!+\!z t)^3}
    +\frac{2z+4z^2+6z^3}{(z+1)^2 (1\!+\!z\!+\!z t)^2}
    +\frac{6z+12z^2+12z^3}{(z+1)^3 (1\!+\!z\!+\!z t)}\biggr] 
    \ln\biggl(\frac{(1+t)(1+z)}{1\!+\!z\!+\!z t}\biggr)\nn
    \\
    &\qquad\quad
    +\frac{2z+2z^2+3z^3}{(z+1) (1\!+\!z\!+\!z t)^3}
    +\frac{6z^2+9z^3+5z^4}{(z+1)^2 (1\!+\!z\!+\!z t)^2}
    -\frac{z-7z^2}{1\!+\!z\!+\!z t}
    +\frac{1-7z}{t+1}
    +\frac{2}{(t+1)^2}
    \biggr\}\nn
    \\
    &\quad
    +2C_A C_F\biggl\{
    +\frac{4+8z+8z^2}{(z+1)^3}\biggl[\frac{\ln t}{t}\biggr]_+
    +\biggl[\frac{4+8z+8z^2}{(z+1)^3} \log (z+1)-\frac{2}{(z+1)^3}\biggr]
        \biggl[\frac{1}{t}\biggr]_+\nn
    \\
    &\qquad\quad
    +\frac{1+2z+2z^2}{(z+1)^3}\,\frac{1}{t}
    \biggl[4\ln\biggl(\frac{1\!+\!z\!+\!z t}{1+z}\biggr)-\ln(1+t)-7\ln(1+z t)\biggr]\nn
    \\
    &\qquad\quad
    +\biggl[
    -\frac{7z+7z^2+4z^3}{(z+1) (1\!+\!z\!+\!z t)^3}
    +\frac{2t}{(1\!+\!z\!+\!z t)^3}
    -\frac{12+24z+12z^2+4z^3}{(z+1)^2 (1\!+\!z\!+\!z t)^2}
    +\frac{-8t-8t^2}{(1\!+\!z\!+\!z t)^2}\nn
    \\
    &\qquad\qquad\quad
    +\frac{20+56z+52z^2+12z^3}{(z+1)^3 (1\!+\!z\!+\!z t)}
    +\frac{34t+36t^2+22t^3}{1\!+\!z\!+\!z t}\nn
    \\
    &\qquad\qquad\quad
    -\frac{4}{(1+z t)^4}
    +\frac{4-8t}{(1+z t)^3}
    -\frac{8+14t^2}{(1+z t)^2}
    -\frac{20t+14t^2+22t^3}{1+z t}
    \biggr]
    \ln\biggl(\frac{t(1\!+\!z\!+\!z t)}{(1+z t)^2}\biggr)\nn
    \\
    &\qquad\quad
    +\biggl[
    -\frac{1}{(z+1) (1\!+\!z\!+\!z t)^3}
    +\frac{3+5z+z^2}{(z+1)^2 (1\!+\!z\!+\!z t)^2}
    +\frac{t}{(1\!+\!z\!+\!z t)^2}\nn
    \\
    &\qquad\qquad\quad
    -\frac{4+11z+10z^2+2z^3}{(z+1)^3 (1\!+\!z\!+\!z t)}
    -\frac{3t+t^2}{1\!+\!z\!+\!z t}
    +\frac{2+2t+t^2}{1+z t}
    \biggr]
    \ln\biggl(\frac{1+t}{1+z t}\biggr)\nn
    \\
    &\qquad\quad
    +\biggl[
    +\frac{1+2z+z^2}{(1\!+\!z\!+\!z t)^3}
    +\frac{1+z-t}{(1\!+\!z\!+\!z t)^2}
    -\frac{t-t^2}{1\!+\!z\!+\!z t}
    -\frac{2-2t+t^2}{1+z t}
    \biggr]
    \ln\biggl(\frac{t(1+z)}{1+z t}\biggr)\nn
    \\
    &\qquad\quad
    +\frac{5z+5z^2+2z^3}{(z+1) (1\!+\!z\!+\!z t)^3}
    -\frac{4t}{(1\!+\!z\!+\!z t)^3}
    +\frac{2+4z-2z^2-2z^3}{(z+1)^2 (1\!+\!z\!+\!z t)^2}
    -\frac{20t+20t^2}{(1\!+\!z\!+\!z t)^2}\nn
    \\
    &\qquad\quad
    -\frac{4+10z+12z^2+4z^3}{(z+1)^3 (1\!+\!z\!+\!z t)}
    -\frac{44t+100t^2+60t^3}{1\!+\!z\!+\!z t}\nn
    \\
    &\qquad\quad
    +\frac{12}{(1+z t)^4}
    -\frac{12-24 t}{(1+z t)^3}
    -\frac{-40 t^2-2}{(1+z t)^2}
    -\frac{-60 t^3-40 t^2-4 t}{1+z t}
    \biggr\}\nn
    \\&\nn\\
    &{^5}\!K^{(1)}_{q\to qgg}(z,t)
    \\
    &=
    2C_F^2\biggl\{
    -\frac{12}{z t}\ln\biggl(\frac{(1+z)(1+z t)}{1\!+\!z\!+\!z t}\biggr)
    +\frac{24+30z+12z^2}{(z+1)^3}
        \frac{1}{t}\ln\biggl(\frac{(1+z)(1+z t)}{1\!+\!z\!+\!z t}\biggr)\nn
    \\
    &\qquad\quad
    +\biggl[\frac{1+2z}{(z+1) (1\!+\!z\!+\!z t)^3}
    +\frac{1+4z+2z^2}{(z+1)^2 (1\!+\!z\!+\!z t)^2}
    -\frac{2+3t+t^2}{1\!+\!z\!+\!z t}\nn
    \\
    &\qquad\qquad\quad
    +\frac{t}{(1+z t)^2}
    +\frac{2t+t^2}{1+z t}
    \biggr] 
    \ln\biggl(\frac{t(1\!+\!z\!+\!z t)}{(1+z t)^2}\biggr)\nn
    \\
    &\qquad\quad
    +\biggl[\frac{6+4z+2z^2}{(z+1) (1\!+\!z\!+\!z t)^3}
    +\frac{6+4z+2z^2}{(z+1)^2 (1\!+\!z\!+\!z t)^2}
    +\frac{12+12z+6z^2}{(z+1)^3 (1\!+\!z\!+\!z t)}
    \biggr] 
    \ln\biggl(\frac{(1\!+\!z)(1\!+\!z t)}{1\!+\!z\!+\!z t}\biggr)\nn
    \\
    &\qquad\quad
    +\frac{3+2z+2z^2}{(z+1) (1\!+\!z\!+\!z t)^3}
    -\frac{1-z}{(z+1)^2 (1\!+\!z\!+\!z t)^2}
    -\frac{5+5t}{(1\!+\!z\!+\!z t)^2}
    +\frac{3+10t+7t^2}{1\!+\!z\!+\!z t}\nn
    \\
    &\qquad\quad
    -\frac{2 t}{(1+z t)^2}
    -\frac{3t+7t^2}{1+z t}
    \biggr\}\nn
    \\
    &\quad
    +2C_A C_F\biggl\{\biggl[\frac{1}{z}\biggr]_+
    \biggl[8\biggl[\frac{\ln t}{t}\biggr]_+
    -\frac{10+18t+14t^2+2t^3}{(t+1)^4}\ln t\nn
    \\
    &\qquad\qquad\quad
    -\frac{14+40t+60t^2+40t^3+14t^4}{(t+1)^4}\frac{\ln(1+t)}{t}
    +\frac{2-8t+2t^2}{(t+1)^4}
    \biggr]\nn
    \\
    &\qquad\quad
    -\frac{16+20z+8z^2}{(z+1)^3}\biggl[\frac{\ln t}{t}\biggr]_+
    +\biggl[\frac{8+8z+4z^2}{(z+1)^3}\,\frac{\ln(1+z)}{z}-\frac{2z}{(z+1)^3}\biggr]
        \biggl[\frac{1}{t}\biggr]_+\nn
    \\
    &\qquad\quad
    +\frac{8}{z t}\ln\biggl(\frac{1\!+\!z\!+\!z t}{1+z}\biggr)
    -2\frac{\ln(1+z t)}{z t}
    +\frac{2}{t+1}\,\frac{1}{z}\ln\biggl(\frac{1+z t}{1+z}\biggr)
    +\frac{8+12z+8z^2}{(t+1)^4}\,\frac{\ln(1\!+\!z\!+\!z t)}{z}\nn
    \\
    &\qquad\quad
    +\frac{4+5z+2z^2}{(z+1)^3}
    \frac{1}{t}\biggl[7\ln(1+t)+\ln(1+z t)
        -4\ln\biggl(\frac{1\!+\!z\!+\!z t}{1+z}\biggr)\biggr]
    -\frac{18}{t+1}\ln(1\!+\!z\!+\!z t)\nn
    \\
    &\qquad\quad
    +\biggl[
    -\frac{3+z+5z^2+2z^3}{(1\!+\!z\!+\!z t)^3}
    -\frac{3+2z+9z^2+8z^3}{(1\!+\!z\!+\!z t)^2}
    -\frac{6+10z+17z^2+22z^3}{1\!+\!z\!+\!z t}\nn
    \\
    &\qquad\qquad\quad
    +\frac{8}{(t+1)^3}
    -\frac{8+14z}{(t+1)^2}
    +\frac{10+17z+22z^2}{t+1}
    \biggr]
    \ln\biggl(\frac{t}{1\!+\!z\!+\!z t}\biggr)\nn
    \\
    &\qquad\quad
    +\biggl[
    +\frac{z^3}{(z+1) (1\!+\!z\!+\!z t)^3}
    -\frac{2z^2-z^4}{(z+1)^2 (1\!+\!z\!+\!z t)^2}\nn
    \\
    &\qquad\qquad\quad
    +\frac{2z-z^3+z^4+z^5}{(z+1)^3 (1\!+\!z\!+\!z t)}
    +\frac{2-z}{t+1}
    \biggr]
    \ln\biggl(\frac{t^4 (1+t)}{(1\!+\!z\!+\!z t)^4 (1+z t)}\biggr)\nn
    \\
    &\qquad\quad
    +\biggl[
    +\frac{1+2z+z^2}{(1\!+\!z\!+\!z t)^3}
    +\frac{1+2z+z^2}{(1\!+\!z\!+\!z t)^2}
    +\frac{2+2z+z^2}{(1\!+\!z\!+\!z t)}
    -\frac{2+z}{t+1}
    \biggr]
    \ln\biggl(\frac{(1+z)(1\!+\!z\!+\!z t)}{1+t}\biggr)\nn
    \\
    &\qquad\quad
    +\frac{2+5z+9z^2+8z^3+4z^4}{(z+1) (1\!+\!z\!+\!z t)^3}
    -\frac{2+2z-20z^2-54z^3-56z^4-20z^5}{(z+1)^2 (1\!+\!z\!+\!z t)^2}\nn
    \\
    &\qquad\quad
    +\frac{4z+54z^2+192z^3+304z^4+220z^5+60z^6}{(z+1)^3 (1\!+\!z\!+\!z t)}\nn
    \\
    &\qquad\quad
    -\frac{24}{(t+1)^3}
    +\frac{40 z+24}{(t+1)^2}
    +\frac{-60 z^2-40 z-4}{t+1}
    \biggr\}\nn
\end{align}

\subsubsection*{$\boldsymbol{g\to gq\bar{q}}$}

\begin{align}
    &{^1}\!K^{(1)}_{g\to gq\bar{q}}(z,t)
    \\
    &=4C_F T_F\biggl\{
    8\frac{1+z^2}{(1+z)^4}\biggl[\frac{\ln t}{t}\biggr]_+
    +8\biggl[\frac{1+z^2}{(1+z)^4}\ln(1+z)-\frac{z}{(1+z)^4}\biggr]
        \biggl[\frac{1}{t}\biggr]_+\nn
    \\
    &\qquad\quad
    +8\frac{1+z^2}{(1+z)^4}
    \frac{1}{t}\ln\biggl(\frac{1\!+\!z\!+\!z t}{(1+z)(1+t)(1+z t)}\biggr)\nn
    \\
    &\qquad\quad
    +\biggl[
    -\frac{4z+4z^3}{(z+1) (1\!+\!z\!+\!z t)^4}
    +\frac{6z^2-4z^3-2z^4}{(z+1)^2 (1\!+\!z\!+\!z t)^3}
    -\frac{3z+10z^3+8z^4+3z^5}{(z+1)^3 (1\!+\!z\!+\!z t)^2}\nn
    \\
    &\qquad\qquad\quad
    -\frac{3z+14z^3+20z^4+15z^5+4z^6}{(z+1)^4 (1\!+\!z\!+\!z t)}
    -\frac{1}{(t+1)^2}
    -\frac{1-4z}{t+1}
    \biggr]
    \ln\biggl(\frac{t(1\!+\!z\!+\!z t)}{(1+t)^2}\biggl)\nn
    \\
    &\qquad\quad
    +\biggl[
    -\frac{4z+4z^3}{(z+1) (1\!+\!z\!+\!z t)^4}
    -\frac{2+4z-6z^2}{(z+1)^2 (1\!+\!z\!+\!z t)^3}
    -\frac{5+10z+3z^3-3t}{(z+1)^3 (1\!+\!z\!+\!z t)^2}\nn
    \\
    &\qquad\qquad\quad
    +\frac{-9+10t-4t^2-14z-3z^3}{(z+1)^4 (1\!+\!z\!+\!z t)}
    +\frac{10-4 t}{(z+1)^4}
    +\frac{6-4 t}{(z+1)^3}
    +\frac{2-4 t}{(z+1)^2}
    +\frac{-4 t-2}{z+1}
    \nn
    \\
    &\qquad\qquad\quad
    +\frac{t}{(1+z t)^2}
    +\frac{4 t^2+2 t}{1+z t}
    \biggr]
    \ln\biggl(\frac{t(1\!+\!z\!+\!z t)}{(1+z t)^2}\biggr)\nn
    \\
    &\qquad\quad
    -\frac{8z+8z^2+8z^3}{(z+1) (1\!+\!z\!+\!z t)^4}
    -\frac{3+6z-2z^2+6z^3+3z^4}{(z+1)^2 (1\!+\!z\!+\!z t)^3}
    +\frac{8 z^2}{(z+1)^3 (1\!+\!z\!+\!z t)^2}\nn
    \\
    &\qquad\quad
    +\frac{2-2z^2+2t}{(1\!+\!z\!+\!z t)^2}
    +\frac{8 z^2}{(z+1)^4 (1\!+\!z\!+\!z t)}
    +\frac{1-z^2-t^2}{1\!+\!z\!+\!z t}\nn
    \\
    &\qquad\quad
    -\frac{t-t^2}{1+z t}
    -\frac{t}{(1+z t)^2}
    +\frac{1}{(t+1)^2}
    +\frac{z}{t+1}
    \biggr\}\nn
    \\
    &\quad
    +4C_A T_F\biggl\{
    -4\frac{1+z^2}{(1+z)^4}\biggl[\frac{\ln t}{t}\biggr]_+
    -4\biggl[\frac{1+z^2}{(1+z)^4}\ln(1+z)-\frac{z}{(1+z)^4}\biggr]
        \biggl[\frac{1}{t}\biggr]_+\nn
    \\
    &\qquad\quad
    +\frac{1+z^2}{(1+z)^4}\frac{1}{t}
    \ln\biggl(\frac{(1\!+\!z\!+\!z t)^2(1+t)(1+z t)}{(1+z)^2}\biggr)\nn
    -\frac{z-z^2}{(1+z)(1\!+\!z\!+\!z t)^3}
        \ln\biggl(\frac{t(1\!+\!z\!+\!z t)}{(1+z t)^2}\biggr)
    \\
    &\qquad\quad
    +\biggl[\frac{4z+4z^3}{(z+1) (1\!+\!z\!+\!z t)^4}
    -\frac{2z+2z^2+2z^3}{(z+1)^2 (1\!+\!z\!+\!z t)^3}
    +\frac{6z+2z^2+6z^3}{(z+1)^3 (1\!+\!z\!+\!z t)^2}\nn
    \\
    &\qquad\qquad\quad
    +\frac{z+z^3}{(z+1)^4 (1\!+\!z\!+\!z t)}
    \biggr]
    \ln\biggl(\frac{t(1+z)^2}{1\!+\!z\!+\!z t}\biggr)\nn
    \\
    &\qquad\quad
    +
    \biggl[
    \frac{z-2z^2-z^3}{(1\!+\!z)^2 (1\!+\!z\!+\!z t)^3}
    -\frac{2z^2}{(1\!+\!z)^3 (1\!+\!z\!+\!z t)^2}
    +\frac{z+z^3}{(1\!+\!z)^4 (1\!+\!z\!+\!z t)}
    \biggr]
    \ln\biggl(\frac{t(1\!+\!z\!+\!z t)}{(1\!+\!t)(1\!+\!z t)}\biggr)\nn
    \\
    &\qquad\quad
    -\frac{4z^2}{(1\!+\!z) (1\!+\!z\!+\!z t)^4}
    -\frac{z-4z^2+z^3}{(1\!+\!z)^2 (1\!+\!z\!+\!z t)^3}
    +\frac{2z-16z^2+2z^3}{(1\!+\!z)^3 (1\!+\!z\!+\!z t)^2}
    -\frac{2z-8z^2+2z^3}{(1\!+\!z)^4 (1\!+\!z\!+\!z t)}
    \biggr\}\nn
    \\&\nn\\
    &{^3}\!K^{(1)}_{g\to gq\bar{q}}(z,t)
    \\
    &=4C_F T_F\biggl\{\biggl[\frac{4z^2+2z^3}{(z+1) (1\!+\!z\!+\!z t)^3}
    +\frac{4z+8z^2+8z^3+3z^4}{(z+1)^2 (1\!+\!z\!+\!z t)^2}\nn
    \\
    &\qquad\qquad\quad
    +\frac{4 z^2+2 z}{1\!+\!z\!+\!z t}
    +\frac{1}{(t+1)^2}
    +\frac{-4z-2}{t+1}\biggr]
    \ln\biggl(\frac{t(1\!+\!z\!+\!z t)}{(1+t)^2}\biggr)\nn
    \\
    &\qquad\quad
    +2\biggl[
    \frac{4 z^2+8 z+8}{(1\!+\!z\!+\!z t)^4}
    -\frac{8+10z+4z^2}{(z+1) (1\!+\!z\!+\!z t)^3}
    +\frac{4+6z+3z^2}{(z+1)^2 (1\!+\!z\!+\!z t)^2}\biggr] 
    \ln\biggl(\frac{1\!+\!z\!+\!z t}{(1+z)(1+t)}\biggr)\nn
    \\
    &\qquad\quad
    +\frac{8 z^2+8 z+8}{(1\!+\!z\!+\!z t)^4}
    -\frac{8+10z+2z^2-3z^3}{(z+1) (1\!+\!z\!+\!z t)^3}
    +\frac{3z^3+2z^4}{(z+1)^2 (1\!+\!z\!+\!z t)^2}
    +\frac{z^2-z}{1\!+\!z\!+\!z t}\nn
    \\
    &\qquad\quad
    -\frac{1}{(t+1)^2}
    +\frac{1-z}{t+1}
    \biggr\}\nn
    \\
    &\quad
    +4C_A T_F\biggl\{\biggl[\frac{4 z^2+8 z+8}{(1\!+\!z\!+\!z t)^4}
    +\frac{-10 t-3 z-10}{(1\!+\!z\!+\!z t)^3}
    +\frac{22 t^2+24 t+8}{(1\!+\!z\!+\!z t)^2}
    +\frac{-42 t^3-64 t^2-27 t-5}{1\!+\!z\!+\!z t}\nn 
    \\
    &\qquad\qquad\quad
    +\frac{4}{(1+z t)^4}
    +\frac{8 t-4}{(1+z t)^3}
    +\frac{20 t^2+2}{(1+z t)^2}
    +\frac{42 t^3+22 t^2+5 t}{1+z t}\biggr] 
    \ln\biggl(\frac{t(1\!+\!z\!+\!z t)}{(1+z t)^2}\biggr)\nn
    \\
    &\qquad\quad
    +\frac{4t^2}{1+z t}\ln(1+z t)
    +\biggl[+\frac{4+2z}{(1\!+\!z\!+\!z t)^3}
    +\frac{-4-4t}{(1\!+\!z\!+\!z t)^2}
    +\frac{4t^2+4t}{1\!+\!z\!+\!z t}\biggr]
    \ln\biggl(\frac{1+t}{1+z t}\biggr)\nn
    \\
    &\qquad\quad
    +\biggl[\frac{2 t-z-2}{(1\!+\!z\!+\!z t)^3}
    +\frac{2-2 t^2}{(1\!+\!z\!+\!z t)^2}
    +\frac{2 t^3-t+1}{1\!+\!z\!+\!z t}
    +\frac{-2 t^3+2 t^2-t}{1+z t}\biggr]
    \ln\biggl(\frac{(1+z)(1+t)}{1\!+\!z\!+\!z t}\biggr)\nn
    \\
    &\qquad\quad
    +\frac{-4 z-4}{(1\!+\!z\!+\!z t)^4}
    +\frac{10 t-z+4}{(1\!+\!z\!+\!z t)^3}
    +\frac{-36 t^2-36 t-2}{(1\!+\!z\!+\!z t)^2}
    +\frac{94 t^3+152 t^2+62 t+4}{1\!+\!z\!+\!z t}\nn
    \\
    &\qquad\quad
    -\frac{16}{(1+z t)^4}
    +\frac{16-32 t}{(1+z t)^3}
    +\frac{-58 t^2-2}{(1+z t)^2}
    +\frac{-94 t^3-58 t^2-4 t}{1+z t}
    \biggr\}\nn
    \\&\nn\\
    &{^5}\!K^{(1)}_{g\to gq\bar{q}}(z,t)
    \\
    &=4C_F T_F\biggl\{\biggl[\frac{8 z^3+8 z^2+4 z}{(1\!+\!z\!+\!z t)^4}
    +\frac{-8 z^2-2 z+2}{(1\!+\!z\!+\!z t)^3}
    +\frac{-3 t+4 z-1}{(1\!+\!z\!+\!z t)^2}
    +\frac{4 t^2+3 t-1}{1\!+\!z\!+\!z t}\nn 
    \\
    &\qquad\qquad\quad
    -\frac{t}{(1+z t)^2}
    +\frac{t-4 t^2}{1+z t}\biggr]
    \ln\biggl(\frac{t(1\!+\!z\!+\!z t)}{(1+z t)^2}\biggr)\nn
    \\
    &\qquad\quad
    -\biggl[\frac{8z^3+8z^2+4z}{(1\!+\!z\!+\!z t)^4}
    -\frac{4z+10z^2+8z^3}{(z+1) (1\!+\!z\!+\!z t)^3}
    +\frac{3z+6z^2+4z^3}{(z+1)^2 (1\!+\!z\!+\!z t)^2}\biggr] 
    \ln\biggl(\frac{t(1+z)^2}{1\!+\!z\!+\!z t}\biggr)\nn
    \\
    &\qquad\quad
    -\frac{-8 z^3-8 z^2-8 z}{(1\!+\!z\!+\!z t)^4}
    +\frac{3-2z+10z^2-8z^3}{(z+1) (1\!+\!z\!+\!z t)^3}
    -\frac{1+2z}{(z+1)^2 (1\!+\!z\!+\!z t)^2}
    -\frac{2 t+2}{(1\!+\!z\!+\!z t)^2}\nn
    \\
    &\qquad\qquad\quad
    -\frac{-t^2-t}{1\!+\!z\!+\!z t}
    +\frac{t}{(1+z t)^2}
    -\frac{t^2}{1+z t}
    \biggr\}\nn
    \\
    &\quad
    +4C_A T_F\biggl\{
    \biggl[\frac{1}{z}\biggr]_+
    \biggl[2\frac{1+t^2}{(t+1)^4}\log\biggl(\frac{t}{(1+t)^2}\biggr)
        -\frac{2-12t+2t^2}{(t+1)^4}\biggr]
    +2\frac{1+t^2}{(t+1)^4}\frac{\ln(1\!+\!z\!+\!z t)}{z}\nn
    \\
    &\qquad\quad
    +\biggl[\frac{8 z^3+8 z^2+4 z}{(1\!+\!z\!+\!z t)^4}
    +\frac{10 z^3-3 z}{(1\!+\!z\!+\!z t)^3}
    +\frac{22 z^3+10 z^2+6 z}{(1\!+\!z\!+\!z t)^2}
    +\frac{42 z^3+18 z^2+5 z}{1\!+\!z\!+\!z t}\nn
    \\
    &\qquad\qquad\quad
    -\frac{8}{(t+1)^3}
    +\frac{20 z+8}{(t+1)^2}
    +\frac{-42 z^2-18 z-5}{t+1}\biggr]
    \ln\biggl(\frac{t(1\!+\!z\!+\!z t)}{(1+t)^2}\biggr)\nn
    \\
    &\qquad\quad
    +\biggl[\frac{z+4z^2+2z^3}{(1\!+\!z\!+\!z t)^3}
    +\frac{2z^2+2z^3}{(1\!+\!z\!+\!z t)^2}
    +\frac{z+2z^2+2z^3}{1\!+\!z\!+\!z t}
    -\frac{1+2z+2z^2}{t+1}\biggr]
    \log\biggl(\frac{1\!+\!z\!+\!z t}{(1+z)(1+z t)}\biggr)\nn
    \\
    &\qquad\quad
    +\biggl[\frac{2z+4z^2}{(1\!+\!z\!+\!z t)^3}
    +\frac{4z^2}{(1\!+\!z\!+\!z t)^2}
    +\frac{4z^2}{1\!+\!z\!+\!z t}
    -\frac{4z}{t+1}\biggr]
    \ln\biggl(\frac{1+z t}{1+t}\biggr)\nn
    \\
    &\qquad\quad
    +\frac{-4 z^3-4 z^2}{(1\!+\!z\!+\!z t)^4}
    +\frac{-10 z^3-6 z^2-z}{(1\!+\!z\!+\!z t)^3}
    +\frac{-36 z^3-26 z^2-2 z}{(1\!+\!z\!+\!z t)^2}
    +\frac{-94 z^3-58 z^2-4 z}{1\!+\!z\!+\!z t}\nn
    \\
    &\qquad\quad 
    +\frac{32}{(t+1)^3}
    +\frac{-58 z-32}{(t+1)^2}
    +\frac{94 z^2+58 z+4}{t+1}
    \biggr\}\nn
\end{align}

\subsubsection*{$\boldsymbol{g\to ggg}$}

\begin{align}\label{eq:gg_full}
    &{^1}\!K^{(1)}_{g\to ggg}(z,t)
    \\
    &=
    \frac{\textcolor{magenta}{4}\textcolor{OliveGreen}{C_A^2}}{\textcolor{magenta}{3}}\biggl\{\nn
    \textcolor{magenta}{ \biggl[\frac{1}{z}\biggr]_+ }
    \biggl(
    \textcolor{magenta}{4\biggl[\frac{\ln t}{t}\biggr]_+ 
            -\frac{\ln t}{1+t}-\frac{7\ln(1+t)}{t}}
    -\frac{4+6t+4t^2}{(1+t)^4}\ln\biggl(\frac{t}{(1+t)^2}\biggr)
        +\frac{1-4t+t^2}{(1+t)^4}\biggr)\nn
    \\
    &\qquad\quad
    -\frac{8+12z+8z^2}{(1+z)^4}\biggl[\frac{\ln t}{t}\biggr]_+
    +\textcolor{magenta}{\frac{4\ln(1+z)}{z}\biggl[\frac{1}{t}\biggr]_+}
    -\frac{8+12z+8z^2}{(1+z)^4}\ln(1+z)\biggl[\frac{1}{t}\biggr]_+\nn
    \\
    &\qquad\quad
    +\textcolor{magenta}{\frac{1}{z t}\biggl[10\ln(1\!+\!z\!+\!z t)-10\ln(1+z)-7\ln(1+z t)\biggr]}
    -\frac{4+6t+4t^2}{(1+t)^4}\frac{\ln(1\!+\!z\!+\!z t)}{z}\nn
    \\
    &\qquad\quad
    +\frac{2+3z+2z^2}{(1+z)^4}\frac{1}{t}\biggl[7\ln\bigl((1+t)(1+z t)\bigr)
        -10\ln\Bigl(\frac{1\!+\!z\!+\!z t}{1+z}\Bigr)\biggr]\nn
    \\
    &\qquad\quad
    \textcolor{magenta}{ -\frac{\ln t}{(t+1) (t z+1)} }
    \textcolor{magenta}{+\biggl[\frac{1}{(t+1) (t z+1)}+\frac{1}{(t+1) (z+1)}\biggr]\ln (1+t) }\nn
    \\
    &\qquad\quad
    \textcolor{magenta}{ +\biggl[\frac{1}{(t+1) (z+1)}+\frac{1}{(z+1) (t z+1)}-\frac{1}{t z+1}-\frac{1}{(t+1) z}\biggr]
   \ln (1+z) }\nn
   \\
   &\qquad\quad
   \textcolor{magenta}{ +\biggl[\frac{1}{(t+1) (t z+1)}+\frac{1}{(z+1) (t z+1)}-\frac{1}{t z+1}+\frac{1}{(t+1) z}\biggr] \ln (1+t z) }\nn
   \\
   &\qquad\quad
   \textcolor{magenta}{ +\biggl[\frac{1}{t z+1}-\frac{1}{(t+1) (t z+1)}-\frac{1}{(t+1)
   (z+1)}-\frac{1}{(z+1) (t z+1)}\biggr] \ln (1+z+t z) }
   \nn
    \\
    &\qquad\quad
    -\biggl[
    \frac{8+16z+24z^2+16z^3+8z^4}{(z+1) (1\!+\!z\!+\!z t)^4}
    -\frac{9+22z+9z^2-9z^3-22z^4-10z^5}{(z+1)^2 (1\!+\!z\!+\!z t)^3}
    -\frac{t}{(1\!+\!z\!+\!z t)^3}\nn
    \\
    &\qquad\qquad\quad
    +\frac{17+59z+110z^2+131z^3+126z^4+78z^5+22z^6}{(z+1)^3 (1\!+\!z\!+\!z t)^2}
    +\frac{2t+t^2}{(1\!+\!z\!+\!z t)^2}\nn
    \\
    &\qquad\qquad\quad
    -\frac{2+2z-67z^2-220z^3-366z^4-352z^5-188z^6-42z^7}{(z+1)^4 (1\!+\!z\!+\!z t)}
    -\frac{3t+2t^2+t^3}{1\!+\!z\!+\!z t}\nn
    \\
    &\qquad\qquad\quad
    +\frac{2t+t^2+t^3}{1+z t}
    -\frac{8}{(t+1)^3}
    +\frac{8+20z}{(t+1)^2}
    -\frac{20+20z+42z^2}{t+1}
    \biggr]
    \ln\biggl(\frac{t(1\!+\!z\!+\!z t)}{(1+t)(1+z t)}\biggr)\nn
    \\
    &\qquad\quad
    +\frac{1}{2}\biggl[
    \frac{8+16z+24z^2+16z^3+8z^4}{(z+1) (1\!+\!z\!+\!z t)^4}
    -\frac{7+17z+21z^2+14z^3+4z^4-z^5}{(z+1)^2 (1\!+\!z\!+\!z t)^3}
    -\frac{t}{(1\!+\!z\!+\!z t)^3}\nn
    \\
    &\qquad\qquad\quad
    +\frac{15+37z+56z^2+43z^3+22z^4+4z^5+z^6}{(z+1)^3 (1\!+\!z\!+\!z t)^2}
    +\frac{t^2}{(1\!+\!z\!+\!z t)^2}\nn
    \\
    &\qquad\qquad\quad
    -\frac{2+10z+9z^2-5z^3-16z^4-12z^5-5z^6-z^7}{(z+1)^4 (1\!+\!z\!+\!z t)}
    -\frac{t+t^3}{1\!+\!z\!+\!z t}\nn
    \\
    &\qquad\qquad\quad
    +\frac{2t-t^2+t^3}{1+z t}
    -\frac{2+z+z^2}{t+1}
    \biggr]
    \ln\biggl(\frac{t(1+z)^2}{1\!+\!z\!+\!z t}\biggr)\nn
    \\
    &\qquad\quad
    +\frac{1}{2}\biggl[
    \frac{-1+9z^2+19z^3+9z^4}{(z+1) (1\!+\!z\!+\!z t)^3}
    +\frac{9t}{(1\!+\!z\!+\!z t)^3}
    -\frac{7-8z-13z^2-21z^3+20t+21t^2}{(1\!+\!z\!+\!z t)^2}\nn
    \\
    &\qquad\qquad\quad
    +\frac{18+18z+21z^2+41z^3+37t+60t^2+41t^3}{1\!+\!z\!+\!z t}
    -\frac{4}{(1+z t)^4}
    +\frac{4-8t}{(1+z t)^3}
    -\frac{8+20t^2}{(1+z t)^2}\nn
    \\
    &\qquad\qquad\quad
    -\frac{18t+19t^2+41t^3}{1+z t}
    -\frac{8}{(t+1)^3}
    +\frac{8+20z}{(t+1)^2}
    -\frac{18+21z+41z^2}{t+1}
    \biggr]
    \ln\biggl(\frac{t(1\!+\!z\!+\!z t)}{(1+z t)^2}\biggr)\nn
    \\
    &\qquad\quad
    +\frac{3z+6z^2+6z^3+3z^4+z^5}{(z+1)^2 (1\!+\!z\!+\!z t)^3}
    -\frac{t}{(1\!+\!z\!+\!z t)^3}\nn
    \\
    &\qquad\quad
    +\frac{1+3z+17z^2+32z^3+46z^4+31z^5+8z^6}{(z+1)^3 (1\!+\!z\!+\!z t)^2}
    +\frac{8t+8t^2}{(1\!+\!z\!+\!z t)^2}\nn
    \\
    &\qquad\quad
    -\frac{2+5z-11z^2-108z^3-228z^4-240z^5-127z^6-27z^7}{(z+1)^4 (1\!+\!z\!+\!z t)}
    -\frac{21t+46t^2+27t^3}{1\!+\!z\!+\!z t}\nn
    \\
    &\qquad\quad
    +\frac{6}{(1+z t)^4}
    -\frac{6-12t}{(1+z t)^3}
    +\frac{1+19t^2}{(1+z t)^2}
    +\frac{2t+19t^2+27t^3}{1+z t}\nn
    \\
    &\qquad\quad
    -\frac{12}{(t+1)^3}
    +\frac{12+19z}{(t+1)^2}
    -\frac{2+19z+27z^2}{t+1}
    \biggr\}\nn
\end{align}
To highlight that $\alpha_s C_A$ (i.e. including the factor of $C_A$) corresponds to the $\mathcal{N}=4$ 't Hooft coupling $a$, these factors are shown in green. The terms highlighted in magenta correspond to the NLO kernel $K^{(1)}_{1\to3}$ (\Eq{eq:N4_K1to3}) in $\mathcal{N}=4$ SYM, as will be discussed in \Sec{sec:compare}.\footnote{For \Eq{eq:N4_K1to3}, the expression of $K^{(1)}_{1\to3}$ is written in a more compact form.} The expressions for both the LO and NLO kernels are available in a Mathematica notebook attached with the submission of this article. 



\newpage

\begin{figure}
\begin{center}
\subfloat[]{
\includegraphics[scale=0.38]{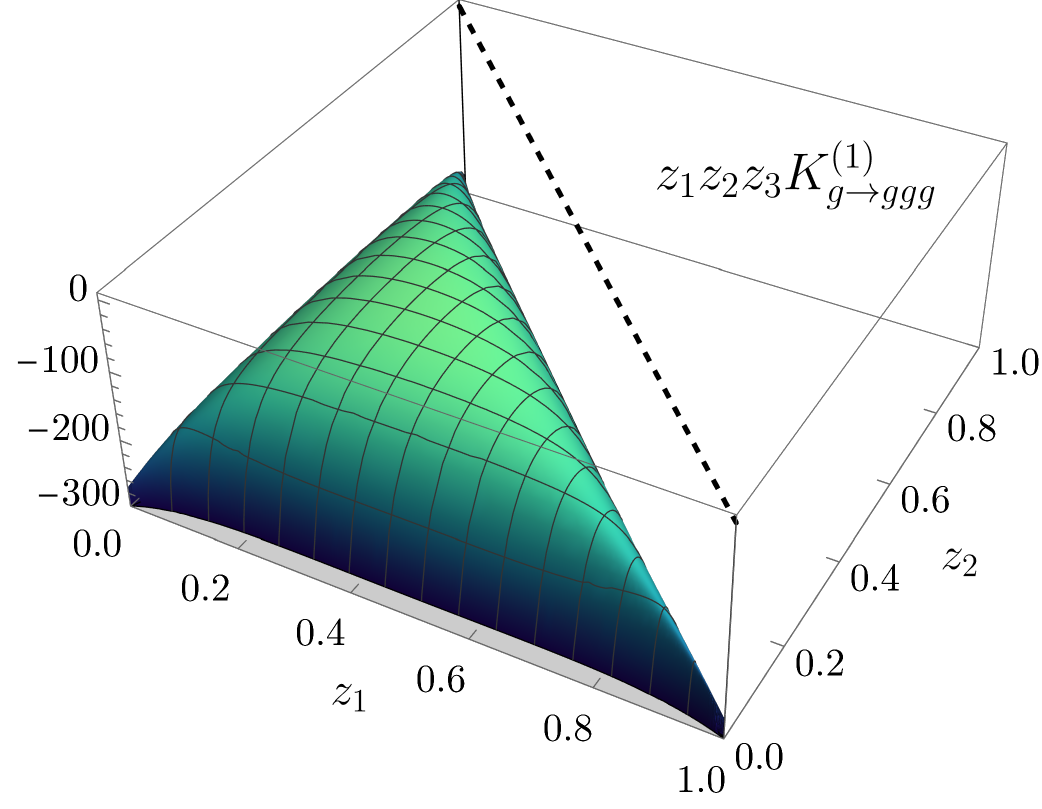}\label{fig:eflow_a}
}\qquad 
\subfloat[]{
\includegraphics[scale=0.38]{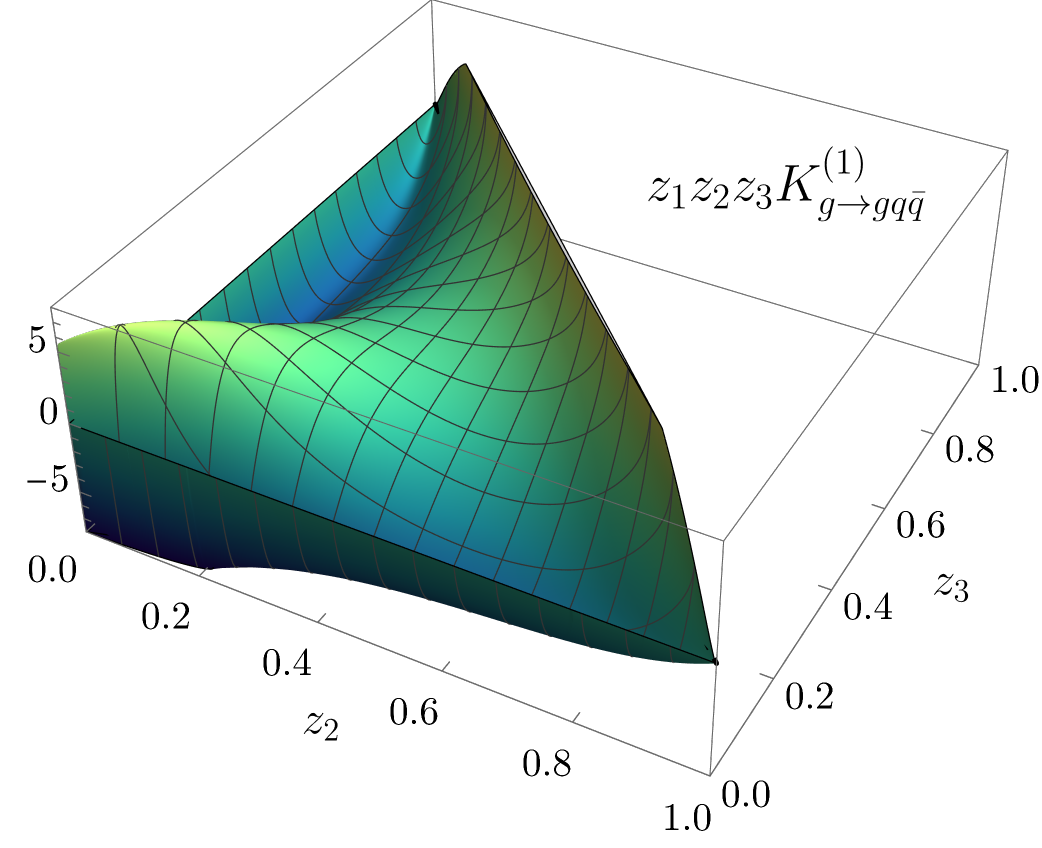}\label{fig:eflow_b}
}\\
\subfloat[]{
\includegraphics[scale=0.38]{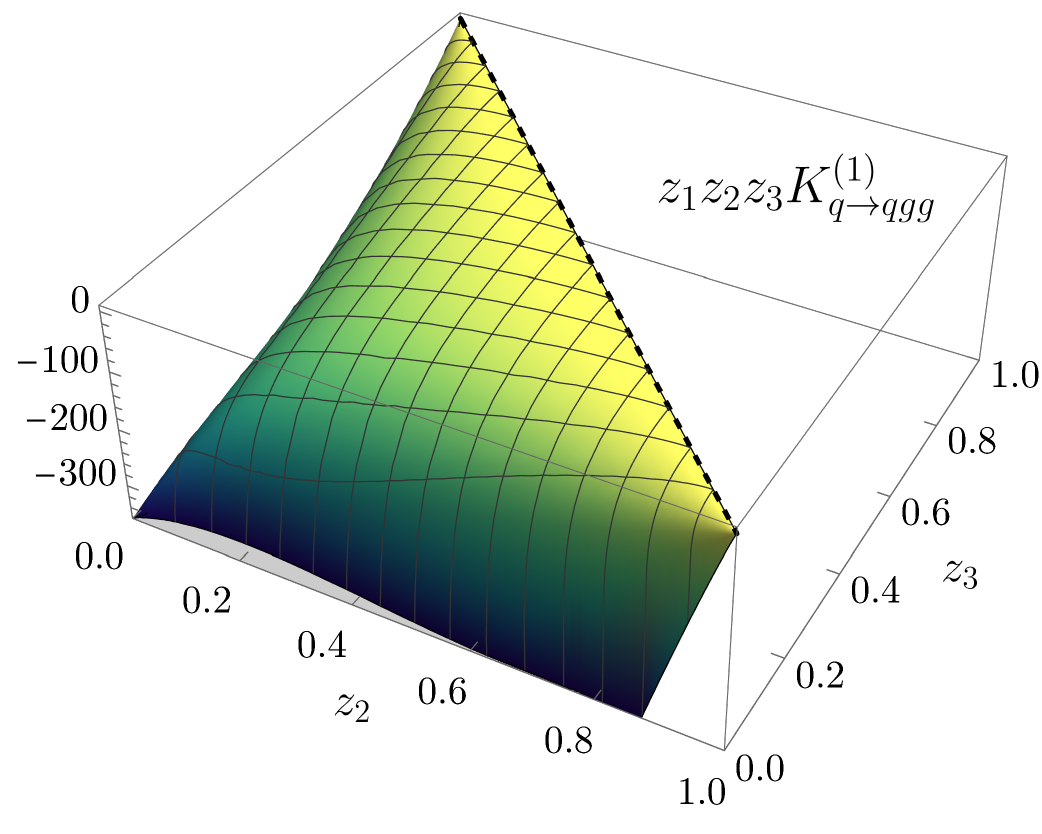}\label{fig:eflow_a}
}\qquad
\subfloat[]{
\includegraphics[scale=0.38]{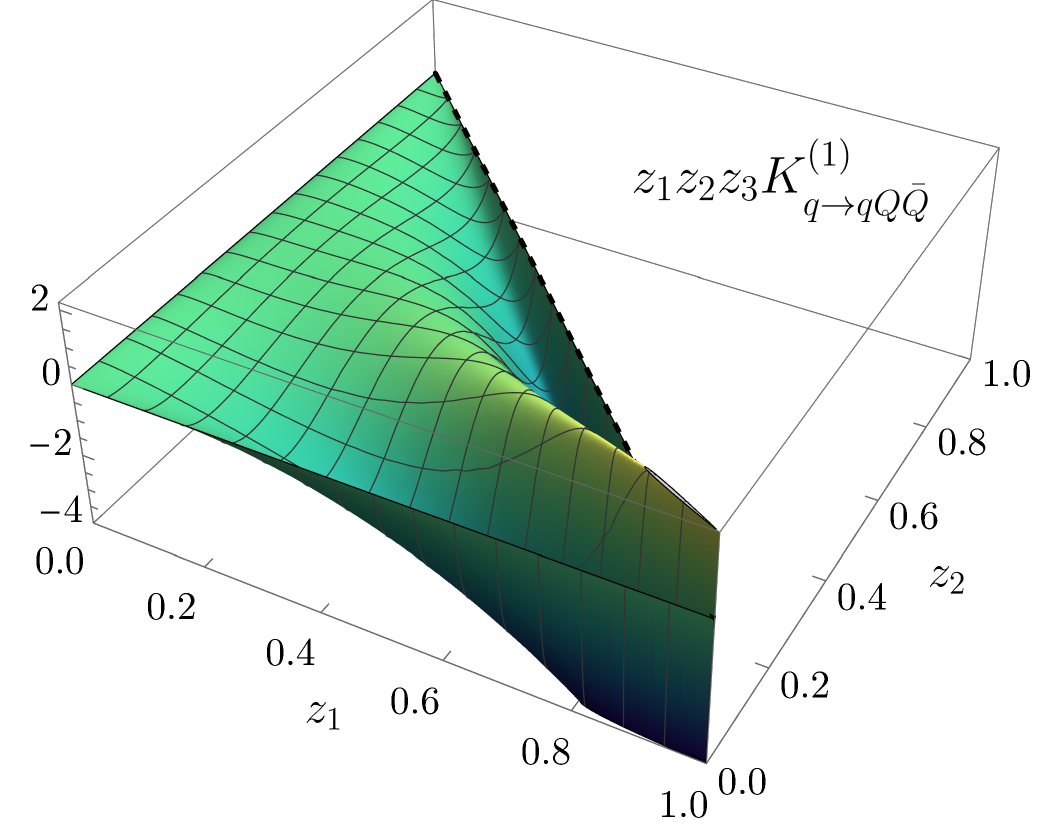}\label{fig:eflow_b}
}
\\
\subfloat[]{
\includegraphics[scale=0.38]{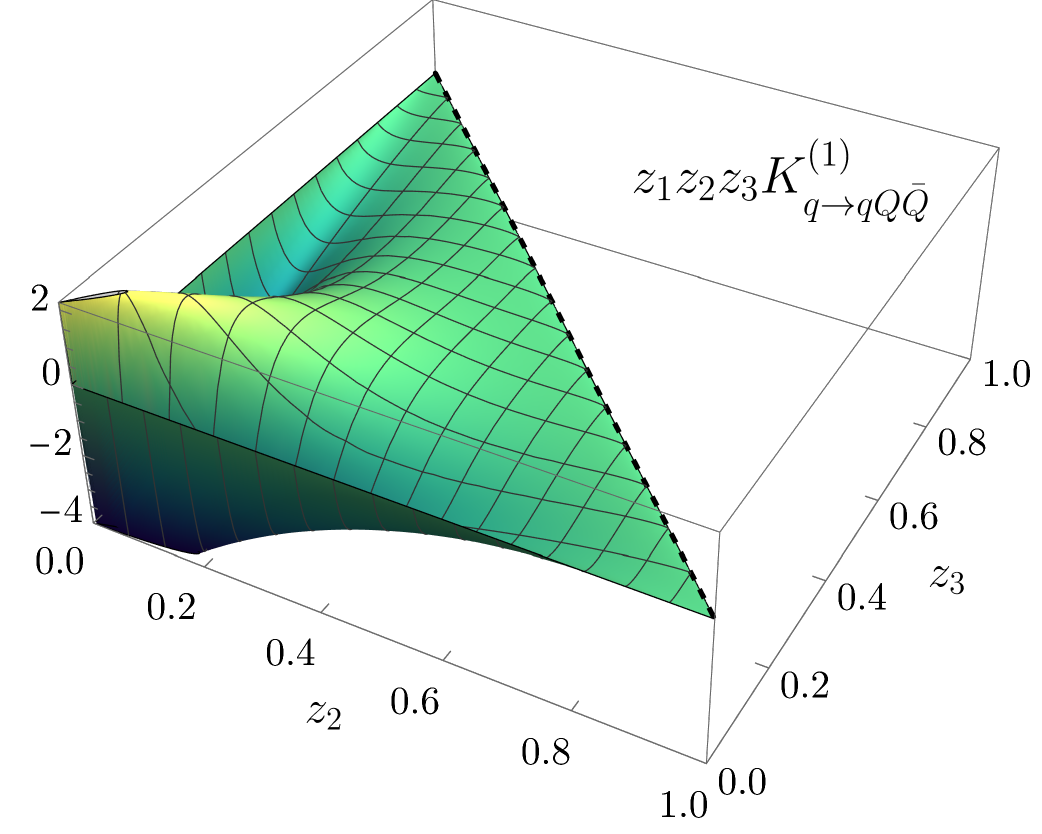}\label{fig:eflow_a}
}\qquad
\subfloat[]{
\includegraphics[scale=0.38]{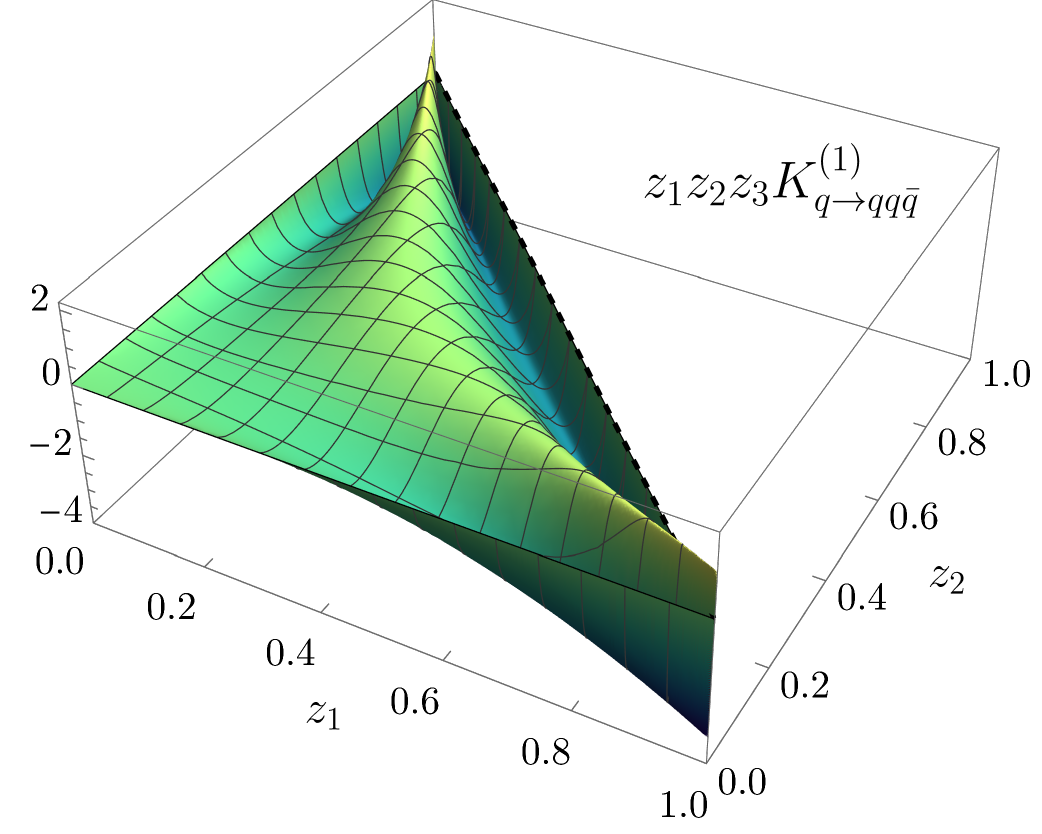}\label{fig:eflow_b}
}
\end{center}
\caption{Two-dimensional plots of the NLO kernels, $K^{(1)}_{i\to i_1i_2i_3}$ appearing in the track function evolution. These exhibit a highly non-trivial structure reflecting how energy is distributed amongst correlated fragmenting partons.}
\label{fig:plots_2d}
\end{figure}

\subsection{Consistency Checks}\label{sec:checks}

Due to the highly non-trivial nature of the kernels, we have performed a number of cross checks on our results, in addition to the already strong cross check that we have obtained an infrared finite evolution equation. Most importantly, we have previously computed the evolution equations for the moments of the track functions \cite{Li:2021zcf,Jaarsma:2022kdd}. We have checked that by directly taking moments of the evolution equations derived here we reproduce the evolution equations for the moments up to the $6$th moment. Additionally, for the case of $\cN=4$, we have analytically checked that by integrating over the kernels, we reproduce NLO DGLAP, as will be described in \Sec{sec:multihadron}. For QCD we have also carried out this check, but only for the first 30 moments.

\subsection{Comparison of the $\cN=4$ and QCD Evolution Equations}\label{sec:compare}

It has been observed that many quantities computed in gauge theories exhibit a ``principle of maximal transcendentality", namely that the terms of highest transcendental weight are universal. Since this was originally discovered in DGLAP \cite{Kotikov:2004er}, it is interesting to understand whether it extends also to the track function evolution. We may expect some version of this principle to hold, since the track function evolution reduces to DGLAP upon integration over momentum fractions, as we describe in \Sec{sec:multihadron}.

It is likely that for a complete analysis we should consider this question in moment space, however, as a preliminary investigation, we simply compare the track function evolution equations for the gluon track function and $\cN=4$ track function. Due to the length of the $T\to TTT$ expressions, here
we focus only on the $T\to T$ and $T\to TT$ kernels.

The NLO kernel in $\cN=4$ is given by
\begin{align}
  \frac{d}{d\ln\mu^2}T_{\cN=4}(x)=\ \textcolor{OliveGreen}{a^2}\
  &\Biggl\{ {\color{magenta}-25 \zeta_3} {\color{red} T(x)}
  +\int_0^1\df x_1\int_0^1\df x_2\int_0^1\df z\ {\color{red}T(x_1)T(x_2)}
  \ \delta\left(x-x_1\frac{1}{1+z}-x_2\frac{z}{1+z}\right)\nn\\
  &\times \left\{ {\color{magenta}16 \zeta_2
  \left[\frac{1}{z}\right]_+  +\frac{32 \ln^2(z+1)}{z}-\frac{16 \ln (z) \ln (z+1)}{z} } \right\}\nn\\
  &+ {\color{magenta} K^{(1)}_{1\to 3}} \otimes {\color{red}TTT}
  \Biggr\}\,,
\end{align}
while the $C_A^2$ component of the NLO evolution of the gluon track function is given by 
\begin{align}\label{eq:N4match_Tg}
&\frac{d}{d\ln\mu^2}T_{g}(x)=\textcolor{OliveGreen}{a_s^2}\Biggl\{
\textcolor{OliveGreen}{C_A^2}\left(\frac{1880}{27}-\frac{44}{3}\zeta_2 -\frac{274}{9}\ln(2)-\frac{44}{3}\ln^2(2)  {\color{magenta}-25 \zeta_3}\right){\color{red}T_g(x)} \nn \\
&+\int_0^1\df x_1\int_0^1\df x_2\int_0^1\df z\ {\color{red}T_g(x_1)T_g(x_2)}
  \ \delta\left(x-x_1\frac{1}{1+z}-x_2\frac{z}{1+z}\right) \nn \\
&\times \textcolor{OliveGreen}{C_A^2}
\Big\{ {\color{magenta} 16 \zeta_2 \left[ \frac{1}{z}\right]_+ } + {\color{magenta}16\left(2\ln(1+z)^2-\ln(z)\ln(1+z)\right)} \left( {\color{magenta}\frac{1}{z}}-\frac{1}{(1+z)^4}+\frac{1}{(1+z)^3}-\frac{2}{(1+z)^2}   \right)\nn \\
&-16 \frac{\zeta_2}{(1+z)^4} +16 \frac{\zeta_2}{(1+z)^3}+\frac{4/3-32 \zeta_2}{(1+z)^2} \Big \}+\ 6\ {^1}\!K^{(1)}_{g\to ggg} \otimes {\color{red}\,T_g T_g T_g}
\Biggr\}\,.
\end{align}
In the gluon kernel, we have highlighted in magenta the terms that appear in the $\cN=4$ kernel. These are terms of maximal weight if one assigns $1/z$ to have weight 1, and these terms will give rise to the terms of maximal weight in the DGLAP kernels, upon integrating out momentum fractions. We observe that the same is true for the full $1\to 3$ kernels, with the corresponding terms highlighted in \Eq{eq:gg_full}. It would be interesting to understand this relation better. 

\subsection{Plots of QCD Evolution Kernels}\label{sec:plots}

Since the QCD kernels are quite complicated multi-variable functions, it is interesting to plot their form as a function of energy fraction variables (as opposed to the sector decomposed variables). 

In \Fig{fig:plots_2d} we collect several plots of the $K^{(1)}_{i\to i_1i_2i_3}$ kernels. These exhibit highly non-trivial features, exhibiting correlations in energy flow in the fragmentation process. While they take a relatively simple shape for $g\to ggg$ and $g\to g q\bar q$, the kernels for $q\to q q \bar q$ and $q \to q q' \bar q'$ exhibit interesting patterns. In particular, we believe that the structure of $q \to q q \bar q$ arises from the non-trivial interference diagrams that contribute to this splitting.

\section{Reduction to DGLAP and Multi-hadron Fragmentation}\label{sec:multihadron}
\begin{figure}
\begin{center}
\includegraphics[scale=0.35]{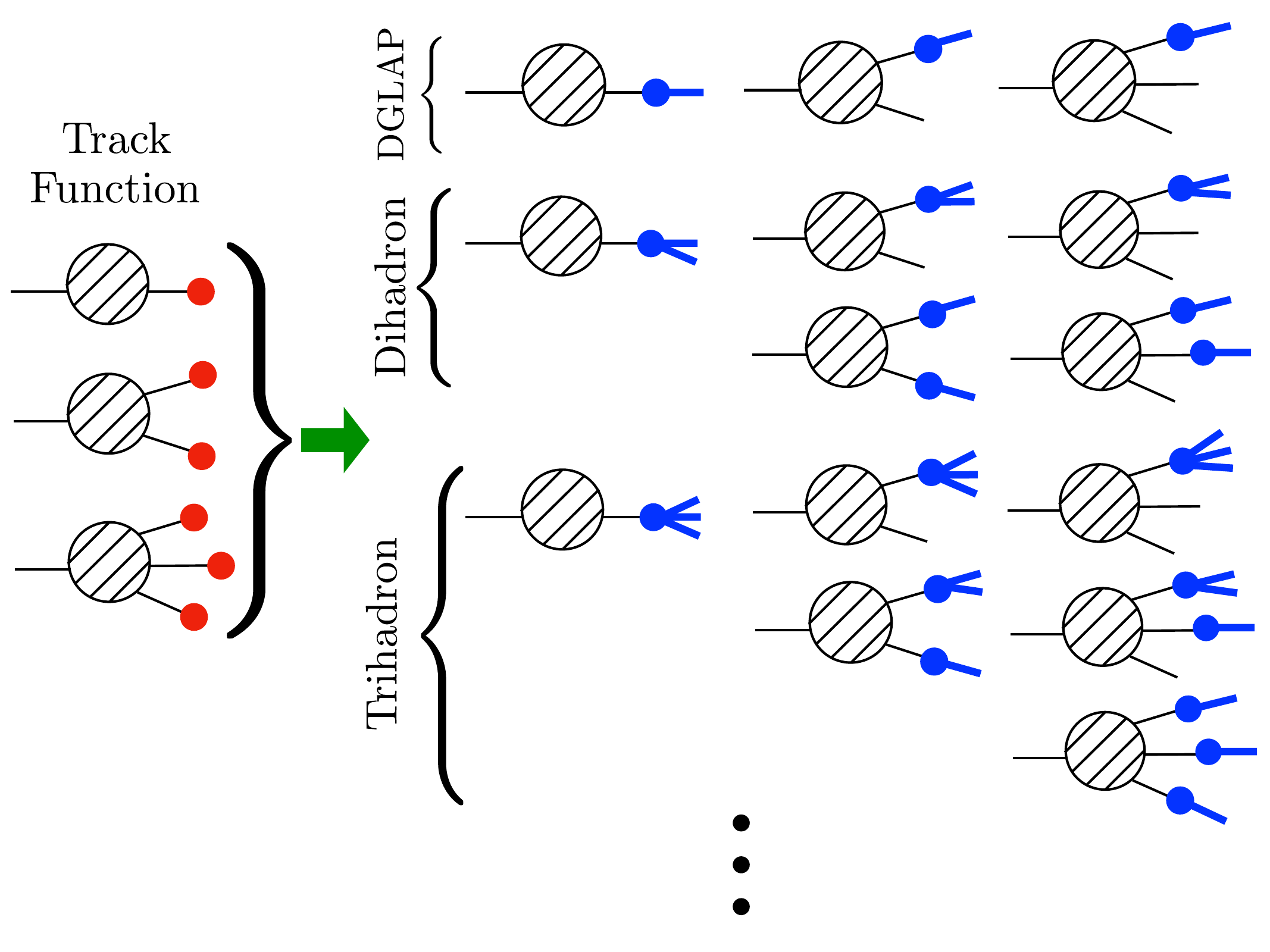}
\end{center}
\caption{A schematic of the reduction of the track function evolution equation into the evolution equations for multi-hadron fragmentation functions. The evolution equations for the $N$-hadron fragmentation functions are encoded in the track function for all values of $N$. From this perspective, the track function can be viewed as a master equation for collinear evolution.}
\label{fig:reduction_all}
\end{figure}

Unlike for multi- or single-hadron fragmentation functions, which study the properties of a fixed number $N$ of hadrons, the track functions measure the total charge in hadrons, and must therefore keep track of an arbitrary number of hadrons. This implies that the evolution equation for track functions must track all partons in a splitting, and therefore the evolution equation conserves momentum. This was exploited in refs.~\cite{Li:2021zcf,Jaarsma:2022kdd} to simplify the calculation of the moments. Since the track function captures correlations between hadrons for any number of hadrons, one expects that there is a way to extract the evolution of the $N$-hadron fragmentation function \cite{Konishi:1979cb,Sukhatme:1980vs,Sukhatme:1981ym,Majumder:2004wh} from that of the track function. This implies that the RG equations of track functions are a form of master equation encompassing all collinear evolution equations.

For any finite $N$, the evolution equations for $N$-hadron fragmentation functions describe radiated energy loss. This is quite distinct from the energy conserving evolution equation for the track functions. Therefore, to reduce the track function evolution equations to the evolution equations for multi-hadron fragmentation functions, one must devise a way of effectively ignoring all but a fixed subset of hadrons in the fragmentation process. This can be done by substituting $T(x_i)\to \delta(x_i)$ for some subset of track functions in the evolution equation, which effectively means that this particle is integrated out. We will now describe this procedure in some detail, starting with the fragmentation of single hadrons in sec.~\ref{sec:singlehadron}, two hadrons in sec.~\ref{sec:di-hadron} and $N$ hadrons in sec.~\ref{sec:N-hadron}. A derivation of this procedure is given in sec.~\ref{sec:derivation}.

\subsection{Single Identified Hadrons and DGLAP}
\label{sec:singlehadron}

We begin by considering the simplest case of the reducing the evolution of the track function to that of the fragmentation function for a single identified hadron. The evolution equation for the single-hadron fragmentation is a linear evolution equation, which is the celebrated DGLAP evolution equation \cite{Gribov:1972ri,Dokshitzer:1977sg,Altarelli:1977zs}.

\begin{figure}
\begin{center}
\includegraphics[scale=0.45]{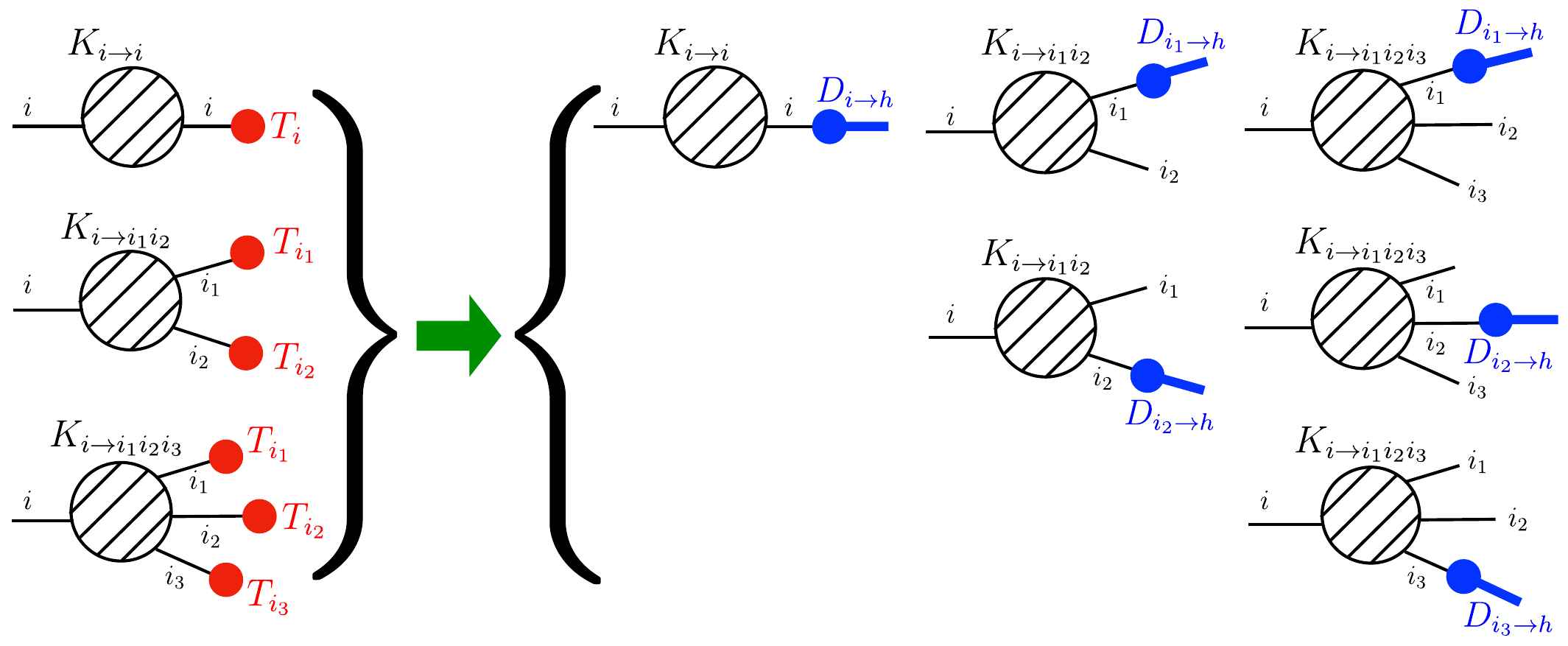}
\end{center}
\caption{A schematic of the reduction of the track function evolution equation into the evolution equation for single-hadron fragmentation functions (DGLAP). }
\label{fig:reduction2singleH}
\end{figure}

For simplicity, we assume that a parton splits into at most three partons (which holds up to NLO), but the discussion below can straightforwardly extend to splitting processes with more than three final-state partons. In \Fig{fig:reduction2singleH}, the left-hand side depicts the track function evolution equation of $T_i$, which corresponds to the process of $i$ to $i$, a branching vertex of $i\to i_1i_2$ and a vertex of $i\to i_1i_2i_3$ followed by charged hadron production which is described by the corresponding track functions. For each splitting process, the total momentum of the charged hadrons observed is the sum of the momentum fractions of the tracks produced by every branch. On the other hand, for single-hadron fragmentation function evolution, the single identified hadron that is observed can come from each branch of each splitting process. So, the evolution for the single-hadron case corresponds to the sum of all the different ways of single hadron production, as is shown on the right of \Fig{fig:reduction2singleH}. 

The track function evolution equation in schematic form reads
\begin{align}\label{eq:schematiceq_sec6}
    \frac{\df}{\df \ln\mu^2}T_i(x)=&\,K_{i\to i}{\color{red}T(x)}
    +\sum_{\{i_1,i_2\}}K_{i\to i_1i_2}\otimes {\color{red}T_{i_1}(x_1)T_{i_2}(x_2)}\nn\\
    &
    +\sum_{\{i_1,i_2,i_3\}}K_{i\to i_1i_2i_3}\otimes {\color{red}T_{i_1}(x_1)T_{i_2}(x_2)T_{i_3}(x_3)}
\,,\end{align}
where to present our idea clearly the momentum fraction arguments of track functions and the sums over splitting processes are displayed explicitly while the renormalization scale $\mu$ is suppressed. 
The reduction of the track function evolution equation to the DGLAP evolution of single-hadron fragmentation functions can be accomplished in a two-step process: 
\begin{itemize}
  \item First, we let $T(x_m)\to \delta(x_m)$ which drops the contribution from that branch. 
  In this way, we can set a cutoff at the number of branches (of splitting processes) that are observed. 
  For the single-hadron case here, one sums over all ways of keeping exactly one $T$: 
  \begin{align}
    \frac{\df}{\df \ln\mu^2}T_i(x)=&\,K_{i\to i}T_i(x)
    +\sum_{\{i_1,i_2\}}\biggl[K_{i\to i_1i_2}\otimes T_{i_1}(x_1)\delta(x_2)+K_{i\to i_1i_2}\otimes \delta(x_1)T_{i_2}(x_2)\biggr]\nn\\
    &+\sum_{\{i_1,i_2,i_3\}}\biggl[K_{i\to i_1i_2i_3}\otimes T_{i_1}(x_1)\delta(x_2)\delta(x_3)
    +K_{i\to i_1i_2i_3}\otimes \delta(x_1)T_{i_2}(x_2)\delta(x_3)\nn\\
    &+K_{i\to i_1i_2i_3}\otimes \delta(x_1)\delta(x_2)T_{i_3}(x_3)\biggr]\,.
  \end{align}
  \item Second, with $\delta(x_m)$ integrated out, we replace $T(x)$ and $T(x_i)$ with the corresponding single-hadron fragmentation functions $D$:
  \begin{align}
    \frac{\df}{\df\ln\mu^2}D_{i\to h}(x)
    &=\,K_{i\to i}{\color{blue} D_{i\to h}(x)}
    +\sum_{ \{i_1,i_2\} }K_{i\to i_1i_2}
    \otimes [{\color{blue} D_{i_1\to h}(x_1)}+{\color{blue}D_{i_2\to h}(x_2) }]\\
    &\quad +\sum_{ \{i_1,i_2,i_3\} }K_{i\to i_1i_2i_3 }
    \otimes [{\color{blue} D_{i_1\to h}(x_1)}+{\color{blue}D_{i_2\to h}(x_2)}+{\color{blue}D_{i_3\to h}(x_3)}]\,.
  \nn\end{align}
\end{itemize}
For example, for the $\mathcal{N}=4$ SYM case, we can derive an expression for the DGLAP evolution in terms of the track function kernels in the sector decomposed coordinates $(z,t)$ as
\begin{align}\label{eq:N4_reduction2single}
  \frac{\df}{\df \ln\mu^2}D(x)=&K_{1\to1}D(x)
  +\int\df x_1\,\df z\, \delta\Bigl(x-x_1\frac{z}{1+z}\Bigr)D(x_1)K_{1\to 2}(z)\\
  &+\int\df x_2\,\df z\, \delta\Bigl(x-x_2\frac{1}{1+z}\Bigr)D(x_2)K_{1\to 2}(z)\nn\\
  &+\int\df x_1\,\df z\,\df t\, \delta\Bigl(x-x_1\frac{zt}{1+z+zt}\Bigr)D(x_1)K_{1\to 3}(z,t)\nn\\
  &+\int\df x_2\,\df z\,\df t\, \delta\Bigl(x-x_2\frac{z}{1+z+zt}\Bigr)D(x_2)K_{1\to 3}(z,t)\nn\\
  &+\int\df x_3\,\df z\,\df t\, \delta\Bigl(x-x_3\frac{1}{1+z+zt}\Bigr)D(x_3)K_{1\to 3}(z,t)
\,.\nn\end{align}
Up to NLO, the kernels $K_{1\to M}=aK^{(0)}_{1\to M}+a^2 K^{(1)}_{1\to M}$ ($M=1,2$) and $K_{1\to 3}=a^2K^{(1)}_{1\to 3}$ given in eqs.~\eqref{eq:N4_K1to1_LO}, \eqref{eq:N4_K1to2_LO} and eqs.~(\ref{eq:N4_K1to1}--\ref{eq:N4_K1to3}). 

To verify this procedure reproduces the standard DGLAP kernels, we convert eq.~\eqref{eq:N4_reduction2single} to Mellin space, where the $N$-th moment of the single-hadron fragmentation function is defined as
\begin{align}
    D(N)=\int_0^1\df x\, x^N D(x)\,.
\end{align}
Then, acting on both sides of eq.~\eqref{eq:N4_reduction2single} with $\int_0^1\df x\, x^N$, we have 
\begin{align}
    \frac{\df D(N)}{\df \ln\mu^2}=
    &
    D(N)\Biggl\{
    K_{1\to1}
    +\int_0^1\df z\biggl[\biggr(\frac{1}{1+z}\biggr)^N+\biggr(\frac{z}{1+z}\biggr)^N\biggr]K_{1\to 2}(z)
    \nn\\
    &
    \!+\!\int_0^1\!\df z\!\int_0^1\!\df t\biggl[\!\biggr(\frac{1}{1+z+zt}\biggr)^N\!+\biggr(\frac{z}{1+z+zt}\biggr)^N\!+\biggr(\frac{zt}{1+z+zt}\biggr)^N\biggr]K_{1\to 3}(z,t)
    \Biggr\}\nn\\
    \equiv
    &
    -D(N)K_{\mathcal{N}=4}(N;a)
    \,,
\end{align}
where we obtain
\begin{align}\label{eq:N4_kernel_moment}
K_{\mathcal{N}=4}(N;a)&=
4a\bigl[\gamma_E+\Psi(N)\bigr]
+4a^2\biggl\{
\frac{8}{N^3}-\frac{2}{(N+1)^3}
+\frac{2\zeta_2}{N}
-\biggl(\frac{4}{N^2}+2\zeta_2\biggr) [\Psi(N+1)+\gamma_E]
\nn\\
& \quad
+\biggl(\Psi(N+1)+\gamma_E-\frac{1}{N}\biggr) \biggl[\Psi^{(1)}\Bigl(\frac{N+1}{2}\Bigr)-4 \Psi^{(1)}(N+1)-\Psi^{(1)}\Bigl(\frac{N+2}{2}\Bigr)\biggr]
\nn\\
& \quad
-2 \Psi^{(2)}(N+1)
+\frac{1}{8}\biggl[\Psi^{(2)}\Bigl(\frac{N+2}{2}\Bigr)-\Psi^{(2)}\Bigl(\frac{N+1}{2}\Bigr)\biggr]
+\Psi^{(2)}(N+2)
\nn\\
& \quad
-3 \zeta_3-4 \tilde{C}(N)
\biggr\}
\end{align}
by first considering positive integer $N$ and analytically continuing \cite{Blumlein:1998if,Kotikov:2005gr,Albino:2009ci,Blumlein:2009ta}. This expression contains Euler's constant $\gamma_E$, the digamma function $\Psi(\nu)=\Gamma^\prime(\nu)/\Gamma(\nu)$ and its $n$th derivative $\Psi^{(n)}(\nu)=\df^n\Psi(\nu)/\df \nu^n$. $\tilde{C}(N)$ denotes the nontrivial part of the harmonic sum $S_{-2,1}(N)$ within the dispersion representation where the domain of $N$ is $\mathbb{C}\backslash \{N\in\mathbb{Z}|N<0\}$\footnote{For a positive integer $N$, $\tilde{C}(N)=(-1)^N\left[S_{-2,1}(N)+\frac{5}{8}\zeta_3\right]$. }, which can be obtained from refs.~\cite{Velizhanin:2022seo, Velizhanin:2022ays},
\begin{align}
    \tilde{C}(N)=\sum_{m=1}^\infty\frac{(-1)^m}{m+N}\left[S_{-2}(m-1)-\zeta_2\right]
\,,\end{align}
with harmonic sums 
\begin{align}
    S_{a_1}(N)&\equiv \sum_{m_1=1}^N\frac{(\text{sign}(a_1))^{m_1}}{m_1^{|a_1|}}\,,\\
    S_{a_1,a_2}(N)&\equiv \sum_{m_1=1}^N\frac{(\text{sign}(a_1))^{m_1}}{m_1^{|a_1|}}S_{a_2}(m_1)\,.
\nn\end{align}
$K_{\mathcal{N}=4}(N;a)$ agrees with the twist-2 spin-$N\!+\!1$ anomalous dimension up to NLO
\begin{align}
    \gamma_{T,\text{uni}}(N+1;a)\equiv -\int_0^1\df z\,z^N P_{T,\text{uni}}(z,a)
\,.\end{align}
This time-like anomalous dimension (as indicated by subscript $T$) can be derived from the universal (space-like) anomalous dimensions listed in ref.~\cite{Kotikov:2004er} or from the Mellin transform of the LO and NLO universal space-like splitting functions, $I^{(0)}(z)$ and $I^{(1)}(z)$, listed in ref.~\cite{Banerjee:2018yrn}, according to the reciprocity relation~\cite{Gribov:1972rt,Basso:2006nk,Dokshitzer:2006nm}. 

For the case of QCD, one must additionally incorporate flavor. We have obtained the DGLAP evolution equations by the same procedure described above and have checked them for the first 30 moments. This provides a direct relation between the track function evolution kernels and the much more studied DGLAP kernels. In particular, in \Sec{sec:compare}, we showed that the track function kernels exhibit some form of maximal transcendentality, which can be partially explained by this relation to the DGLAP kernels. 

\subsection{Di-hadron Fragmentation}
\label{sec:di-hadron}

We can now extend this procedure to the case of di-hadron fragmentation functions. The schematic form of the reduction is shown in \Fig{fig:reduction_all}, where the two hadrons $h_1,h_2$ can be produced by the same partonic branch, or hadron $h_1$ comes from one branch of a splitting while hadron $h_2$ comes from another branch of that splitting. This indicates we must either integrate out all track functions except one, which we replace by a dihadron fragmentation function, $D_{j\to h_1h_2}$, or we must integrate out all track functions, except two, both of which are replaced by single hadron fragmentation functions. All the ways of two hadron production should be added up together.

\begin{figure}
\begin{center}
\includegraphics[scale=0.45]{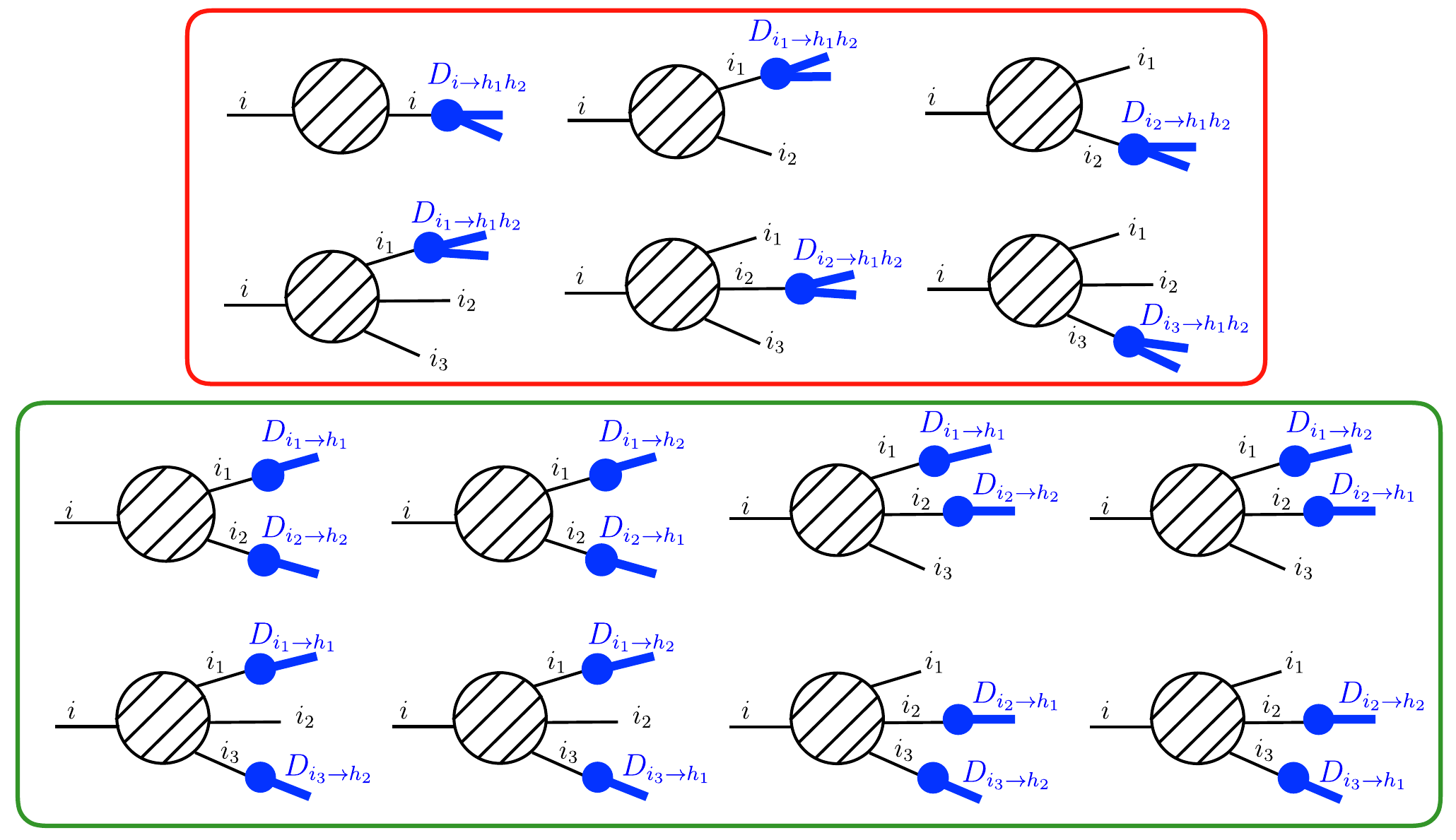}
\end{center}
\caption{The structure of the evolution equation for the di-hadron fragmentation functions, as derived from the track function evolution. The equation involves both $D_{i\to h_1 h_2}$ (shown in the red box) as well as products of the single hadron fragmentation functions $D_{i\to h_1}$ (in the green box).}
\label{fig:reduction2dihadron}
\end{figure}

Explicitly, starting from the evolution equation in \Eq{eq:schematiceq_sec6}, we insert appropriate delta functions, and then we convert these to single or dihadron fragmentation functions as required. After the first step of inserting delta functions, we have
\begin{align}
  \frac{\df}{\df\ln\mu^2}T_i(x)=
  &
  \biggl\{
  K_{i\to i}T_i(x)
  +\sum_{\{i_1,i_2\}}\Bigl[
  K_{i\to i_1i_2}\otimes T_{i_1}(x_1)\delta(x_2)+K_{i\to i_1i_2}\otimes \delta(x_1)T_{i_2}(x_2)\Bigr]
  \nn\\
  &
  +\sum_{\{i_1,i_2,i_3\}}\Bigl[
  K_{i\to i_1i_2i_3}\otimes T_{i_1}(x_1)\delta(x_2)\delta(x_3)
  +K_{i\to i_1i_2i_3}\otimes \delta(x_1)T_{i_2}(x_2)\delta(x_3)
  \nn\\
  &
  +K_{i\to i_1i_2i_3}\otimes \delta(x_1)\delta(x_2)T_{i_3}(x_3)
  \Bigr]
  \biggr\}
  \nn\\
  &
  +\biggl\{
  \sum_{\{i_1,i_2\}}
  K_{i\to i_1i_2}\otimes T(x_1)T(x_2)
  +\sum_{\{i_1,i_2,i_3\}}\Bigl[
  K_{i\to i_1i_2i_3}\otimes T(x_1)T(x_2)\delta(x_3)\nn\\
  &+K_{i\to i_1i_2i_3}\otimes T(x_1)\delta(x_2)T(x_3)
  +K_{i\to i_1i_2i_3}\otimes \delta(x_1)T(x_2)T(x_3)
  \Bigr]
  \biggr\}
\end{align}
where the number of the partonic branches that are observed has been limited.
After the second step, we obtain the evolution equation for dihadron fragmentation functions,
\begin{align}\label{eq:dihadron_final}
    \frac{\df}{\df\ln\mu^2}D_{i\to h_1h_2}(y_1,y_2)=&\,\textcolor{red}{\Biggl\{}
      K_{i\to i}{\color{blue} D_{i\to h_1h_2}(y_1,y_2)}
    +\sum_{ \{i_1,i_2\} }K_{i\to i_1i_2}
    \otimes [{\color{blue} D_{i_1\to h_1h_2}}+{\color{blue}D_{i_2\to h_1h_2} }]\nn\\
    &+\sum_{ \{i_1,i_2,i_3\} }K_{i\to i_1i_2i_3 }
    \otimes [{\color{blue} D_{i_1\to h_1h_2}}+{\color{blue}D_{i_2\to h_1h_2}}
    +{\color{blue}D_{i_3\to h_1h_2}} ]
    \textcolor{red}{\Biggr\}}\nn\\
    &+\textcolor{darkgreen}{\Biggl\{}
    \sum_{ \{i_1,i_2\} }K_{i\to i_1i_2}
    \otimes {\color{blue} [D_{i_1\to h_1} D_{i_2\to h_2}}+{\color{blue}D_{i_1\to h_2} D_{i_2\to h_1}] }
    \nn\\
    &+\sum_{ \{i_1,i_2,i_3\} }K_{i\to i_1i_2i_3 }
    \otimes [{\color{blue} D_{i_1\to h_1}D_{i_2\to h_2}}+{\color{blue}D_{i_1\to h_2}D_{i_2\to h_1}}+{\color{blue}D_{i_1\to h_1}D_{i_3\to h_2}}\nn\\
    &+{\color{blue}D_{i_1\to h_2}D_{i_3\to h_1}}+{\color{blue}D_{i_2\to h_1}D_{i_3\to h_2}}+{\color{blue}D_{i_2\to h_2}D_{i_3\to h_1}}]
    \textcolor{darkgreen}{\Biggr\}}.
\end{align}
The terms in the red brackets are depicted in the red box of fig.~\ref{fig:reduction2dihadron} and those in the green brackets are depicted in the green box of fig.~\ref{fig:reduction2dihadron}. Note that in \Eq{eq:dihadron_final}, the $\otimes$ are not the same as for the track function evolution equation, since in the single-/multi-hadron fragmentation function evolution equation one no longer tracks all the momenta of the splitting. In particular, the momentum conserving delta function, $\delta(x-\sum x_m z_m)$ must be split into the pieces corresponding to the observed hadrons, which is explicitly shown in eq.~\eqref{eq:evo_di-hadron_zt} below. 

In the sector decomposed coordinates we adopt, the evolution equation in QCD reads
\begin{align}
  \frac{\df}{\df\ln\mu^2}T_i(x)=\,&
  K_{i\to i}T_i(x)\nn\\
  &+\sum_{\{i_1,i_2  \}}\sum_{n=1}^2 \int\df x_1\df x_2 \int_0^1\df z\
  {^n}\! K_{i\to i_1 i_2}(z) T_{i_1}(x_1)T_{i_2}(x_2)\delta(x- {^n}\!z_1 x_1- {^n}\!z_2 x_2)\nn\\
  &+\sum_{\{i_1,i_2,i_3\}}\sum_{n=1}^6 \int\df x_1\df x_2\df x_3
  \int_0^1 \df z\int_0^1\df t\ {^n}\!K_{i\to i_1 i_2 i_3}(z,t)\,
  T_{i_1}(x_1)T_{i_2}(x_2)T_{i_3}(x_3)\nn\\
  &\qquad\times 
  \delta(x- {^n}\!z_1 x_1-{^n}\!z_2x_2-{^n}\!z_3 x_3)\,,
\end{align}
where there are no $\delta$-functions in the $K$'s since the soft contributions are included in the terms with fewer track functions. 
This allows us to derive a result for the NLO evolution of the dihadron fragmentation function in terms of our kernels. We find 
\begin{align}\label{eq:evo_di-hadron_zt}
  &\frac{\df}{\df\ln\mu^2}D_{i\to h_1h_2}(y_1,y_2)\nn\\
  &= \,
  a_s^2\textcolor{red}{\Biggl\{ }
  K^{(1)}_{i\to i}{\color{blue}D_{i\to h_1h_2}(y_1,y_2)}\nn\\
  &+\sum_{\{i_1,i_2\}}\sum_n\int_0^1\df z\ {^n}\! K^{(1)}_{i\to i_1 i_2}(z)
  \biggl[
  \int\df y_{11}^\prime\df y_{12}' 
  {\color{blue} D_{i_1\to h_1h_2}(y_{11}',y_{12}') }
  \delta\bigl(y_1-{^n}\!z_1\, y_{11}'\bigr)\delta\bigl(y_2-{^n}\!z_1\, y_{12}'\bigr)\nn\\
  &+\int \df y_{21}'\df y_{22}'
  {\color{blue} D_{i_2\to h_1h_2}(y_{21}',y_{22}') }
  \delta\bigl(y_1-{^n}\!z_2\, y_{21}'\bigr)\delta\bigl(y_2-{^n}\!z_2\, y_{22}'\bigr)
  \biggr]\nn\\
  &+\sum_{\{i_1,i_2,i_3\}}\sum_n\int \df z\df t \ {^n}\!K^{(1)}_{i\to i_1 i_2 i_3}(z,t)
  \biggl[
  \int\df y_{11}'\df y_{12}'
  {\color{blue} D_{i_1\to h_1h_2}(y_{11}',y_{12}') }
  \delta\bigl(y_1- {^n}\!z_1\,y_{11}'\bigr)
  \delta\bigl(y_2-{^n}\!z_1\,y_{12}'\bigr)\nn\\
  &+\int\df y_{21}'\df y_{22}'
  {\color{blue} D_{i_2\to h_1h_2}(y_{21}',y_{22}') }
  \delta\bigl(y_1- {^n}\!z_2\,y_{21}' \bigr)
  \delta\bigl(y_2-{^n}\!z_2\, y_{22}'\bigr)\nn\\
  &+\int\df y_{31}'\df y_{32}'
  {\color{blue} D_{i_3\to h_1h_2}(y_{31}',y_{32}') }
  \delta\bigl(y_1-{^n}\!z_3\,y_{31}'\bigr)
  \delta\bigl(y_2-{^n}\!z_3\,y_{32}'\bigr)
  \biggr]
  \textcolor{red}{ \Biggr\} } \nn\\
  &+a_s^2\textcolor{OliveGreen}{ \Biggl\{ }
  \sum_{\{i_1,i_2 \}} 
  \sum_n\int_0^1\df z\ {^n}\!K^{(1)}_{i\to i_1 i_2}(z)\nn\\
  &\times
  \biggl[\int\df y_{11}'\df y_{22}' {\color{blue} D_{i_1\to h_1}(y_{11}')D_{i_2\to h_2}(y_{22}') }
  \delta(y_1- {^n}\!z_1\,y_{11}')\delta(y_2-{^n}\!z_2\,y_{22}')\nn\\
  &
  +\int\df y_{21}'\df y_{12}' {\color{blue} D_{i_2\to h_1}(y_{21}')D_{i_1\to h_2}(y_{12}') }
  \delta(y_1-{^n}\!z_2\,y_{21}')\delta(y_2-{^n}\!z_1\,y_{12}')
  \biggr]\nn\\
  &+\sum_{\{i_1,i_2,i_3\}}\sum_n\int \df z\df t \ {^n}\!K^{(1)}_{i\to i_1 i_2 i_3}(z,t)
  \nn\\
  &\times \biggl[
    \int\df y_{11}'\df y_{22}' {\color{blue} D_{i_1\to h_1}(y_{11}')D_{i_2\to h_2}(y_{22}') }
  \delta(y_1- {^n}\!z_1\,y_{11}')\delta(y_2-{^n}\!z_2\,y_{22}')\nn\\
  &
  +\int\df y_{21}'\df y_{12}' {\color{blue} D_{i_2\to h_1}(y_{21}')D_{i_1\to h_2}(y_{12}') }
  \delta(y_1-{^n}\!z_2\,y_{21}')\delta(y_2-{^n}\!z_1 y_{12}')
  \nn\\
  &
  +\int\df y_{11}'\df y_{32}' {\color{blue} D_{i_1\to h_1}(y_{11}')D_{i_3\to h_2}(y_{32}') }
  \delta(y_1-{^n}\!z_1\,y_{11}')\delta(y_2- {^n}\! z_3\,y_{32}')\nn\\
  &+\int\df y_{31}'\df y_{12}' {\color{blue} D_{i_3\to h_1}(y_{31}')D_{i_1\to h_2}(y_{12}') }
  \delta(y_1-{^n}\!z_3\,y_{31}')\delta(y_2-{^n}\!z_1\,y_{12}')
  \nn\\
  &+\int\df y_{21}' \df y_{32}' {\color{blue} D_{i_2\to h_1}(y_{21}')D_{i_3\to h_2}(y_{32}') }
  \delta(y_1-{^n}\!z_2\,y_{21}')\delta( y_2-{^n}\!z_3\,y_{32}')\nn\\
  &+\int\df y_{31}' \df y_{22}' {\color{blue} D_{i_3\to h_1}(y_{31}')D_{i_2\to h_2}(y_{22}') }
  \delta(y_1-{^n}\!z_3\,y_{31}')\delta(y_2-{^n}\!z_2\,y_{22}')
  \biggr]
  \textcolor{OliveGreen}{ \Biggr\} }\,.
\end{align}
The structure of this evolution equation is exactly illustrated in \Fig{fig:reduction2dihadron}, which shows the contributions from mixing into both $D_{i\to h_1 h_2}$, as well as products of the single hadron fragmentation functions $D_{i\to h_1}$. To our knowledge, the NLO evolution of the di-hadron fragmentation function has not previously appeared in the literature.


\subsection{Generalized Multi-hadron Fragmentation}
\label{sec:N-hadron}

\begin{figure}
\begin{center}
\includegraphics[scale=0.45]{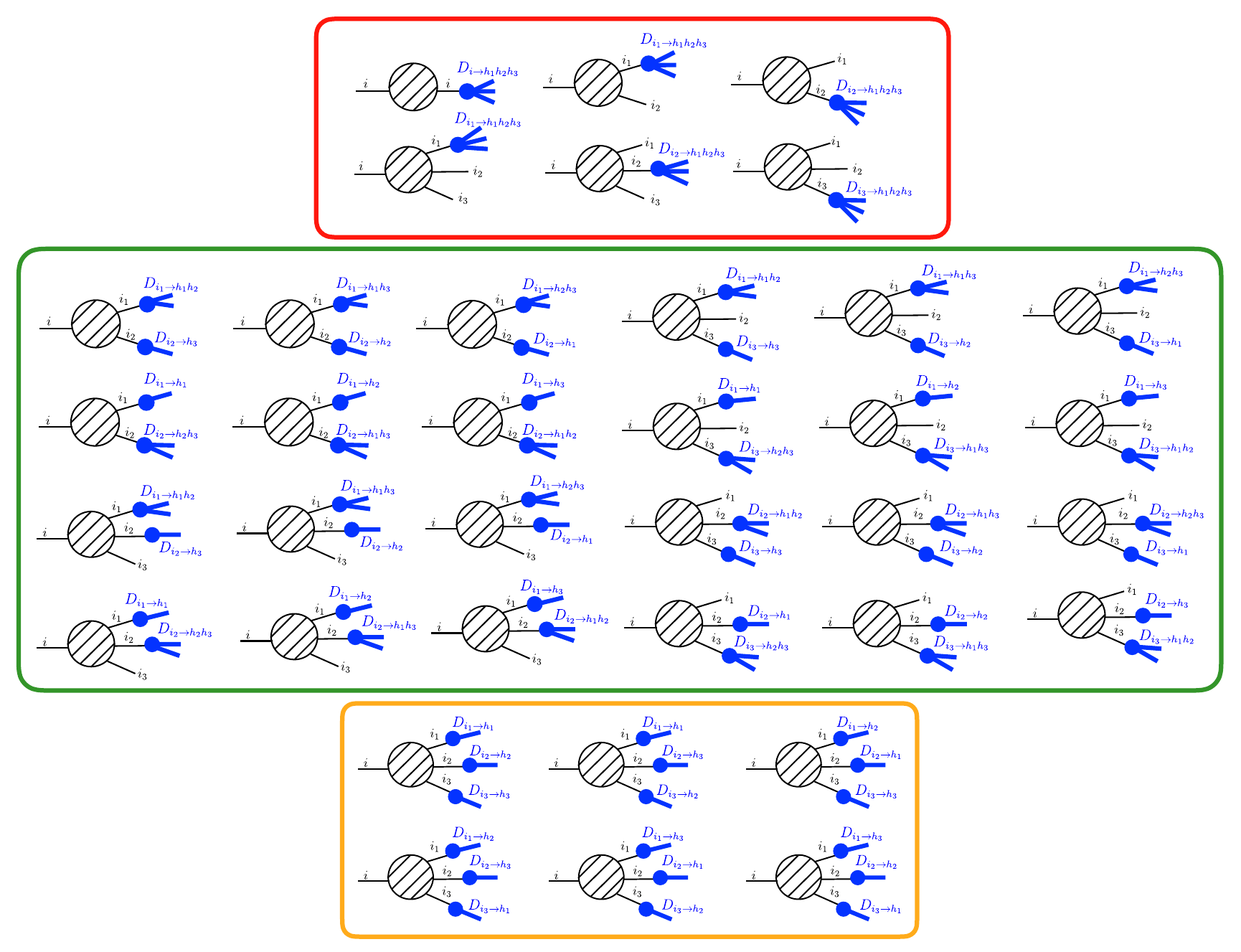}
\end{center}
\caption{The structure of the evolution equation for the tri-hadron fragmentation functions, as derived from the track function evolution. The equation involves the trihadron fragmentation function $D_{i\to h_1 h_2 h_3}$ (shown in the red box), products of the dihadron and single hadron fragmentation functions (in the green box), and products of three single hadron fragmentation functions (in the orange box).}
\label{fig:reduction2trihadron}
\end{figure}

We can now extend this to a general procedure, as is illustrate in \Fig{fig:reduction_all}. We write the  evolution of track functions in the form 
\begin{align}
  \frac{\df}{\df\ln\mu^2}T_i(x)=\sum_M \sum_{ \{i_f\} }K_{i\to i_1i_2\cdots i_M }
  \otimes T_{i_1}T_{i_2}\cdots T_{i_M}(x)\,,
\end{align}
with no Dirac delta function in $K$'s\footnote{In fact, soft contributions (expressed as Dirac delta functions) in the kernels of track function evolution would just lead to $\delta(y_m)$-terms in the RHS of the resulting evolution equation of single-/multi-hadron fragmentation function $D(\{y_m\})$, which can be immediately dropped. In particular, the reduction procedure we present here still works, ``no Dirac delta function in $K$'s'' is not a prerequisite but just a convention.}. 
Then for the evolution equation for the $N$-hadron fragmentation function  (where $N$ is a positive integer), we simply sum over all ways of keeping $\leq N$ $T$'s, and replace the track functions with the appropriate multi-hadron fragmenation functions: 
\begin{align}
  \frac{\df}{\df\ln\mu^2}&T_i(x)\\
  =&\sum_{M\geq 1}\, \sum_{ \{i_f\} }\,\sum_{ f_1\in\{1,2,\cdots,M\} }
  K_{i\to i_1i_2\cdots i_M }\otimes 
  T_{i_{f_1}}(x_{f_1})\delta(x_{f_2})\cdots\delta(x_{f_M})\nn\\
  &+ \sum_{M\geq 2}\, \sum_{ \{i_f\} }\,\sum_{ \{f_1,f_2\}\atop\subseteq\{1,2,\cdots,M\} }
  K_{i\to i_1i_2\cdots i_M }\otimes 
  T_{i_{f_1}}(x_{f_1})T_{i_{f_2}}(x_{f_2})\delta(x_{f_3})\cdots\delta(x_{f_M})\nn\\
  &+ \sum_{M\geq 3}\, \sum_{ \{i_f\} }\,\sum_{ \{f_1,f_2,f_3\}\atop\subseteq\{1,2,\cdots,M\} }
  K_{i\to i_1i_2\cdots i_M }\otimes 
  T_{i_{f_1}}(x_{f_1})T_{i_{f_2}}(x_{f_2})T_{i_{f_3}}(x_{f_3})\delta(x_{f_4})\cdots\delta(x_{f_M})\nn\\
  &+\cdots\nn\\
  &+\sum_{M\geq N}\, \sum_{ \{i_f\} }\,\sum_{ \{f_1,f_2,\cdots,f_N\}\atop\subseteq\{1,2,\cdots,M\} }
  K_{i\to i_1i_2\cdots i_M }\otimes 
  T_{i_{f_1}}(x_{f_1})T_{i_{f_2}}(x_{f_2})\cdots T_{i_{f_N}}(x_{f_N})\delta(x_{f_{N+1}})\cdots\delta(x_{f_M})\nn\\
  &\Downarrow\nn\\
  \frac{\df}{\df\ln\mu^2}&D_{i\to h_1h_2\cdots h_N}(y_1,y_2,\cdots,y_N)\\
  =&\sum_{M\geq 1}\, \sum_{ \{i_f\} }\,\sum_{ f_1\in\{1,2,\cdots,M\} }
  K_{i\to i_1i_2\cdots i_M }\otimes 
  D_{i_{f_1}\to h_1h_2\cdots h_N}(y'_{f_1,1},y'_{f_1,2},\cdots,y'_{f_1,N})\nn\\
  &+ \sum_{M\geq 2}\, \sum_{ \{i_f\} }\sum_{ \{f_1,f_2\}\atop\subseteq\{1,2,\cdots,M\} }
  K_{i\to i_1i_2\cdots i_M }\otimes 
  \sum_{S_2}
  D_{i_{f_1}\to k\text{ hadrons}}(y'_{f_1,1},\cdots,y'_{f_1,k})\nn\\
  &\qquad\qquad\qquad\qquad\qquad\qquad\qquad\qquad\qquad\quad
  \times D_{i_{f_2}\to (N-k)\text{ hadrons}}(y'_{f_2,k+1},\cdots,y'_{f_2,N})
  \nn\\
  &+ \sum_{M\geq 3}\, \sum_{ \{i_f\} }\sum_{ \{f_1,f_2,f_3\}\atop\subseteq\{1,2,\cdots,M\} }
  K_{i\to i_1i_2\cdots i_M }\otimes 
  \sum_{S_3}
  D_{i_{f_1}\to k_1\text{ hadrons}}(y'_{f_1,1},\cdots,y'_{f_1,k_1})\nn\\
  &\qquad\times
  D_{i_{f_2}\to k_2\text{ hadrons}}(y'_{f_2,k_1+1},\cdots,y'_{f_2,k_1+k_2})
  D_{i_{f_3}\to (N-k_1-k_2)\text{ hadrons}}(y'_{f_3,k_1+k_2+1},\cdots,y'_{f_2,N})\nn\\
  &+\cdots\nn\\
  &+\sum_{M\geq N}\, \sum_{ \{i_f\} }\,\sum_{ \{f_1,f_2,\cdots,f_N\}\atop\subseteq\{1,2,\cdots,M\} }
  K_{i\to i_1i_2\cdots i_M }\otimes 
  \sum_{S_N}
  D_{i_{f_1}\to 1\text{ hadron}}(y'_{f_1,1})\times\cdots 
  \times D_{i_{f_N}\to 1\text{ hadron}}(y'_{f_N,N})\nn\,.
\end{align} 
To simplify the notation, we have used $S_n$ to denote all ways of dividing the set of hadrons into $n$ non-empty sets. Note that, for the above equation, the sets obtained by dividing a set are ordered. 
E.g., the division $\{\{h_{j_1}\cdots h_{j_m}\},\{h_{j_{m+1}}\cdots h_{j_N}\}\}$ 
is considered to be different from 
$\{\{h_{j_{m+1}}\cdots h_{j_N}\},\{h_{j_1}\cdots h_{j_m}\} \}\,$; 
$\{\{h_{j_1}\},\{h_{j_2}\},\cdots,\{h_{j_N}\}\}$ is different from 
$\{\{h_{j_2}\},\{h_{j_1}\},\cdots,\{h_{j_N}\}\}$.
An example of the structure of this evolution equation for the case of the tri-hadron fragmentation function is shown in \Fig{fig:reduction2trihadron}.

Although we will not focus on the case of multi-hadron fragmentation functions in this paper, to our knowledge, the NLO corrections to multi-hadron fragmentation have not appeared in the literature. Here we have shown how they can be derived them from our master equation for the track function, and it would be interesting to study their structure in more detail. We leave this to future work.

\subsection{Derivation of reduction}
\label{sec:derivation}

In the previous subsections we presented a reduction algorithm that allows one to obtain the RG evolution of (multi-)hadron fragmentation functions from that of the track function. While this may be intuitively clear, and the explicit NLO check for the DGLAP evolution of the single hadron fragmentation functions is compelling, here we provide a derivation of this reduction based on field theory. We do this by connecting the track jet function in sec.~\ref{sec:jet_func} to the fragmenting jet function in refs.~\cite{Procura:2009vm,Jain:2011xz}, from which the evolution of the track function and fragmentation function can be derived. For simplicity we start by considering the case of single-hadron fragmentation.

The calculation of the (bare) track jet function in sec.~\ref{sec:jet_func} was given by
\begin{align} \label{eq:trj}
  J^{\text{bare}}_{\text{tr},i}(s,x)&=\sum_N\sum_{ \{i_f\} }\int \df\Phi_N^c\,\delta(s-s' )\,
  \sigma^c_{i\to \{i_f\} }\bigl(\{z_f\},\{s_{ff'}\},s'\bigr)\nn \\
  &\quad \times\int\biggl[\prod_{m=1}^N \df x_m T^{(0)}_{i_m}(x_m)\biggr]\delta\biggl(x-\sum_{m=1}^N x_mz_m\biggr)\,.
\end{align}
The corresponding calculation of the fragmenting jet function~\cite{Ritzmann:2014mka}, can be written as
\begin{align} \label{eq:fjf}
  J^{\text{bare}}_{i \to h}(s,x)&=\sum_N\sum_{ \{i_f\} }\int \df\Phi_N^c\,\delta(s-s' )\,
  \sigma^c_{i\to \{i_f\} }\bigl(\{z_f\},\{s_{ff'}\},s'\bigr)\nn \\
  &\quad  \times 
  \sum_{m=1}^N \int \! \df x_m\, D_{i_m \to h}^{(0)}(x_m) \de(x - x_m z_m)
\,.\end{align}
In ref.~\cite{Ritzmann:2014mka}, $h$ was replaced by a quark or gluon, using that $D_{q \to q}^{(0)}(x_m) = \de(1-x_m)$, $D_{q \to g}^{(0)}(x_m) = 0$, etc. However, by keeping $D_{i_m \to h}^{(0)}$ explicit, we directly see that the connection between the track jet function and fragmenting jet function is in accordance with our reduction, i.e.~we obtain \eq{eq:fjf} from \eq{eq:trj} by replacing one track function by a fragmentation function and the other track functions by $\de(x_m)$. These $\de(x_m)$'s eliminate the corresponding $x_m$ integrals and remove the corresponding term from the sum inside $\de(x - \sum_m x_m z_m)$.

The next step in the calculation is renormalization. The UV divergences do not depend on the IR details of whether a single hadron or all charged hadrons are measured, which is why the track jet function and fragmenting jet function both have the same renormalization as the invariant-mass jet function $J(s)$. In the final step of the calculation, the track jet function and fragmenting jet function are matched onto track functions and fragmentation functions, respectively:
\begin{align} 
    J_{\text{tr},i}(s,x,\mu)
    &=\sum_N\sum_{ \{i_f\} } \int\! \prod_{m=1}^N \df z_m\, \mathcal{J}_{i \to \{i_f\}}(s, \{z_f\},\mu)
   \int\biggl[\prod_{m=1}^N \df x_m T_{i_m}(x_m,\mu)\biggr]\delta\biggl(x-\sum_{m=1}^N x_mz_m\biggr)\,,
\nn \\
    J_{i \to h}(s,x,\mu)
    &=\sum_j \int\! \df z\, \mathcal{J}_{ij}(s, z,\mu) \int \! \df x'\, D_{j \to h}(x',\mu)\, \de(x - z x')
\,.\end{align}
While we have already established that the reduction holds for the left-hand side of these equations, the key is to derive it for the $D_{i \to h}$ and $T_i$ that appear on the right-hand side. We will proceed by working order by order in $\alpha_s$. 

At $\mathcal{O}(\alpha_s^0)$, the only non-vanishing matching coefficients are
\begin{align}
   \mathcal{J}^{(0)}_{i \to i}(s,z,\mu) = \de(s) \de(1-z) =    \mathcal{J}_{ii}^{(0)}(s,z,\mu)
\,.\end{align}
Using this, we find that at order $\alpha_s$ 
\begin{align}
  J_{\text{tr},i}^{(1)}(s,x,\mu) &= \de(s)\, T^{(1)}(x,\mu) +   \int\! \df z_1\,\df z_2\, \mathcal{J}_{i \to i_1i_2}^{(1)}(s, z_1,z_2,\mu) 
  \nn \\ & \quad \times
  \int\! \df x_1\, \df x_2\, T_{i_1}^{(0)}(x_1)\, T_{i_2}^{(0)}(x_2)\, \de(x- x_1 z_1 - x_2 z_2)
  \,,\nn \\ 
  J_{i \to h}^{(1)}(s,x,\mu) &= \de(s) D_{i \to h}^{(1)}(x,\mu) +  \sum_j  \int\! \df z\, \mathcal{J}_{ij}^{(1)}(s, z,\mu) 
\int \! \df x'\, D_{j \to h}^{(0)}(x')\, \de(x- x' z)
\,,\end{align}
with summation over channels $i_1,i_2$ implied.
Combining the observation that  $J_{i \to h}^{(1)}$ can be obtained from $ J_{\text{tr},i}^{(1)}$ by our reduction and that the first term on the right-hand side of these equations consist solely of infrared poles while the second terms are infrared finite, we conclude that
$D_{i \to h}^{(1)}$ can also be obtained from $T_i^{(1)}$ by using the reduction
and obtain the following relation between matching coefficients
\begin{align} \label{eq:J_rel}
  \mathcal{J}_{i i_1}^{(1)}(s, z_1,\mu)  \de(1-z_1-z_2)
   = (1+ \de_{i_1 i_2}) \mathcal{J}_{i \to i_1i_2}^{(1)}(s, z_1,z_2,\mu)  
   =  \mathcal{J}_{i i_2}^{(1)}(s, z_2,\mu)  \de(1-z_1-z_2)
\,.\end{align}
The relation between $D_{i \to h}^{(1)}$ and $T_i^{(1)}$ can also be seen from their explicit expressions:
\begin{align} \label{eq:T_D}
 T_i^{(1)}(x,\mu) &= \frac1{1+\de_{i_1i_2}} \int \df z\, \frac{\alpha_s(\mu)}{2\pi} \frac{1}{\eps_{\text{IR}}} P^{(0)}_{i \to i_1 i_2}(z) \int\! \df x_1\, \df x_2\, T_{i_1}^{(0)}(x_1,\mu) T_{i_2}^{(0)}(x_2,\mu) 
  \\ & \quad \times
 \de(x- x_1 z - x_2(1-z))
\,, \nn \\ 
 D_{i\to h}^{(1)}(x,\mu) &= \frac1{1+\de_{i_1i_2}} \int \df z\, \frac{\alpha_s(\mu)}{2\pi} \frac{1}{\eps_{\text{IR}}} P^{(0)}_{i \to i_1 i_2}(z) 
 \nn \\ & \quad \times
 \int\! \df x'\,
 \Bigl[D_{i_1 \to h}^{(0)}(x',\mu) \de(x- x' z) +  D_{i_2\to h}^{(0)}(x',\mu) \de(x- x' (1-z))\Bigr]
\,.\nn\end{align}
Finally, we consider order $\alpha_s^2$: 
\begin{align}
  J_{tr,i}^{(2)}(s,x,\mu) &= \de(s) T_i^{(2)}(x,\mu) +    \int\! \df z_1\,\df z_2\, \mathcal{J}_{i \to i_1i_2}^{(1)}(s, z_1,z_2,\mu) 
  \\ & \quad \times
  \int\! \df x_1\, \df x_2 \Bigl[T_{i_1}^{(1)}(x_1) T_{i_2}^{(0)}(x_2) + T_{i_1}^{(0)}(x_1) T_{i_2}^{(1)}(x_2)\Bigr]\, \de(x- x_1 z_1 - x_2 z_2) + \mathcal{O}(\eps^0)
  \nn \\ 
  J_{i \to h}^{(2)}(s,x,\mu) &= \de(s) D_{i \to h}^{(2)}(x,\mu) +  \sum_j  \int\! \df z\, \mathcal{J}_{ij}^{(1)}(s, z,\mu) 
\int \! \df x'\, D_{j \to h}^{(1)}(x')\, \de(x- x' z) + \mathcal{O}(\eps^0)
\,.\nn \end{align}
Since the track function and fragmentation function consist purely of poles, IR finite terms are irrelevant and dropped in the above.
Using that the reduction has been established for the left-hand side and wanting to derive it for $T_i^{(2)}$ and $D_{i \to h}^{(2)}$, we thus need to show that the reduction holds for the second terms in these equations. This follows from the previous result that $T_i^{(1)}$ reduces to $D_i^{(1)}$ and  \eq{eq:J_rel}. In principle the reduction also yields terms where in $T_{i_1}^{(1)}(x_1) T_{i_2}^{(0)}(x_2)$ both $T^{(0)}$'s in $T_i^{(1)}$ are replaced by a $\delta(x_m)$, but these vanish: These replacements yields $T_i^{(1)}(x,\mu) \propto \de(x)$ (see \eq{eq:T_D}), whose coefficient is fixed to be zero because of the momentum sum rule $\int \df x\, T_i(x,\mu) = 1$ that implies $\int \df x\, T_i^{(1)}(x,\mu) = 0$. 

These arguments can be extended to multi-hadron fragmentation functions, by comparing track jet functions to multi-hadron fragmenting jet functions. We conclude this section by illustrating this in some detail for the dihadron case. That our reduction can be used to obtain the bare dihadron fragmenting jet function from the bare track jet function is again immediate, and the renormalization is also the same. However, the matching equation for dihadron fragmenting jet functions is given by~\cite{Waalewijn:2012sv}
\begin{align}
    J_{i \to h_1h_2}(s,x_1,x_2,\mu)
    &=\sum_j \int\! \df z\, \mathcal{J}_{ij}(s, z,\mu) \int \! \df x_1'\, \df x_2'\, D_{j \to h_1 h_2}(x_1',x_2',\mu)\, \de(x_1 - z x_1') \, \de(x_2 - z x_2')
    \nn \\ & \quad
    + \sum_{j,k} \int\! \df z_1\, \df z_2\, \mathcal{J}_{ijk}(s, z_1,z_2,\mu) \int \! \df x_1'\, D_{j \to h}(x_1',\mu)\, \de(x_1 - z_1 x_1')
    \nn \\ & \quad
\times \int \! \df x_2'\, D_{k \to h}(x_2',\mu)\, \de(x_2 - z_2 x_2')
\,,\end{align}
where $\mathcal{J}_{ij}$ are the same as in the single hadron fragmentation case.
Expanding this to order $\alpha_s$, we get
\begin{align}
    J_{i \to h_1h_2}^{(1)}(s,x_1,x_2,\mu)
    &= D_{i \to h_1 h_2}^{(1)}(x_1,x_2,\mu) +
    \sum_j \int\! \df z\, \mathcal{J}_{ij}^{(1)}(s, z,\mu) \int \! \df x_1'\, \df x_2'\, D^{(0)}_{j \to h_1 h_2}(x_1',x_2',\mu)
    \nn \\ & \quad
     \times \de(x_1 - z x_1') \, \de(x_2 - z x_2')
    + \sum_{j,k} \int\! \df z_1\, \df z_2\, \mathcal{J}_{ijk}^{(1)}(s, z_1,z_2,\mu) 
    \nn \\ & \quad
   \times \int \! \df x_1'\, D^{(0)}_{j \to h}(x_1',\mu)\, \de(x_1 \!-\! z_1 x_1')
   \int \! \df x_2'\, D^{(0)}_{k \to h}(x_2',\mu)\, \de(x_2 \!-\! z_2 x_2')
\,.\end{align}
The first term only consists of IR poles while the other terms are finite, implying that $D_{i \to h_1 h_2}^{(1)}$ can be obtained from $T_i^{(1)}$ by using the reduction and that the new matching coefficient coefficient for the term with two fragmentation functions is equal to that for two track functions,
\begin{align}
   \mathcal{J}_{i i_1 i_2}^{(1)}(s, z_1,z_2,\mu) = \mathcal{J}_{i\to i_1 i_2}^{(1)}(s, z_1,z_2,\mu) 
\,.\end{align}
Finally, at order $\alpha_s^2$,
\begin{align}
    J_{i \to h_1h_2}^{(2)}(s,x_1,x_2,\mu)
    &= D_{i \to h_1 h_2}^{(2)}(x_1,x_2,\mu) +
    \sum_j \int\! \df z\, \mathcal{J}_{ij}^{(1)}(s, z,\mu) \int \! \df x_1'\, \df x_2'\, D^{(1)}_{j \to h_1 h_2}(x_1',x_2',\mu)
    \nn \\ & \quad
     \times \de(x_1 - z x_1') \, \de(x_2 - z x_2')
    + \sum_{j,k} \int\! \df z_1\, \df z_2\, \mathcal{J}_{ijk}^{(1)}(s, z_1,z_2,\mu) 
    \nn \\ & \quad
   \times 
   \biggl[\int \! \df x_1'\, D^{(1)}_{j \to h}(x_1',\mu)\, \de(x_1 \!-\! z_1 x_1')
   \int \! \df x_2'\, D^{(0)}_{k \to h}(x_2',\mu)\, \de(x_2 \!-\! z_2 x_2') +
   \nn \\ & \quad
   \int \! \df x_1'\, D^{(0)}_{j \to h}(x_1',\mu)\, \de(x_1 \!-\! z_1 x_1')
   \int \! \df x_2'\, D^{(1)}_{k \to h}(x_2',\mu)\, \de(x_2 \!-\! z_2 x_2')\biggr]
   +  \mathcal{O}(\eps^0)
\,.\end{align}
We know the reduction allows us to obtain $J_{i \to h_1h_2}^{(2)}$ from  $J_{tr,i}^{(2)}$, but we want to show this also holds for getting $D_{i \to h_1 h_2}^{(2)}$ from $T_i^{(2)}$, such that we can conclude that the reduction holds for the RG equation. Thus we must argue that the replacement holds for the other terms on the right-hand side. For the term in the track jet function where a $T_j^{(0)}$ is replaced by $D_{j \to h_1 h_2}^{(0)}$, this works the same as in the single hadron fragmentation function case. Moving on to the terms involving two single-hadron fragmentation functions,  the $T_{i_1}^{(1)}(x_1) T_{i_2}^{(0)}(x_2)$ term receives a contribution where $T^{(1)}$ and $T^{(0)}$ each yield a $D^{(0)}$ fragmentation function, corresponding to the last two lines. However, the reduction also produces a term where $T^{(1)}$ yields two fragmentation functions and $T^{(0)}$ reduces to a delta function, which is the remaining part of $D_{i \to h_1 h_2}^{(1)}$ on the first line.

\section{Numerical Implementation of the Evolution Equations}\label{sec:num_imp}

Solving the non-linear evolution equation for the track functions is non-trivial, even numerically. We present in some detail several approaches to their solution in sec.~\ref{sec:approaches}, before showing numerical results in sec.~\ref{sec:numerics}. Common to all these approaches is that we express the track function in terms of some basis. We then truncate the basis and solve the resulting finite system of differential equations for the coefficients. The bases we consider are Fourier series, wavelets and moments. The leading-order evolution was studied in ref.~\cite{Chang:2013rca} using moments and discretization. The latter corresponds to writing $T_i(x,\mu) = \sum_{m=1}^M a^i_m(\mu)\, \Theta(\frac{m-1}{M} \leq x \leq \frac{m}{M})$ for some fixed integer $M$ with coefficients $a^i_m$, and is a subset of the wavelets discussed in sec.~\ref{sec:wavelets}.

\subsection{Approaches}
\label{sec:approaches}

Here we summarize our different approaches to solving the RG evolution of the track functions. Numerical implementations of all the techniques can be found at \url{https://github.com/HaoChern14/Track-Evolution}.

\subsubsection{Fourier Series}

The evolution equations for track functions $T_i(x, \mu)$ is a system of integro-differential equations. Truncation to $\mathcal{O}(\alpha_s^2)$ involves a convolution with at most $3$ track functions:
\begin{align}
\frac{\df}{\df \ln \mu^2} T_{i}(x, \mu)
 & = K_{i\to i}(\alpha_s) T_{i}(x, \mu) + \sum_{n=1}^2 \int_{0}^{1}\! \df x_1\, \df x_2 \int_{0}^{1}\! \df z\, {}^n\! K_{i\to i_1 i_2}(z; \alpha_s) \delta\bigl(x\!-\!{^n}\!z_1x_1\!-\!{^n}\!z_2x_2\bigr)
 \nn \\ & \qquad \times
 T_{i_1}(x_1, \mu) T_{i_2}(x_2, \mu) 
+ \sum_{n=1}^6 \int_{0}^{1}\! \df x_1\, \df x_2\, \df x_3 \int_{0}^{1}\! \df z\, \df t\, {}^n\!K_{i\to i_1 i_2 i_3}(z, t; \alpha_s)\, 
\nn \\ & \qquad \times
\delta\bigl(x\!-\!{^n}\!z_1x_1\!-\!{^n}\!z_2x_2\!-\!{^n}\!z_3x_3\bigr) T_{i_1}(x_1, \mu) T_{i_2}(x_2, \mu) T_{i_3}(x_3, \mu)\,,
\end{align}
with summation over all possible splitting (i.e.~$i_1, i_2,i_3$) implied.
We remind the reader that ${^n}\!z_j$ only depends on $z$ (and $t$) for ${}^n\! K_{i\to i_1 i_2}$ (${}^n\! K_{i\to i_1 i_2 i_3}$).
To avoid directly dealing with evolution of functions, we expand the track function in terms of an orthonormal function basis $\left\{ f_m(x)\right\}$:
\begin{align}
T_i(x, \mu) &= \sum_{m} b^i_{m}(u) f_{m}(x)\,, \\
b^i_{m}(u) &= \int_{0}^{1} \df x f_m^*(x) T_i(x,\mu)\,,
\nn\end{align}
where we introduce the abbreviation $u \equiv \ln \mu^2$.

In this way, the evolution equations become a system of ordinary differential equations for the coefficients $\left\{b_{m}^i(u)\right\}$:
\begin{align}\label{eq: coeff_ODE}
\frac{\df}{\df u}\, b_m^i(u) &= K_{i \to i}(\alpha_s)\, b_m^i(u) 
+ \sum_{n=1}^2 \sum_{m_1, m_2} {^n}\!A^{\{m, i\}}_{\{m_1, i_1\}, \{m_2, i_2\}}\, b^{i_1}_{m_1}(u)\, b^{i_2}_{m_2}(u) \\[-2ex]
&\quad + \sum_{n=1}^6 \!\sum_{m_1, m_2, m_3} {^n}\!B^{\{m, i\}}_{\{m_1, i_1\}, \{m_2, i_2\}, \{m_3, i_3\}}\, b^{i_1}_{m_1}(u)\, b^{i_2}_{m_2}(u)\, b^{i_3}_{m_3}(u)\,,
\nn \end{align}
with the coefficients $A,B$ given by
\begin{align} \label{eq:ODE_coeff}
{^n}\!A^{\{m, i\}}_{\{m_1, i_1\}, \{m_2, i_2\}} &= \int_{0}^{1}\! \df x_1\, \df x_2 \int_{0}^{1}\! \df z\, {}^n\! K_{i\to i_1 i_2}(z;\alpha_s)\, f_m^*({^n}\!z_1x_1\!+\!{^n}\!z_2x_2)\, f_{m_1}(x_1)\, f_{m_2}(x_2)
\,, \nn \\
{^n}\!B^{\{m, i\}}_{\{m_1, i_1\}, \{m_2, i_2\}, \{m_3, i_3\}} &= 
\int_{0}^{1} \df x_1\, \df x_2\, \df x_3 \int_{0}^{1} \df z\, \df t\, {}^n\! K_{i\to i_1 i_2 i_3}(z, t;\alpha_s) 
\nn \\ & \quad
f_m^*({^n}\!z_1x_1\!+\!{^n}\!z_2x_2+\!{^n}\!z_3x_3)\, f_{m_1}(x_1)\, f_{m_2}(x_2)\, f_{m_3}(x_3) \,. 
\end{align}

Though any orthonormal basis $\left\{ f_m(x)\right\}$ in principle serves the purpose, we prefer those for which as many integrals as possible can be carried out analytically. This is the case when using Fourier series $\{f_{m}(x) = e^{2 \pi \img\, m\, x }\}$ with $m\in\mathbb{Z}$, because all $x_j$ integrals take the form $\int_{0}^{1} \df x\, \exp(2\pi \img\, a\, x) = \exp(\img \pi a)\, \mathrm{sinc}(\pi a)$. For this basis the coefficients in \eq{eq:ODE_coeff} are given by
\begin{align}
{^n}\!A^{\{m, i\}}_{\{m_1, i_1\}, \{m_2, i_2\}}  &= (-1)^{m_1+m_2-m}\!\int_{0}^{1}\!\df z\, {}^n\! K_{i\to i_1 i_2}(z;\alpha_s)\, \mathrm{sinc}(m_1\pi\!-\!m\pi\, {^n}\!z_1)\, \mathrm{sinc}(m_2\pi\!-\!m\pi\, {^n}\!z_2)\,,  \nn \\
{^n}\!B^{\{m, i\}}_{\{m_1, i_1\}, \{m_2, i_2\}, \{m_3, i_3\}} &= (-1)^{m_1+m_2+m_3-m}\int_{0}^{1}\!\df z\, \df t\, {}^n\! K_{i\to i_1 i_2 i_3}(z, t;\alpha_s) \label{eq: fourier_ode_coeff_2}\nn \\
& \quad \times 
\mathrm{sinc}(m_1\pi - m\pi\, {^n}\!z_1)\, \mathrm{sinc}(m_2\pi- m\pi\, {^n}\!z_2)\, \mathrm{sinc}(m_3\pi - m\pi\, {^n}\!z_3)\,. 
\end{align}
The remaining integrals can be evaluated numerically for any given set of integers $\{m, m_1, m_2, m_3\}$. Since we are unable to deal with infinite systems for numerical integration and ODE solving, we will truncate to a finite set of Fourier modes $\{f_{m}(x) = e^{2 \pi \img\,m\, x }\}$ with $m=-M, -M+1, \dots M$ as an approximation. Using the reality condition $(b_{m}^i)^* = b_{-m}^i$ for Fourier series and the normalization condition $b_0^i=1$ for track functions, the ranges of integer labels in (\ref{eq: coeff_ODE}) are $1\leq m\leq M,\, -M\leq m_1, m_2, m_3 \leq M$. We use the Julia~\cite{bezanson2017julia} package HCubature\footnote{\url{https://github.com/JuliaMath/HCubature.jl}} for numerical integration in \eq{eq: fourier_ode_coeff_2} and the package DifferentialEquations \cite{rackauckas2017differentialequations} for solving \eq{eq: coeff_ODE}. We also use GNU Parallel \cite{Tange2011a} to handle the large amount of numerical integrations.

We briefly comment on the Fourier series method. First, due to Parseval's theorem -- the Euclidean distance in the coefficient space equalling the distance in the $L^2$ function space, the errors in these two spaces should have the same order. Second, as a typical property of Fourier approximation, the error is relatively large near the end points and we can see the wiggles there in the approximation function especially when is is quite flat. The error can be reduced when we use large truncation number $M$. Third, NLO evolution requires rank-4 arrays in the coefficients evolution that make it impractical to extend to a very large $M$. Therefore, we will employ mixed truncation parameters at LO $(M_1)$ and NLO $(M_2)$. We provide a public docker image for the Fourier method on the website \url{https://hub.docker.com/r/haochern/qcd-track-evolution-fourier}, whose truncation parameters are $M_1=100$ and $M_2 = 40$.

\subsubsection{Wavelets}
\label{sec:wavelets}

Another useful choice of basis is to use wavelets, that piecewise approximate a function. In this paper, we consider the Legendre wavelet, for reasons that will soon be clear. We divide the track function domain $0 \leq x \leq 1$ uniformly into $2^{K-1}$ intervals that are indexed by $m = 1, 2, \dots, 2^{K-1}$, and on each interval we define appropriately transformed Legendre polynomials:
\begin{equation}
\psi_{m,\ell}(x) = \Theta\Bigl( \frac{m-1}{2^{K-1}} <x\leq \frac{m}{2^{K-1}}\Bigr) \ 2^{K/2} \widetilde{P}_{\ell}\bigl(2^K x -(2m-1)\bigr)\,.
\end{equation}
The $\Theta$-function restricts $x$ to the interval indexed by $m$ and 
$\widetilde{P}_\ell$ is the normalized Legendre polynomial $\widetilde{P}_{\ell}(x) = \sqrt{\ell + 1/2}\, P_{\ell}(x)$. The Legendre wavelets form an orthonormal basis:
\begin{equation}
\int_0^1 \df x \, \psi_{m_1, \ell_1}(x)\, \psi_{m_2, \ell_2}(x) = \delta_{m_1, m_2} \delta_{\ell_1,\ell_2}\,.
\end{equation}

We approximate the track function $T_i(x,\mu)$ with 
\begin{equation}
T_i(x,\mu) \backsimeq \sum_{m=1}^{2^{K-1}} \sum_{\ell=0}^{L-1} c^i_{m,\ell}(u) \psi_{m,\ell}(x)\,, 
\end{equation}
with again $u\equiv \ln \mu^2$. The orthonormal relation allows the inversion
\begin{equation}
c^i_{m, \ell} (u) = \int_0^1\!\df x\, \psi_{m,\ell}(x) T_i(x,\mu) = 2^{-K/2}\int_{-1}^{1} \df y\, \widetilde{P}_\ell(y)\, T_i\bigl(2^{-K}(y+2m-1),\mu\bigr)\,,
\end{equation}
which leads to 
\begin{align} \label{eq:wavelet_RGE}
\frac{\df }{\df u}\, c^i_{m,\ell}(u)
 &= K_{i\to i}(\alpha_s) c^i_{m,\ell}(u)+ 
 \sum_{n=1}^2 
 \int_{0}^{1}\! \df x_1 \df x_2 \int_{0}^{1}\! \df z\, {}^n\! K_{i \to i_1 i_2}(z; \alpha_s)\, \psi_{m,\ell}({^n}\!z_1x_1\!+\!{^n}\!z_2x_2)\, 
  \\ & \quad \times
 T_{i_1}(x_1, \mu)\, T_{i_2}(x_2, \mu)  + 
  \sum_{n=1}^6 
  \int_{0}^{1} \!\df x_1 \df x_2 \df x_3  \int_{0}^{1} \! \df z\, \df t\, {}^n\! K_{i \to i_1 i_2 i_3}(z, t; \alpha_s)\, 
  \nn \\ & \quad \times
  \psi_{m, \ell}({^n}\!z_1x_1\!+\!{^n}\!z_2x_2+\!{^n}\!z_3x_3) T_{i_1}(x_1, \mu) T_{i_2}(x_2, \mu) T_{i_3}(x_3, \mu) \nn \\
&\approx C^i(\alpha_s) c^i_{m,\ell}(u) 
+  \sum_{n=1}^2\, \sum_{m_1,m_2 =1}^{2^{K-1}}\, \sum_{\ell_1,\ell_2=0}^{L-1}\, c^{i_1}_{m_1,\ell_1}(u)\, c^{i_2}_{m_2,\ell_2}(u) \; {}^n\!M_{i \to i_1 i_2}\!\!
\begin{pmatrix}
m & m_1 & m_2\\
\ell & \ell_1 & \ell_2
\end{pmatrix} \nn \\
&\quad + 
  \sum_{n=1}^6\,
\sum_{m_1,m_2,m_3=1}^{2^{K-1}}\, \sum_{\ell_1,\ell_2,\ell_3=0}^{L-1}\, c^{i_1}_{m_1,\ell_1}(u)\, c^{i_2}_{m_2,\ell_2}(u)\, c^{i_3}_{m_3,\ell_3}(u) 
 {}^n\!M_{i \to i_1 i_2 i_3}\!\!
\begin{pmatrix}
m & m_1 & m_2 & m_3\\
\ell & \ell_1 & \ell_2 & \ell_3
\end{pmatrix}
.\nn \end{align}
The coefficient array ${}^n\!M_{i\to i_1 i_2}$ is 
\begin{align}
{}^n\!M_{i \to i_1 i_2}\!\!
\begin{pmatrix}
m & m_1 & m_2\\
\ell & \ell_1 & \ell_2
\end{pmatrix} &= 2^{-K/2}\int_{-1}^{1} \df y\, \df y_1\, \df y_2 \int_0^1 \df z\, \widetilde{P}_{\ell}(y)\widetilde{P}_{\ell_1}(y_1) \widetilde{P}_{\ell_2}(y_2)\\
&\quad \times \delta\bigl(y-{}^n\!z_1 (y_1+2(m_1-m)) - {}^n\!z_2 (y_2+2(m_2-m))\bigr) {}^n\!K_{i \to i_1 i_2}(z; \alpha_s)\,,
\nn \end{align}
and more complicated ${}^n\!M_{i\to i_1 i_2 i_3 }$ has a similar form
\begin{align}
{}^n\! M_{i \to i_1 i_2 i_3}\!\!
\begin{pmatrix}
m & m_1 & m_2 & m_3\\
\ell & \ell_1 & \ell_2 & \ell_3
\end{pmatrix} &= 2^{-K}\!\!\int_{-1}^{1}\! \df y\, \df y_1\, \df y_2\, \df y_3\! \int_0^1\! \df z\, \df t\, 
\widetilde{P}_{\ell}(y)\widetilde{P}_{\ell_1}(y_1) \widetilde{P}_{\ell_2}(y_2) \widetilde{P}_{\ell_3}(y_3)
\\ & \quad \times
{}^n\! K_{i\to i_1 i_2 i_3}(z, t; \alpha_s)\,
 \delta\bigl[y- {}^n\!z_1 (y_1+2(m_1-m))
\nn \\ & \qquad  \quad
 - {}^n\!z_2 (y_2+2(m_2-m)) - {}^n\!z_3(y_3+2(m_3-m))\bigr].
\nn \end{align}
From these $\delta$ functions, we see that only the differences $\{m-m_i\}$ are important, which is the discretized version of shift symmetry in the wavelet interval label. 

The $\delta$ function of $y$ in the above expressions were used to eliminate the integral over $y$ in the initial lines of \eq{eq:wavelet_RGE}. However, we write them explicitly to highlight that the coefficients $M$ are nonvanishing if and only if the solution of $\delta$-function lie inside the domain $-1<y<1$:
\begin{equation}
\int_{-1}^{1}\df y \, \widetilde{P}_{\ell}(y) \delta(y-a) = \Theta(-1<a<1) \, \widetilde{P}_{\ell}(a). 
\end{equation}
Thus, for a given value of  $z, t$, the $y_i$ integrals are integration of polynomials on polytopes which can algorithmically be carried out exactly. This is our reason for using Legendre wavelets: the dual basis is polynomial and the orthogonality does not involve a non-trivial weight function. For example, the public software LattE \cite{de2004effective} (Lattice point Enumeration) contains a program named integrale \cite{baldoni2011integrate, de2012software} that does this job. In this paper, we choose a more pedestrian method with the help of function \verb|Reduce| in Mathematica. Let's illustrate this with an example for the $n=1$ sector:
\begin{equation}
I^{m\, m_1 m_2}_{\ell\, \ell_1 \ell_2}(z) =\! \int_{0}^{1} \! \df y\, \df y_1\, \df y_2\, \widetilde{P}_{\ell_1}(y_1) \widetilde{P}_{\ell_2}(y_2) \widetilde{P}_{\ell}(y)\, \delta \biggl(y- \frac{(y_1+2(m_1-m)) + z(y_2+2(m_2-m))}{1+z} \biggr)
\,,\end{equation}
integrates the polynomial
\begin{equation}
\widetilde{P}_{\ell_1}(y_1) \widetilde{P}_{\ell_2}(y_2) \widetilde{P}_{\ell}\biggl(\frac{(y_1+2(m_1-m)) + z(y_2+2(m_2-m))}{1+z} \biggr)
\,,\end{equation}
on the polytope $\Omega$ formed by constraints:
\begin{equation}
-1 < \frac{(y_1+2(m_1-m)) + z(y_2+2(m_2-m))}{1+z} < 1, \quad -1 < y_1 < 1,\quad -1 < y_2 < 1\,.
\end{equation}
If integers $m, m_1, m_2$ and the value $z$ are given, \verb|Reduce[...,{y1, y2}]| will cut $\Omega$ into non-overlapping subregions $\Omega_r$
\begin{equation}
\Omega_r = \{(y_1,y_2)\,|\, a_1^{(r)}< y_1 < a_2^{(r)}, a_3^{(r)} y_1+a_4^{(r)} < y_2 < a_5^{(r)} y_1 + a_6^{(r)} \},
\end{equation}
on each of which the polynomial integration becomes straightforward.
We therefore choose to keep all $y_i$ integrations exact while using the midpoint-rule approximation for $z, t$ integrals, e.g.
\begin{equation} \label{eq:M_12}
{}^1\!M_{i \to i_1 i_2}\!\!
\begin{pmatrix}
m & m_1 & m_2\\
\ell & \ell_1 & \ell_2
\end{pmatrix} \approx 2^{-K/2}\sum_{s=1}^{N} {}^1\!K_{i \to i_1 i_2}\Bigl(z=\frac{s-1/2}{N}; \alpha_s\Bigr) I^{m\, m_1 m_2}_{\ell\, \ell_1 \ell_2}\Bigl(z = \frac{s-1/2}{N}\Bigr) \,,
\end{equation}
where the choice of $N$ is in principle independent of $K$.
When the kernels ${}^n\!K_{i \to i_1 i_2}(z; \alpha_s)$ or ${}^n\! K_{i \to i_1 i_2 i_3}(z, t; \alpha_s)$ diverge or involve plus distributions near the boundary, we use the averaging value rather than the midpoint value to improve the accuracy. For example, the kernel ${}^n\!K_{i \to i_1 i_2}(z; \alpha_s)$ may contain $\log^p z$ that is relatively steep near the boundary $ z\sim 0$. Here, we take the following  modification to reduce the error:
\begin{equation}
{}^n\!K_{i \to i_1 i_2}\Bigl(z=\frac{s-1/2}{N};\alpha_s\Bigr) \to N \int_{(s-1)/N}^{s/N} \df z\; {}^n\!K_{i \to i_1 i_2}(z;\alpha_s) \,.
\end{equation}
When there is a plus distribution, we employ the replacement in the summand:
\begin{equation}
    \left[\frac{f(z)}{z}\right]_+ I_{\ell\, \ell_1 \ell_2}^{m\, m_1 m_2}(z) \Bigg|_{z=\frac{s-1/2}{N}} \to \frac{N}{s-1/2} \left[I^{m\, m_1 m_2}_{\ell\, \ell_1 \ell_2}\left( \frac{s-1/2}{N}\right) - I^{m\, m_1 m_2}_{\ell\, \ell_1 \ell_2}\left( 0\right)\right]  N \int_{(s-1)/N}^{s/N} \df z\; f(z) \, ,
\end{equation}
noting that the boundary term of the plus distribution (not included) will drop out in the sum over all bins $s$ in \eq{eq:M_12}.

Though wavelet approximation is not continuous, its central value is still trustable when away from the boundaries of the sub-intervals. This offers a cross check for other approximation methods. The wavelet method docker image with $K=5$ ($16$ intervals) and $L=3$ can be found on \url{https://hub.docker.com/r/haochern/qcd-track-evolution-wavelet}.

\subsubsection{Moments}

One of the ways the LO evolution equations for the track function were solved in ref.~\cite{Chang:2013rca}, was by mapping to moment space, evolving in moment space, and then mapping back to functions.  In refs.~\cite{Jaarsma:2022kdd,Li:2021zcf} we have derived the general structure of the RG equations for the first six moments of the track functions. To achieve reliable results requires more than six moments, which can be obtained from the full $x$-dependent evolution calculated in this paper, and we find that high precision is needed to avoid numerical noise in the inversion. Here we elaborate on this method, as well as some of its challenges.

Recall that the $N$-th moment of a track function $T_i(x,\mu)$ is defined as 
\begin{align}
  T(N,\mu)=\int_0^1\df x\ x^NT_i(x,\mu)\,
\end{align}
with the zeroth moment fixed by the normalization condition, $T(0,\mu)=1$. In principal, if one is able to get analytic expressions for the evolution equation of $T_i(N,\mu)$ for generic values of $N$, as well as derive the solution of the renormalization group equations given some initial condition at the scale $\mu_0$, then this solves the evolution problem. All the information about the track function at a certain scale $\mu_1$ in momentum-fraction space, $T_i(x,\mu_1)$, 
is contained in the solution $T_i(N,\mu_1)$, by taking the inverse Mellin transform. 
However, in contrast to the DGLAP case, the track function evolution only has a simple moment space decomposition for integer moments: only in this case does the evolution involve a finite sum of moments (obtained by using a multinomial expansion). Furthermore it is not easy to obtain the $N$-th moment of the track function evolution at NLO, since one has to work out the $(n_1,n_2,N-n_1-n_2)$-th moment of the $1\to 3$ evolution kernel $K_{i\to i_1i_2i_3}$ for generic $n_1,n_2,N$,
which multiplies the $T_{i_1}(n_1)T_{i_2}(n_2)T_{i_3}(N-n_1-n_2)$ term.\footnote{Although 
we are able to work out the generating functions for the moment-space kernels in $\mathcal{N}=4$ and QCD, 
to extract the kernels, $K_{i\to i_1i_2}(n_1,N-n_1),K_{i\to i_1i_2i_3}(n_1,n_2,N-n_1-n_2)$ for generic $n_1,n_2,N$, is still cumbersome; we leave this to future work.} 

Here we take an alternative approach, deriving the evolution equations for a finite set of integer moments, 
solving them numerically and then mapping back to a function by supposing the track function can be approximated by a polynomial.
To simplify the process of mapping back, we identified a polynomial basis of degree  $\ell$:
\begin{align}
  P_m(x) =
  \sum_{k=0}^\ell \frac{(1+\ell)^2}{1+m+k}f(m)f(k)\,x^k\,,
\end{align}
with $m = 0, 1, \dots \ell$ and
\begin{align}
    f(k)=
    \begin{cases}
    1\,, & \mbox{if }k=0\,,\\
    \prod _{j=1}^k\left(1-\frac{(1+\ell)^2}{j^2}\right)\,, &\mbox{if }k\geq 1, k\in\mathbb{N}\,.
    \end{cases}
\end{align}
These polynomials have the property such that for integer $N$ with $0\leq N\leq \ell$
\begin{align}
  \int_0^1\! \df x\, x^N\, P_m(x) = \de_{N,m}
\,,\end{align}
so we can directly reconstruct the approximate $T_i(x,\mu_1)$ in terms of the moments $T_i(m,\mu)$ as
\begin{align}
  T_i(x,\mu) \backsimeq \sum_{m=0}^\ell T_i(m,\mu) P_m(x)
\,.\end{align}

\begin{figure}
\begin{center}
\subfloat[]{
\includegraphics[scale=0.35]{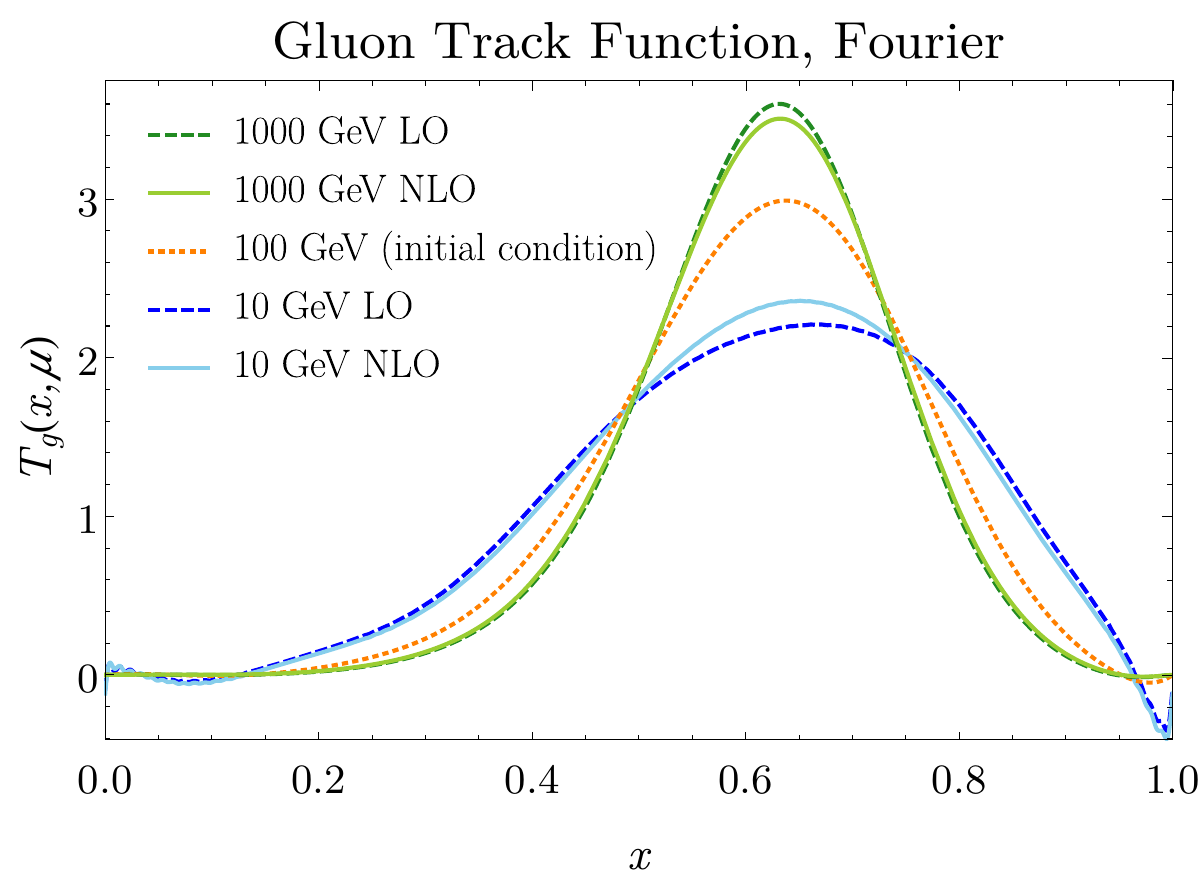}\label{fig:eflow_a}
}\quad
\subfloat[]{
\includegraphics[scale=0.35]{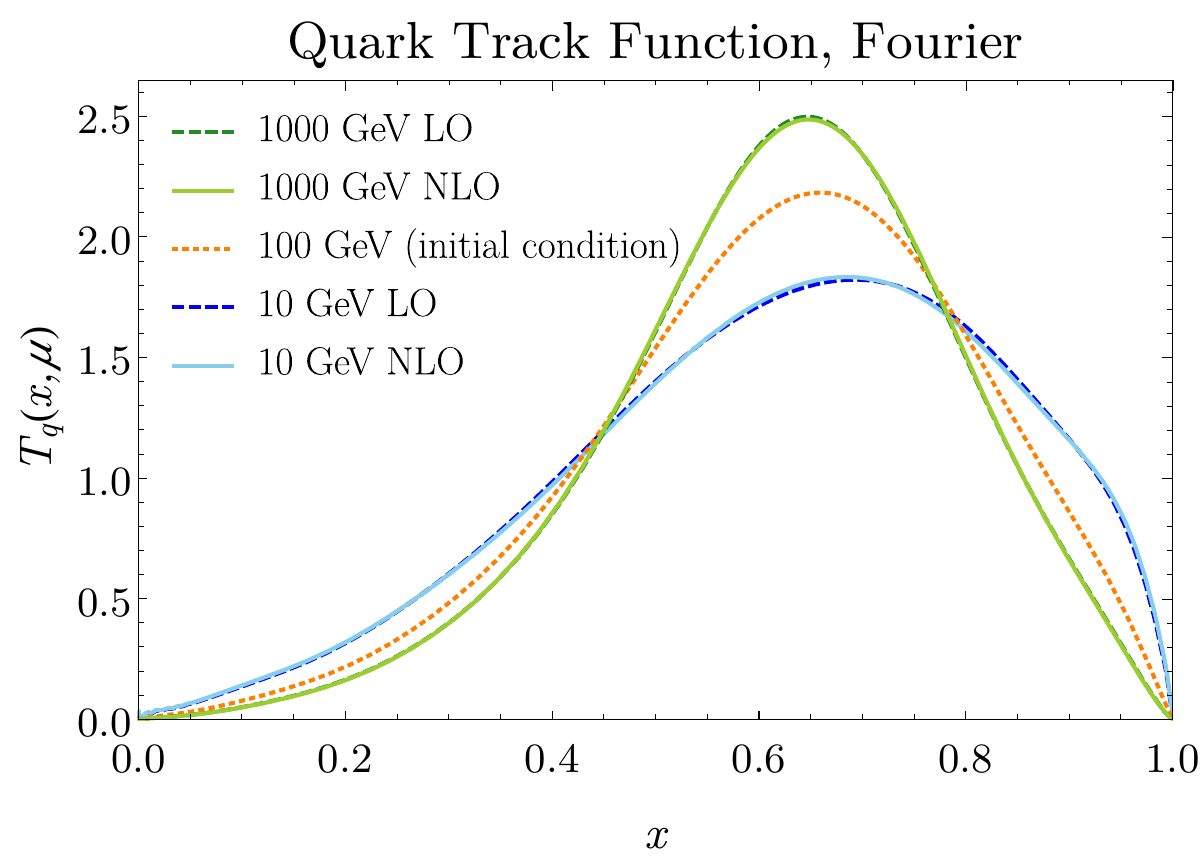}\label{fig:eflow_b}
}\qquad\\
\subfloat[]{
\includegraphics[scale=0.50]{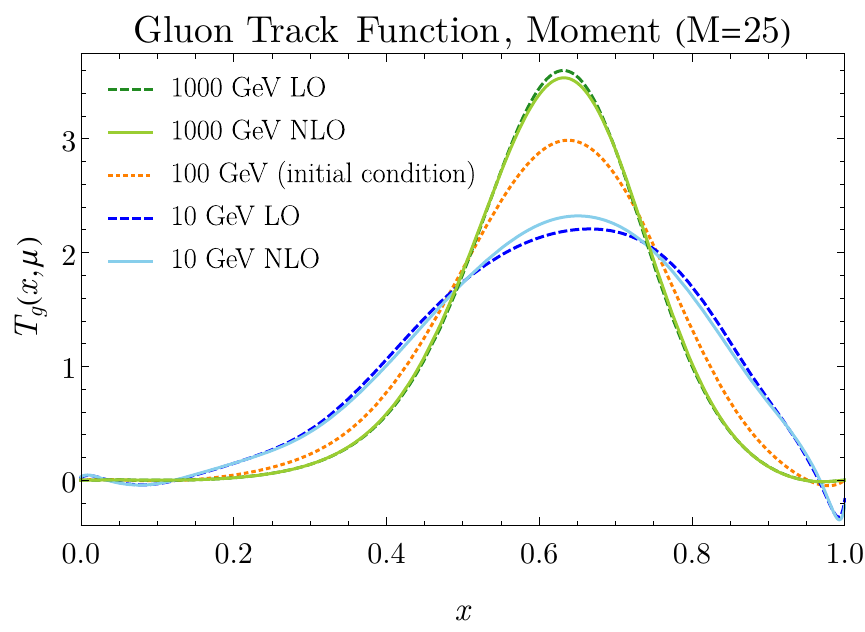}\label{fig:eflow_a}
}\quad
\subfloat[]{
\includegraphics[scale=0.50]{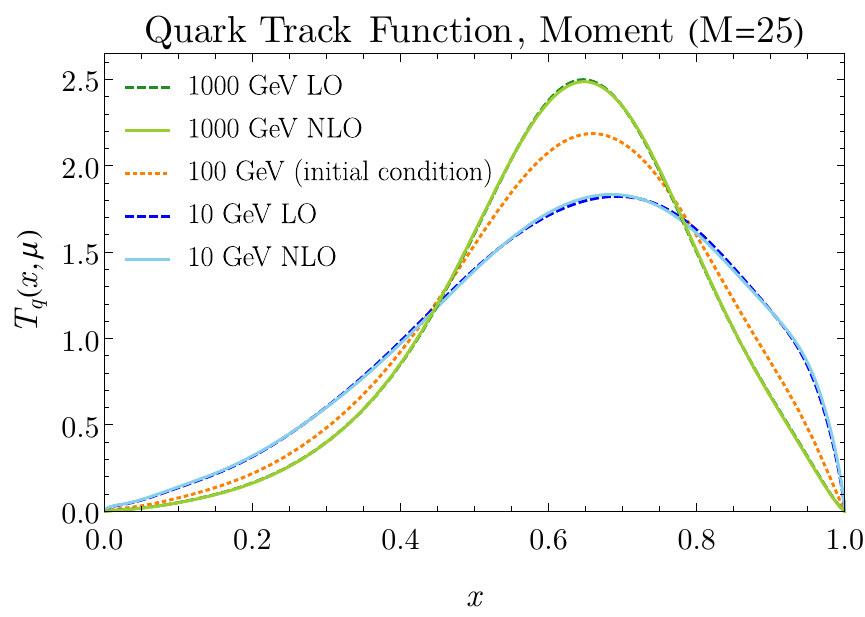}\label{fig:eflow_b}
}\qquad
\end{center}
\caption{The evolution of the quark and gluon track functions at NLO and LO, evolved using the Fourier transform approach (top row) and the moment approach (bottom row). The initial conditions are taken from the Pythia parton shower Monte Carlo~\cite{Chang:2013rca}. The Fourier series approach results are made with $M_1=100$ and $M_2=40$, while the moment method uses first $M=25$ moments.}
\label{fig:evolution_numerics}
\end{figure}
 
The moment method is applied on the basis of the Weierstrass approximation theorem 
that states that any real-valued continuous function defined on a real interval 
can be arbitrarily well approximated by a polynomial on that interval. 
Of course the finite number of moments we use to solve the evolution equations numerically
limits the degree of the polynomial, and thus how well it approximates the real $T_i(x,\mu_1)$. 
Naively, one can always add more moments, and more terms to the polynomial,
to obtain a better approximation to $T_i(x,\mu)$. 
However, in practice, we have found that beyond a certain point adding more moments don't necessarily improve the result: 
At higher degrees, the coefficients of the terms in the polynomial increase significantly and there are large cancellations between terms. Also, the errors of moment values are amplified when converted to the $x$-distribution, especially for large $N$. This can be seen from 
\begin{equation}
    T(N)+\delta T(N)=\int_0^1 \df x\, x^N [T(x) +\delta T(x) ]\,
\end{equation}
which shows that large errors at values of $x$ away from $x=1$ will lead to small errors of $\delta T(N)=\int_0^1\df x\, x^N\delta T(x)$ for large $N$.
Therefore, higher-degree polynomials require higher accuracy of the numerical solutions of track function moments, which makes the process of numerically solving the RGEs less efficient. Despite the effort to achieve that, the high-degree polynomials just take primary effect for the range near $x=1$, which is not surprising since large integer moments $N$ are dominated by the behavior of the track function near $x=1$, due to the $x^N$ factor. 

\begin{figure}
\begin{center}
\subfloat[]{
\includegraphics[scale=0.35]{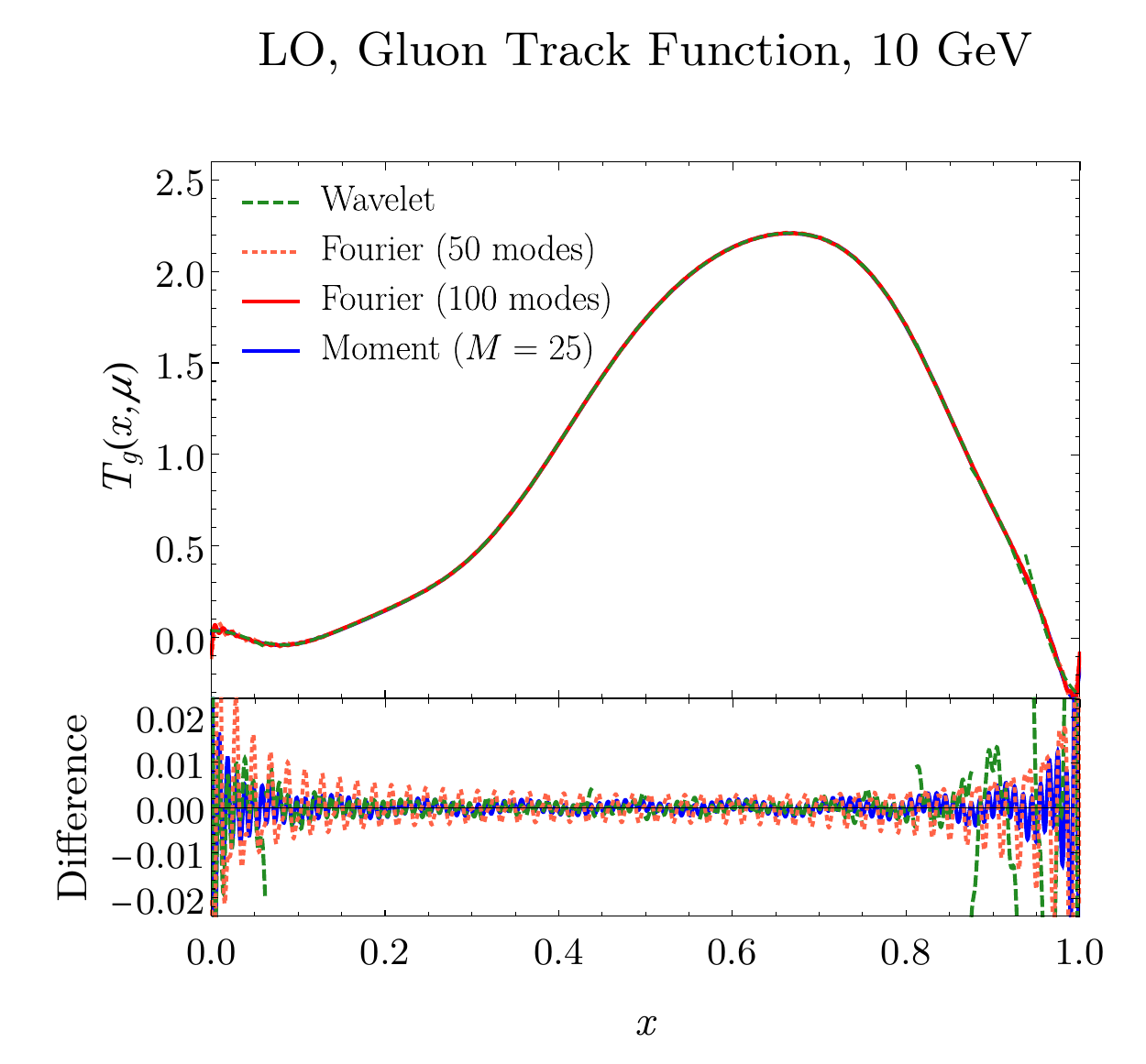}\label{fig:eflow_a}
}
\subfloat[]{
\includegraphics[scale=0.35]{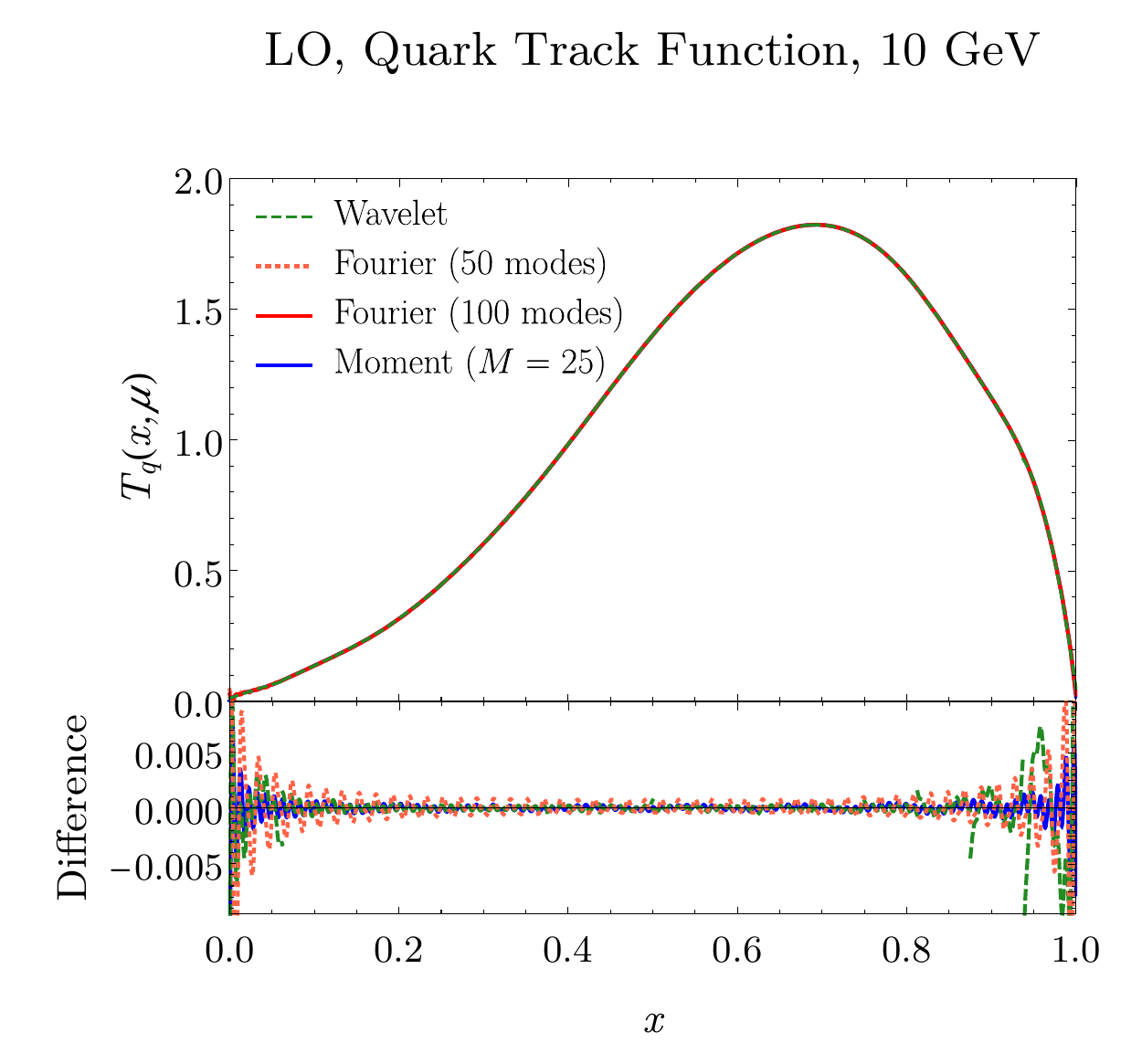}\label{fig:eflow_b}
}\qquad\\
\subfloat[]{
\includegraphics[scale=0.35]{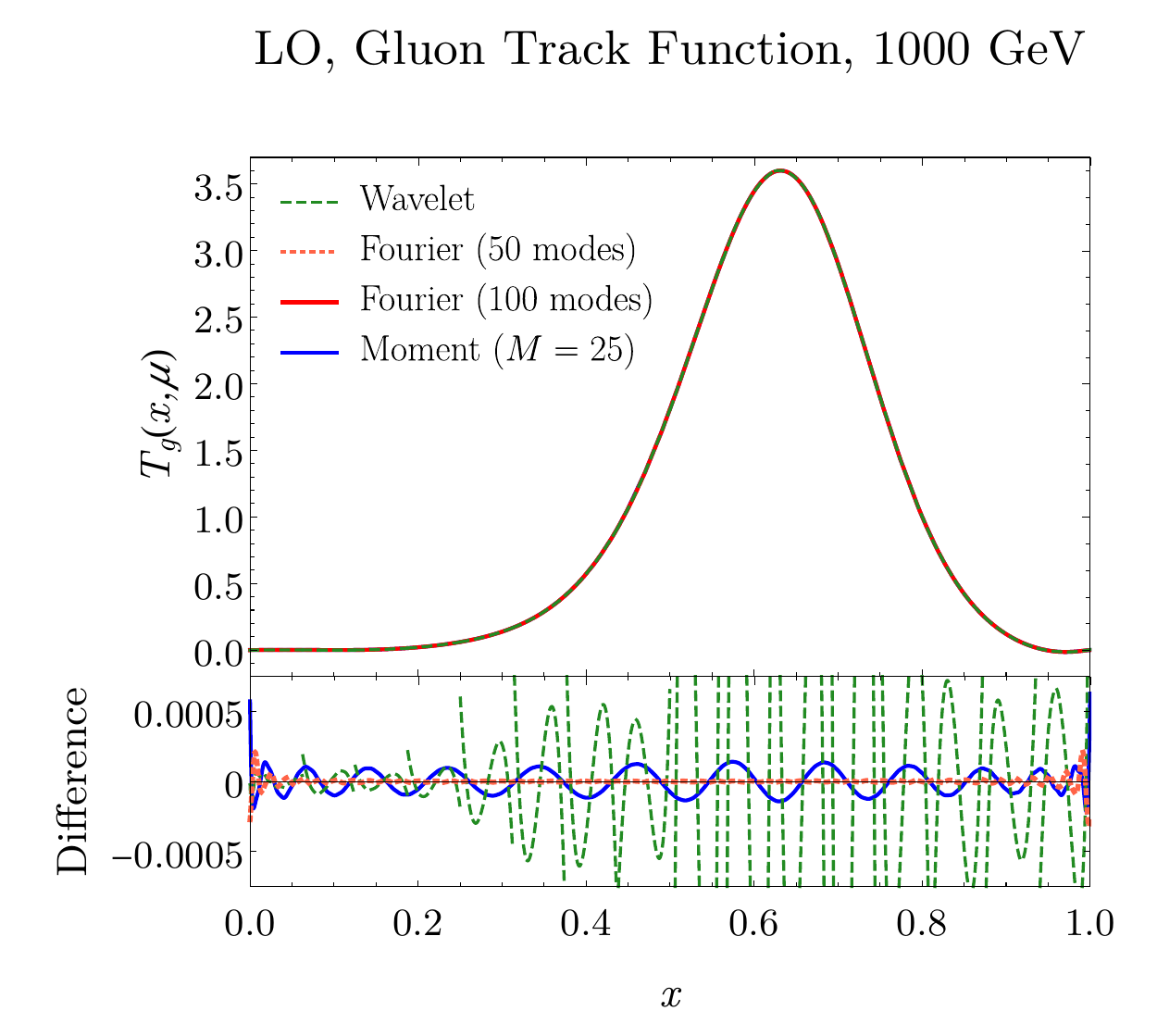}\label{fig:eflow_c}
}
\subfloat[]{
\includegraphics[scale=0.35]{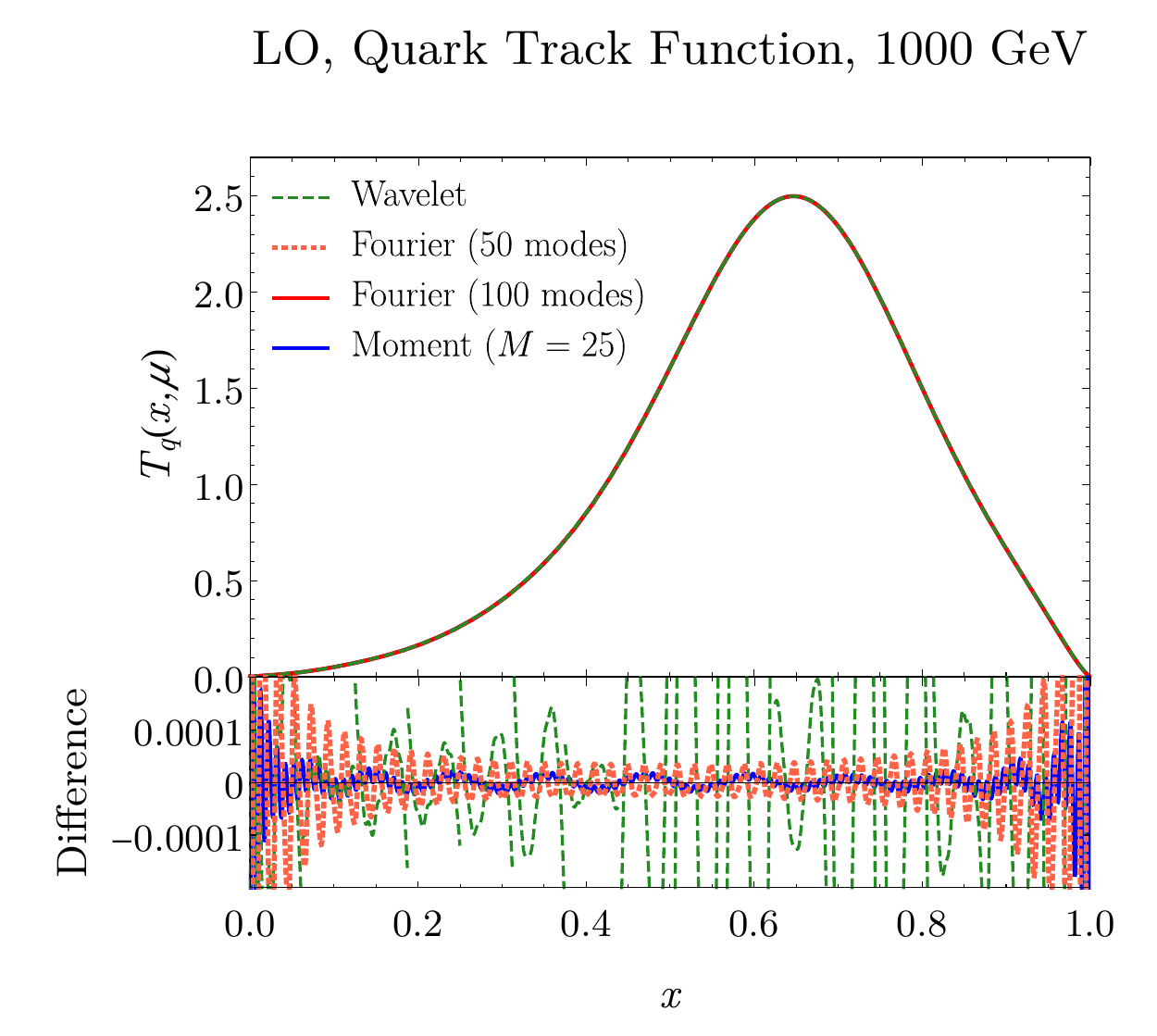}\label{fig:eflow_d}
}\qquad
\end{center}
\caption{A comparison of the different approaches to performing the LO track function evolution. As in \Fig{fig:evolution_numerics}, the boundary conditions are taken from the Pythia parton shower Monte Carlo. We uses 16 intervals and quadratic polynomials on each of them for wavelet approximation. Two Fourier series results correspond to 50 modes and 100 modes truncation respectively. The moment method employs first 25 moments. The difference plotted in the subfigure is computed relative to the Fourier result with 100 modes.}
\label{fig:evolution_numerics_compare}
\end{figure}

The Fourier series method doesn’t have the above disadvantage. 
The coefficients of the Fourier series are of the same order, 
which can be easily understood given the stable integral of $e^{2\pi \img\, m\, x}$.
Additionally, the coefficient is smaller for a harmonic with a larger integer index $m$. 
Both the Fourier series method and the moment method 
result in a smooth function (consisting of a finite number of Fourier modes or monomials) on the interval $0 \leq x \leq 1$ that approximates the track function at some scale.

\subsection{Numerical Results for Track Function Evolution}
\label{sec:numerics}

\begin{figure}
\begin{center}
\subfloat[]{
\includegraphics[scale=0.315]{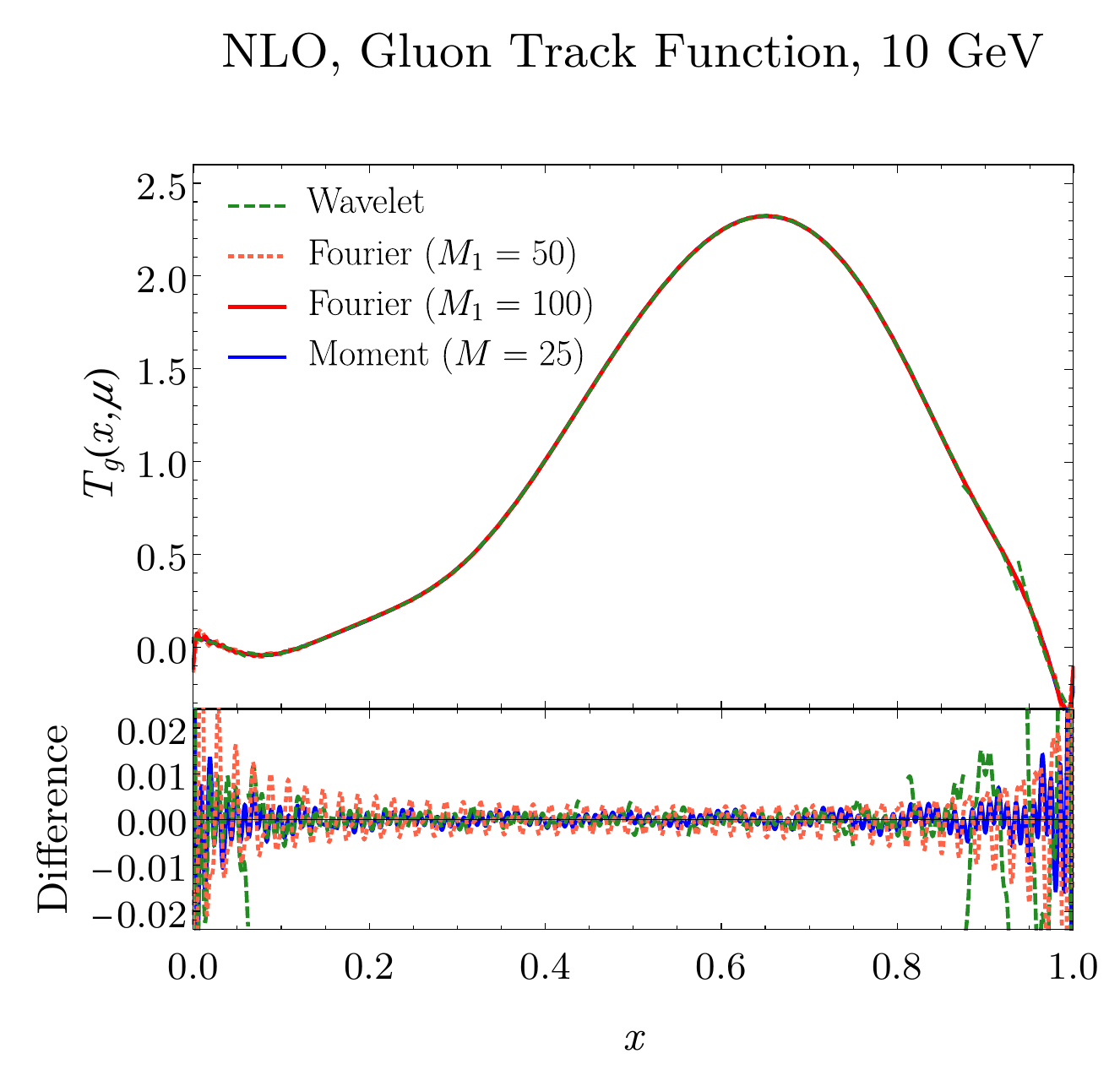}\label{fig:eflow_a}
}
\subfloat[]{
\includegraphics[scale=0.35]{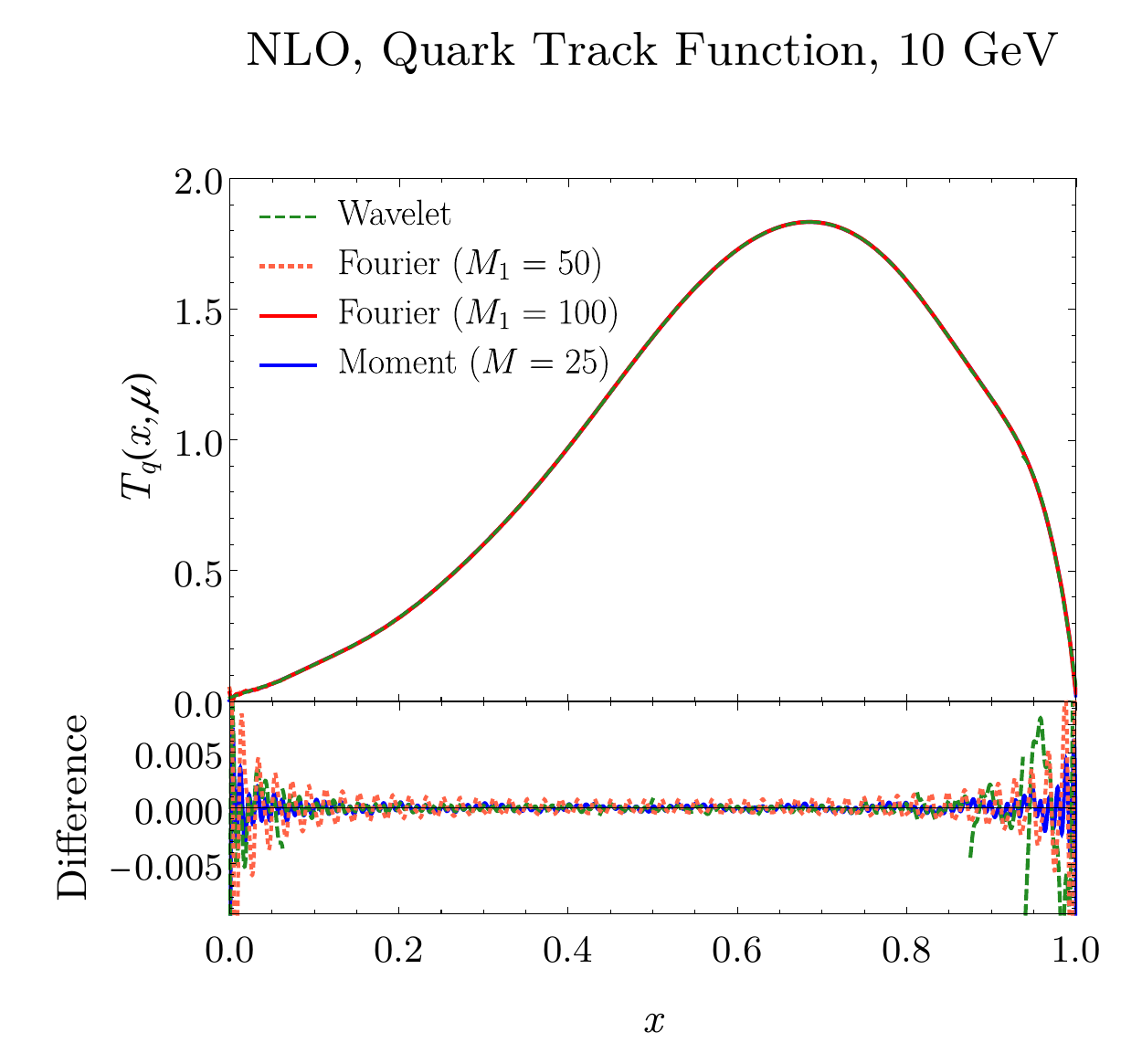}\label{fig:eflow_b}
}\qquad\\
\subfloat[]{
\includegraphics[scale=0.35]{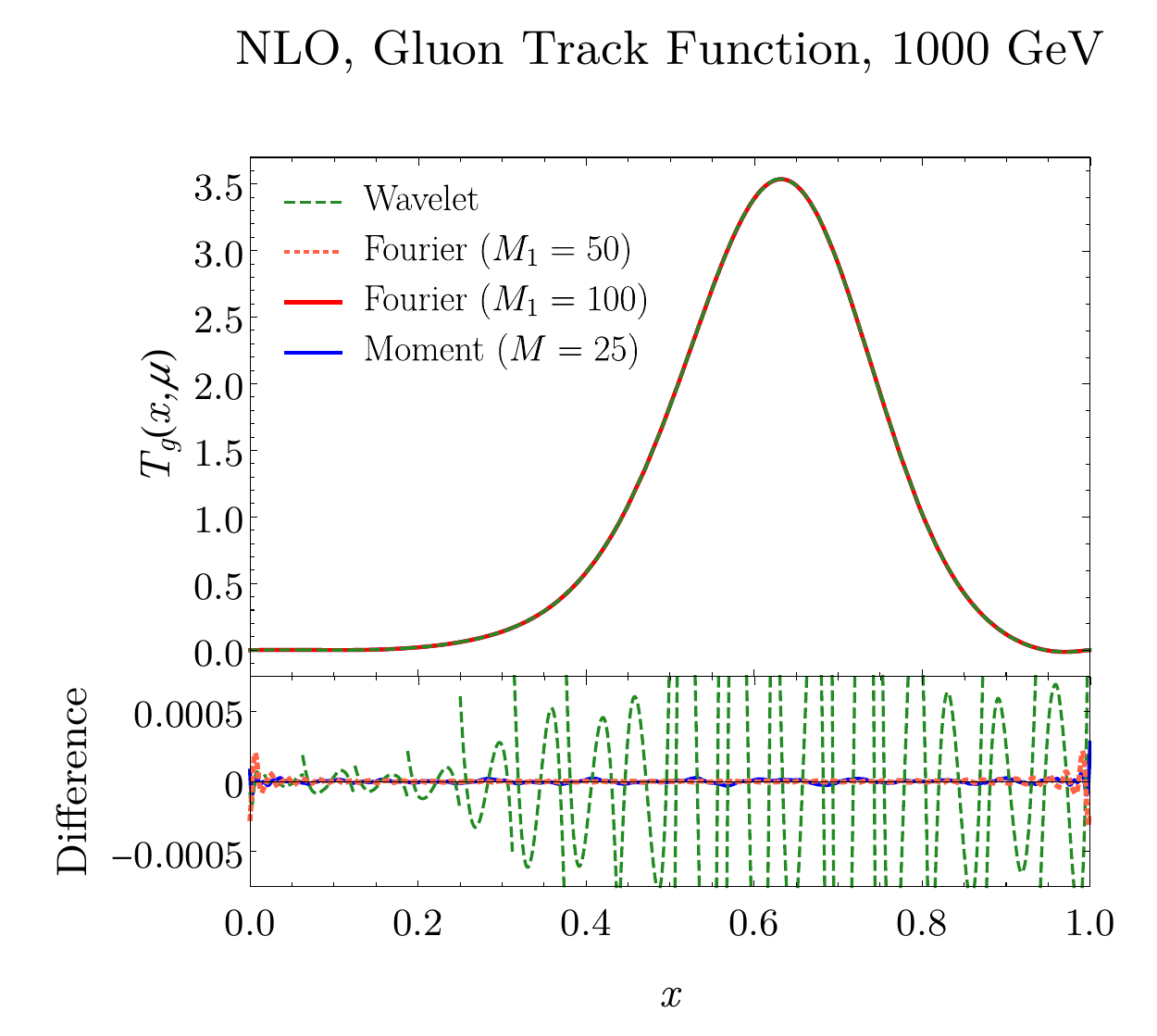}\label{fig:eflow_c}
}
\subfloat[]{
\includegraphics[scale=0.35]{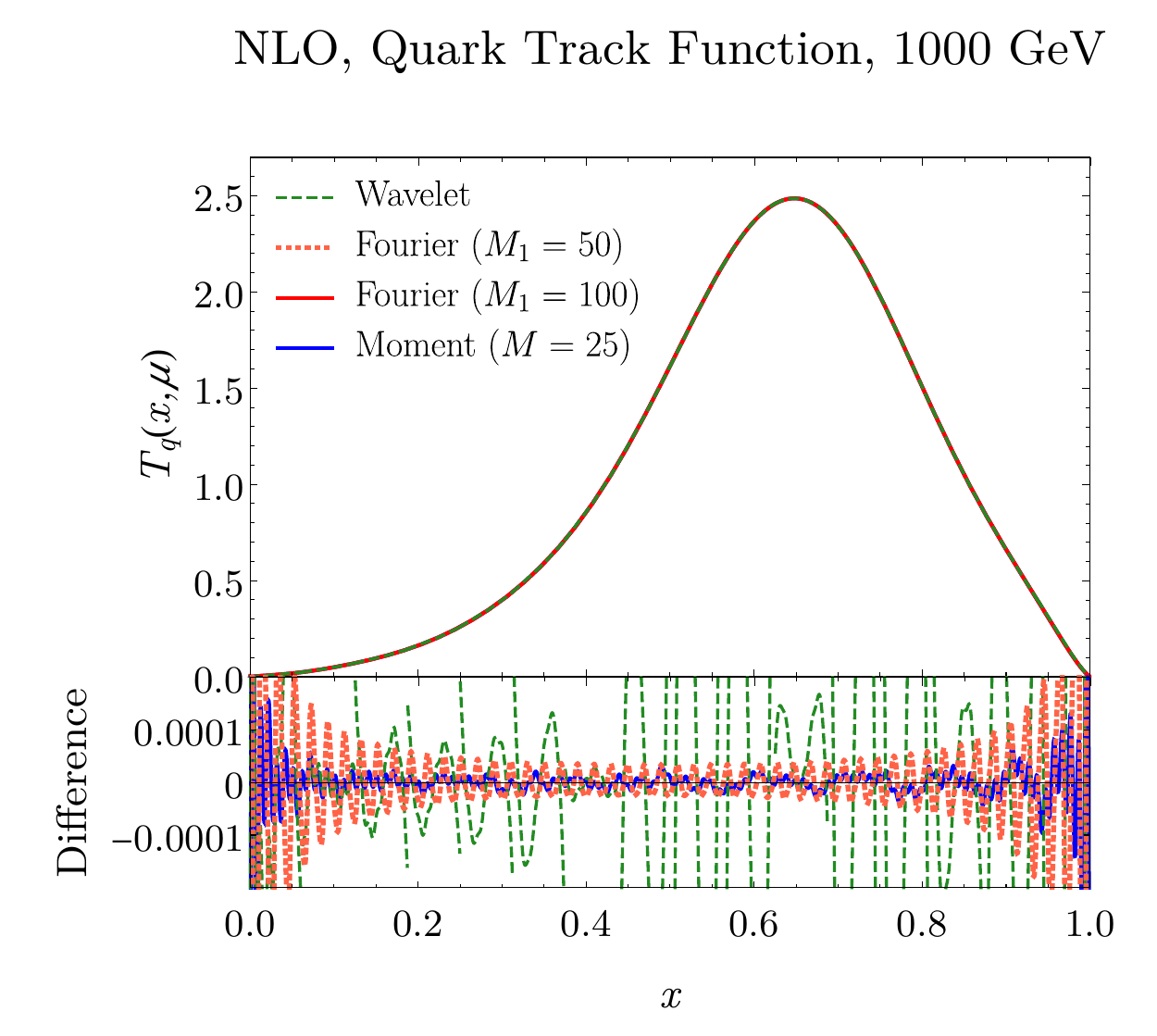}\label{fig:eflow_d}
}\qquad
\end{center}
\caption{A comparison of the different approaches to performing the NLO track function evolution. As in \Fig{fig:evolution_numerics}, the boundary conditions are taken from the Pythia parton shower Monte Carlo. The wavelet and moment methods here use the same parameters as that described in \Fig{fig:evolution_numerics_compare}. The Fourier series approach uses mixed truncation $M_1$ at LO and $M_2$ at NLO. Since we choose $M_2=40$ for all Fourier curves here, we make it implicit in the legend labels. The difference plotted in the subfigure is computed relative to the Fourier result with 100 modes.}
\label{fig:NLO_evolution_numerics_compare}
\end{figure}

Having developed a number of approaches for solving the evolution of the track functions, we now present some numerical results for the evolution of the track functions, and comparisons of the different approaches. Here we will focus only on the evolution of the track functions themselves. In a companion paper we have also used these track functions to compute a physical observable, namely the fraction of energy in charged hadrons in $e^+e^-\to$ hadrons.

In \Fig{fig:evolution_numerics}, we show the evolution of the quark and gluon track functions using an initial condition at 100 GeV obtained \cite{Chang:2013rca} from the Pythia parton shower Monte Carlo \cite{Sjostrand:2014zea}. For convenience, we assume that the quark track functions are flavor-independent, and we take an average of the 100 GeV track functions of the 5 quark flavors as the initial condition for $T_q(x,\mu)$. 
The top row shows the the evolution obtained using the Fourier approach and  the bottom row the moment method.
The effect of the NLO evolution is moderate in the peak region for gluon jets, and is small for quark jets. We emphasize that even if the numerical effect is relatively moderate, the NLO evolution is extremely important to be able to predict the structure of IR divergences in perturbative calculations involving tracks.

Next we investigate the differences between the various techniques in \Fig{fig:evolution_numerics_compare} and \Fig{fig:NLO_evolution_numerics_compare}, to ensure that the approximations used do not modify the final result. We compare the evolution performed using the wavelet, Fourier, and moment approaches, for both quark and gluon jets, using the same initial condition at 100 GeV. We see excellent agreement between all approaches when we evolve to either higher energy (1000 GeV), or lower energy (10 GeV). This provides a strong test of our numerical approaches.

\section{Conclusions}\label{sec:conclusions}

In this paper we have given an extended presentation of the derivation and solution of the NLO RG evolution equations for track functions. We derived the kernels in QCD, for both quark and gluon jets, as well as in $\cN=4$ SYM. We showed how one can incorporate triple collinear splitting functions in evolution equations, and how to systematically treat overlapping singularities occurring in the evolution variables. Our approach is systematic, and therefore in principle could be extended to higher orders, since all the required perturbative amplitudes are known  \cite{Badger:2004uk,Badger:2015cxa,Bern:2004cz,Czakon:2022fqi,Catani:2003vu,DelDuca:2019ggv,DelDuca:2020vst}.

Beyond the particular application to track function evolution considered here, as well as in our companion paper, our evolution equation can be viewed as a master equation for collinear evolution at NLO. In particular, we showed how one can integrate out information from the track function RGE, and obtain the DGLAP equations, as well as the RG evolution for multi-hadron fragmentation functions. For this reason, it would be interesting to understand some of the features of our equation more  systematically, such as the hints of uniform transcendentality in the kernels. 

We believe that our evolution equations takes a concrete step towards improving the description of the collinear dynamics of jets, and in particular, of understanding correlations in the shower/ fragmentation process and their RG evolution. There is significant current work to implement higher order corrections to parton showers, including $1\to3$ splitting functions \cite{Hoche:2017iem,Hoche:2017hno,Gellersen:2021eci}. These implementations are currently based on the standard DGLAP formalism. It would be interesting to understand if our $1\to 3$ evolution equation can used to constrain these implementations, or could itself be the basis of a parton shower.

With the prominent role that jets and their substructure are currently playing at the LHC and other colliders, we believe that our results lay the groundwork for an improved description of measurements involving tracks, as well an improved description of the collinear dynamics of jets more generally.

\acknowledgments
We thank Duff Neill for his exceptional knowledge of the literature.
H.C., Y.L.~and H.X.Z.~are supported by the National Natural Science Foundation of China under contract No.~11975200.
M.J.~is supported by NWO projectruimte 680-91-122. 
I.M.~is supported by start up funds from Yale University.
W.W.~is supported by the D-ITP consortium, a program of NWO that is funded by the Dutch Ministry of Education, Culture and Science (OCW).

\appendix

\section{Sector Decomposition and Plus Distributions}\label{sec:app}

In deriving the track function evolution kernels one needs to subtract terms that involve a double convolution. To subtract these terms one first needs to write these double convolutions as a single convolution. Given two kernels $f$ and $g$, the goal is to find an effective kernel $E[f,g]$ such that
\begin{align}\label{eq:start}
    f\otimes T_{i_1} \bigl(g\otimes T_{i_2} T_{i_3}\bigr)
    &=
    E[f,g]\otimes T_{i_1} T_{i_2} T_{i_3} \ ,
\end{align}
where possible $1\to2$ and $1\to1$ terms have been absorbed into a single $1\to3$ kernel. The convolution on the left-hand-side, in standard momentum fraction, is
\begin{align}
    f\otimes T_{i_1} \bigl(g\otimes T_{i_2} T_{i_3}\bigr)
    &=
    \int\! \df x_1\, \df x_2\, \df x_3\, T_{i_1}(x_1) T_{i_2}(x_2) T_{i_3}(x_3)
    \\
    &\qquad\times
    \int_0^1\! \df v\, \df w \,
    \delta\bigl[x-v x_1-(1-v)w x_2-(1-v)(1-w)x_3\bigr] \ 
    f(v) g(w) .\nn
\end{align}
The convolution on the right-hand side is written in sector-decomposed coordinates as follows,
\begin{align}
    E[f,g]\otimes T_{i_1} T_{i_2} T_{i_3}
    &=
    \sum_n 
    \int\! \rmd x_1\, \rmd x_2\, \rmd x_3\, T_{i_1}(x_1) T_{i_2}(x_2) T_{i_3}(x_3)
    \\
    &\qquad\times
    \int_0^1\! \rmd z\, \rmd t\,
    \delta\bigl[x-{^n}\!z_1 x_1-{^n}\!z_2 x_2-{^n}\!z_3 x_3\bigr]\,{^n}\!E(z,t)\,,
    \nonumber
\end{align}
where the sum on $n$ runs over sectors.

\begin{figure}
\begin{center}
\includegraphics[scale=0.27]{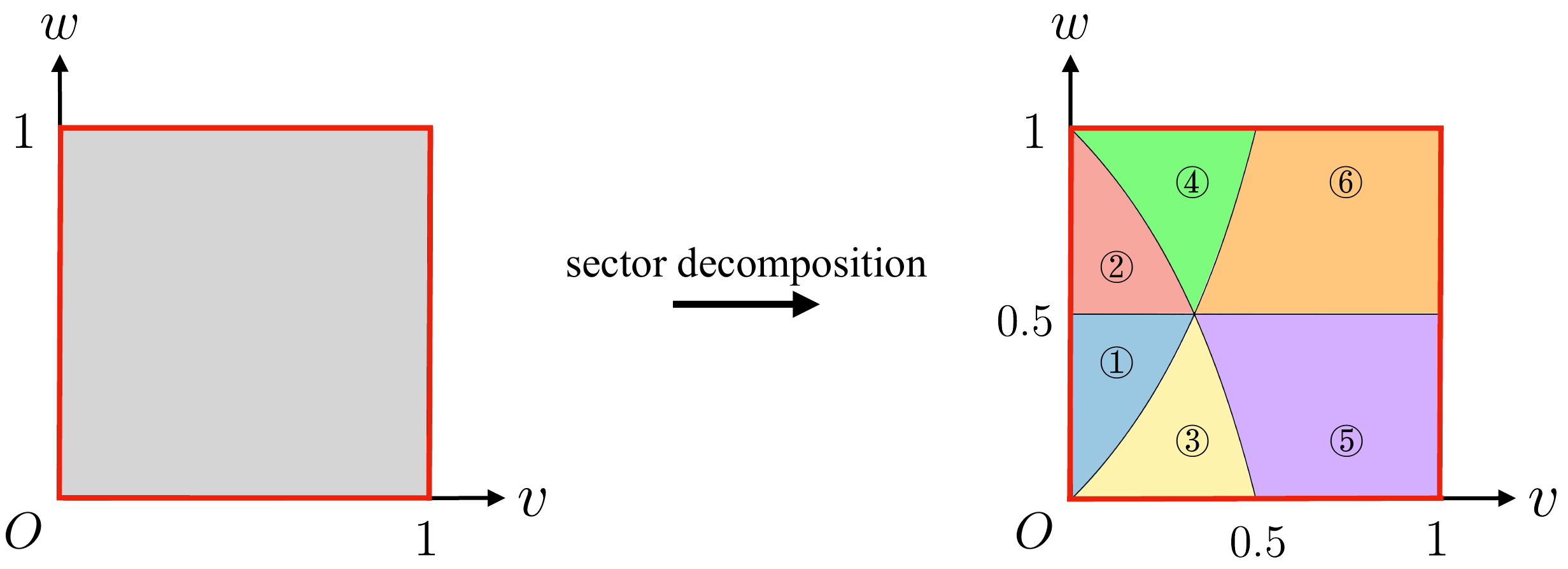}
\end{center}
\caption{The two-dimensional $(v,w)$ space for a 3-particle final state is divided into six sectors. Each maps into a unit square after the coordinate transformations listed in Table~\ref{tab:app_tab}. Red edges represent boundaries of phase space where soft divergences can be present.}
\label{fig:sector_decomposition_appx}
\end{figure}

To apply sector decomposition, we first identify $z_1$, $z_2$ and $z_3$ with  respectively $v$, $(1-v)w$ and $(1-v)(1-w)$. An overview of the change of coordinates in the different sectors is provided in Table~\ref{tab:app_tab}, which is illustrated in Fig.~\ref{fig:sector_decomposition_appx} as well. Obtaining $E[f,g]$, however, is not as trivial as simply changing coordinates as this would lead to multi-variable plus distributions. 

\renewcommand{\arraystretch}{2.0}
\begin{table}
\centering
\begin{tabular}{|c|c|c|c|}
\hline
\multicolumn{1}{|c|}{\textbf{sector}} & \multicolumn{1}{c|}{\textbf{ordering}} & \multicolumn{1}{c|}{${^n}\!v(z,t)$} & \multicolumn{1}{c|}{${^n}\!w(z,t)$} \\ \hline\hline
sector 1                              & $z_1<z_2<z_3$                          & $\frac{z t}{1+z+z t}$  & $\frac{z}{1+z}$        \\ \hline
sector 2                              & $z_1<z_3<z_2$                          & $\frac{z t}{1+z+z t}$  & $\frac{1}{1+z}$        \\ \hline
sector 3                              & $z_2<z_1<z_3$                          & $\frac{z}{1+z+z t}$    & $\frac{z t}{1+z t}$        \\ \hline
sector 4                              & $z_3<z_1<z_2$                          & $\frac{z}{1+z+z t}$    & $\frac{1}{1+z t}$        \\ \hline
sector 5                              & $z_2<z_3<z_1$                          & $\frac{1}{1+z+z t}$    & $\frac{t}{1+t}$        \\ \hline
sector 6                              & $z_3<z_2<z_1$                          & $\frac{1}{1+z+z t}$    & $\frac{1}{1+t}$        \\ \hline
\end{tabular}
\caption{An overview of the change of coordinates in the different sectors.}
\label{tab:app_tab}
\end{table}
\renewcommand{\arraystretch}{1.0}

Instead of directly applying a change of coordinates, we will construct $E[f,g]$ by a matching procedure. In this procedure, we require that both sides of Eq. \eqref{eq:start} agree when integrated over a specific region $\mathcal{R}$ of the integration space,
\begin{align}\label{eq:regions}
    \int_{\mathcal{R}} \df v \df w \ f(v) g(w)
    =\sum_{n=1}^6\int_{{^n}\!\mathcal{R}} \df z \df t \ {^n}\!E(z,t) \ .
\end{align}
Here, ${^n}\!\mathcal{R}$ is the region in $(z,t)$-space that $\mathcal{R}$ maps to in sector $n$. The above restriction ensures that one obtains the correct behaviour around singular points.

Before we proceed to carry out the matching procedure, we write the kernels $f(v)$ and $g(w)$ in the general form
\begin{align}
    f(v)
    &=
    \tilde{f}(v) 
    +\sum_{p=0}^\infty \Bigl({_0}f_p\mathcal{L}_p(v) 
        + {_1}f_p\mathcal{L}_p(1-v)\Bigr)
    +{_0}f_{-1}\delta(v)
    +{_0}f_{-1}\delta(1-v) \ ,
\end{align}
with a similar expression for $g(w)$. Here we take $\tilde{f}(v)$ to be an integrable function that is at most logarithmically divergent as $v\to0$ or $v\to1$. All $1\to2$ kernels considered in this work can be written in this form. Similarly, any kernel in $(z,t)$-space can be written as
\begin{align}
    {^n}\!E(z,t)
    &=
    {^n}\!\tilde{E}(z,t)
    +\sum_{p=0}^\infty\Bigl({^n}\!\tilde{E}^z_p(t) \mathcal{L}_p(z)
        +{^n}\!\tilde{E}^t_p(z) \mathcal{L}_p(t)\Bigr)
    +\sum_{p=0}^\infty \sum_{q=0}^\infty 
        {^n}\!E_{p q}\mathcal{L}_p(z)\mathcal{L}_q(t)
    \nonumber
    \\
    &\qquad
    +\biggl({^n}\!\tilde{E}^t_{-1}(z) 
        +\sum_{p=0}^\infty {^n}\!E_{p,-1} \mathcal{L}_p(z)\biggr) \delta(t)
    +{^n}\!E_{-1,-1}\delta(z)\delta(t) \ .
\end{align}
Again, the notation $\tilde{E}$ implies an integrable function that is at most logarithmically divergent as its arguments go to $0$. Additionally, due to the form of the delta function in the convolution, the $\delta(z)$ term needs no further $t$-dependence.

Let us now carry out the matching procedure. First we consider a region away from the boundaries, such that all delta functions in $f$ and $g$ can be discarded. Translating to $(z,t)$-space, the conditions $0<v<1$ and $0<w<1$ automatically lead to $z>0$ and $t>0$, such that all delta functions in $E[f,g]$ can be ignored as well. As now there are no long plus distributions in the integral, we can safely change variables to obtain four of the seven coefficient functions,
\begin{align}\label{eq:Ebody}
    &{^n}\!\tilde{E}(z,t) 
    +\sum_{p=0}^\infty{^n}\!\tilde{E}^z_p(t)\frac{\log^p z}{z}
    +\sum_{q=0}^\infty{^n}\!\tilde{E}^t_q(z)\frac{\log^q t}{t}
    +\sum_{p=0}^\infty\sum_{q=0}^\infty
        {^n}\!E_{pq} \frac{\log^p z}{z} \frac{\log^q t}{t}
    \\
    &\qquad=
    \frac{z}{(1+z+z t)^3} \frac{1}{1-{^n}\!v} f({^n}\!v) g({^n}\!w) 
    \qquad
    \text{for} \ z,t>0, \ .
    \nonumber
\end{align}
The individual coefficient functions can then be found by expanding the right-hand-side in $z$ and $t$ and matching the coefficients of the pieces that diverge as $z\to0$ and $t\to0$.

Next, let us derive ${^n}\!\tilde{E}^t_{-1}(z)$ and ${^n}\!E_{p,-1}$, the coefficient functions that multiply $\delta(t)$. To derive these functions we need to choose the region of integration such that $\delta(t)$ contributes to the integral. Moreover, we wish to find these coefficient functions for each sector separately, and so we must construct six regions of integration that are each restricted to one. The six regions of integration that satisfy both points are
\begin{alignat}{2}
    R_1:\quad & 0<w<\epsilon    &&,\qquad v_1<v<v_1'       \ ,\nn\\
    R_2:\quad & 1-\epsilon<w<1  &&,\qquad v_2<v<v_2'       \ ,\nn\\
    R_3:\quad & w_3<w<w_3'      &&,\qquad 0<v<\epsilon     \ ,\nn\\
    R_4:\quad & w_4<w<w_4'      &&,\qquad 1-\epsilon<v<1   \ ,\nn\\
    R_5:\quad & w_5<w<w_5'      &&,\qquad 0<v<\epsilon     \ ,\nn\\
    R_6:\quad & w_6<w<w_6'      &&,\qquad 1-\epsilon<v<1   \ ,\nn
\end{alignat}
where $\epsilon$ is taken to be infinitesimally small such that the contributions of regular functions can be discarded. Additionally, the bounds on the regions are such that they lie within a single sector. The condition in Eq. \eqref{eq:regions} then leads to
\begin{align}
    {^1}\!\tilde{E}^t_{-1}(z) + {^1}\!E_{p,-1} \frac{\log^p z}{z}
    &=\biggl\{{_0}f_{-1}
    +\sum_{p=0}^\infty {_0}f_p \frac{1}{1+p}\log^{1+p}\Bigl(\frac{z}{1+z}\Bigr)
    \biggr\}\frac{1}{(1+z)^2}g\Bigl(\frac{z}{1+z}\Bigr)\ ,
    \\&\nn\\
    {^2}\!\tilde{E}^t_{-1}(z) + {^2}\!E_{p,-1} \frac{\log^p z}{z}
    &=\biggl\{{_0}f_{-1}
    +\sum_{p=0}^\infty {_0}f_p \frac{1}{1+p}\log^{1+p}\Bigl(\frac{z}{1+z}\Bigr)
    \biggr\}\frac{1}{(1+z)^2}g\Bigl(\frac{1}{1+z}\Bigr)\ ,
    \\&\nn\\
    {^3}\!\tilde{E}^t_{-1}(z) + {^3}\!E_{p,-1} \frac{\log^p z}{z}
    &=\biggl\{{_0}g_{-1}+\sum_{k=0}^\infty {_0}g_k \frac{\log^{1+k} z}{1+k}\biggr\}
    \frac{1}{(1+z)^2}f\Bigl(\frac{z}{1+z}\Bigr)\ ,
    \\&\nn\\
    {^4}\!\tilde{E}^t_{-1}(z) + {^4}\!E_{p,-1} \frac{\log^p z}{z}
    &=\biggl\{{_1}g_{-1}+\sum_{k=0}^\infty {_1}g_k \frac{\log^{1+k} z}{1+k}\biggr\}
    \frac{1}{(1+z)^2}f\Bigl(\frac{z}{1+z}\Bigr)\ ,
    \\&\nn\\
    {^5}\!\tilde{E}^t_{-1}(z) + {^5}\!E_{p,-1} \frac{\log^p z}{z}
    &={_0}g_{-1}\frac{1}{(1+z)^2}f\Bigl(\frac{1}{1+z}\Bigr)\ ,
    \\&\nn\\
    {^6}\!\tilde{E}^t_{-1}(z) + {^6}\!E_{p,-1} \frac{\log^p z}{z}
    &={_1}g_{-1}\frac{1}{(1+z)^2}f\Bigl(\frac{1}{1+z}\Bigr)\ ,
\end{align}
which holds for $z>0$. Again, the individual coefficient functions can be solved for by expanding the right-hand-side in $z$ and matching the coefficients of the pieces that diverge as $z\to0$.

Finally, we need to derive an expression for ${^n}\!E_{-1,-1}$, which is the coefficient of the $\delta(z)\delta(t)$ term. To extract this term, we choose the region to be the sector itself, resulting in
\begin{align}
    {^n}\!E_{-1,-1}&=\int_{s_n} \df v \df w \ f(v) g(w)
    -\int_0^1 \df z \df t \ {^{1+3}}\tilde{E}(z,t)
    -\int_0^1 \df z \ {^{1+3}}\tilde{E}^t_{-1}(z) \ .
\end{align}
It should be noted that one should be careful about the points $(v,w)=(0,0)$ and $(v,w)=(0,1)$, as the border of some sector cuts this point. This can be solved by simply considering the combined total of the two sectors. For example, the point $(0,0)$ lies on the border of sector $1$ and $3$, and so instead of considering ${^1}\!E_{-1,-1}$ and ${^3}\!E_{-1,-1}$ separately one considers only the sum ${^{1+3}}\!E_{-1,-1}$. This does not lead to any trouble in the evolution equation, as both these terms accompany the same combination of track functions.

As a cross check on this method, we also considered matching the moments of Eq. \eqref{eq:start}. In this case, one starts by requiring that the moments of the two sides agree,
\begin{align}\label{eq:matchmoments}
    \bigl\{\mathcal{M} f\otimes T_{i_1} (g\otimes T_{i_2} T_{i3})\bigr\}(N+1)
    =
    \bigl\{\mathcal{M} E\otimes T_{i_1} T_{i_2} T_{i3}\bigr\}(N+1)
    \qquad
    \forall \ N\in\mathbb{N}
    \ ,
\end{align}
with $\{\mathcal{M}\dotso\}$ denoting the Mellin moment. Treating all moments of the track functions as independent variables, one arrives at the following constraint
\begin{align}\label{eq:tosolve}
    \sum_n\int_0^1 \df z \df t \ 
    {^n}\!z_1^{k_1}\, {^n}\!z_2^{k_2}\, {^n}\!z_3^{k_3} \ {^n}\!E(z,t)
    =\int_0^1 \df v \df w\ v^{k_1} (1-v)^{N-k_1} w^{k_2} (1-w)^{k_3} \ 
    f(v) g(w) \ ,
\end{align}
for all $k_1+k_2+k_3=N$. We have checked that this condition is satisfied for all $k_1+k_2+k_3=N$ up to $N=30$ for all the kernels that are considered in this work. Although in principle possible, we have not been able to solve the above equation exactly as non-trivial changes of variables need to be found.


\bibliographystyle{JHEP}
\bibliography{spinning_gluon.bib}

\end{document}